\documentclass{article}
\usepackage{LaThuileFPSpro}
\usepackage{epsfig}
\begin{document}
\title{ An introduction to  Cosmic Rays and 
 Gamma-Ray Bursts,\\ and to their simple understanding}
\author{
Alvaro De R\'ujula    \\
  {\em CERN \& Boston University} \\
 }
\maketitle

\baselineskip=11.6pt

\begin{abstract}
I review the subjects of non-solar cosmic rays (CRs) and long-duration gamma-ray 
bursts (GRBs). Of the various interpretations of these phenomena, the one 
best supported by the data is the following.
  Accreting compact objects, such as black holes, are seen to emit relativistic 
 puffs of plasma: `cannonballs' (CBs).  The inner domain of a rotating star 
 whose core has collapsed resembles such an accreting system. This suggests 
 that core-collapse supernovae (SNe) emit CBs, as
 SN1987A did.  The fate of a CB as it exits a SN and travels in space
 can be studied as a function of the CB's mass and  energy,
  and of `ambient' properties: the encountered matter- and light- distributions,
 the composition of the former, and the location of intelligent observers. 
  The latter may conclude that the interactions of CBs with ambient matter and light
generate CRs and GRBs, all of whose
properties can be described by this  `CB model'
 with few parameters and simple physics. GRB data are still being taken in  
 unscrutinized  domains of energy and timing. They agree accurately
 with the model's predictions. 
 CR data are centenary. Their precision will improve, but new
striking predictions are unlikely. 
Yet, a one-free-parameter description of all CR data works very well. This is a bit as if 
 one discovered QED today and only needed to fit $\alpha$.
\end{abstract}
%
%

\section{Introduction}
This is a version of an introductory talk to high-energy physicists. 
Cosmic rays (CRs) were the first item in their field,
and will remain the energy record-breakers for the foreseeable 
future. I shall argue that nothing `besides the standard model'
is required to understand CRs of any energy, subtracting from
their interest. `Long' gamma-ray bursts (GRBs) are flashes of
mainly sub-MeV photons, originating in supernova (SN) explosions. 
The $\gamma$-rays are highly collimated. Hence, 
GRBs are not the publicized `highest-energy explosions after the 
big bang', but more modest torches occasionally pointing to
the observer. GRBs are of interest because their understanding
is intimately related to that of CRs. It might have been more 
precise to say `my understanding' of GRBs and CRs,
for the work of my coauthors and I is viewed as
unorthodox.

What a start! I have already admitted that our stance is
not trendy and that the subject is of no post-standard  interest.
But our claims are based on clear hypothesis, which may
be proven wrong, and very basic physics, which is precise enough,
very pretty, understandable to undergraduates, and successful.

The information about GRBs and CRs
is overwhelming. GRBs are known since the late 60's and
CRs since 1912. Surprisingly, no theories have
arisen that are both accepted (`standard')
and acceptable (transparent, predictive and successful).
I cannot refer to a representative subset of
the $\sim\!70\!+\!70$ kilo-papers on CRs and GRBs. For reviews
of the standard views on CRs, see e.g.~Hillas\cite{Hillas} or Hoerandel\cite{Hoer2}.
For the accepted `fireball' model of GRBs see e.g.~Meszaros\cite{Mes} or Piran\cite{Pir}.
Fewer self-citations and
many more references, particularly to data, appear in 
DDD02\cite{AGoptical}, DDD03\cite{AGradio}, DD04\cite{DDGRB}and DD06\cite{DDCR}.

\section{Most of what you may want to recall about Cosmic Rays}

\begin{figure}
\vspace {-0.4cm}
\begin{center}
\hbox{\epsfig{file=test-plot.epsf,width=7.4cm,height=5.2cm}
\epsfig{file=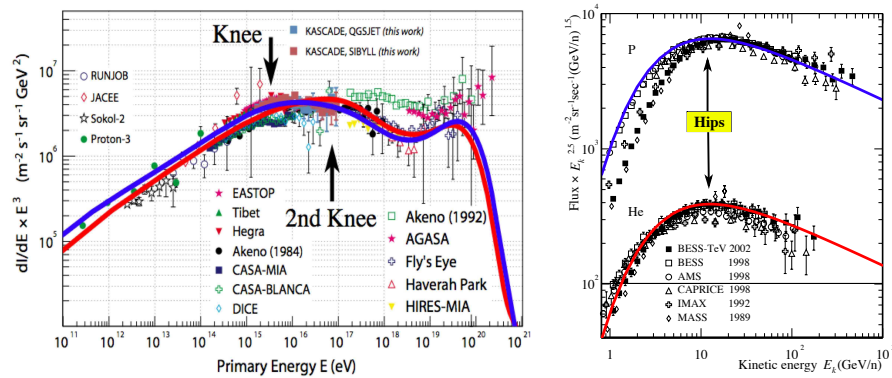,width=4.8cm}}
\end{center}
\vspace{-1.2cm}
\caption{{\bf Left (a)} The weighed all-particle CR spectrum $E^3\,dF/dE$. 
Some of the high-energy data  disagree with others.
{\bf Right (b)} The low-energy H an He flux versus the kinetic energy per nucleon,
multiplied by $E_{\rm kin}^{2.5}$. In (a) the lines are two extreme CB-model
predictions. In (b) both lines are predictions. The data coincide best
with the prediction at solar minimum. The normalizations, shown here to coincide 
with the data, are predicted to within a factor of $\sim\!3$.
}
\label{AllPart}
\end{figure}

In CR physics `all-particle' refers to nuclei: all charged CRs
but electrons.
The CR spectra being fairly featureless, it
is customary to weigh them with powers of energy, to
over-emphasize their features. The $E^3\,dF/dE$ 
all-particle spectrum is shown 
in fig.~\ref{AllPart}a, not updated for recent data at the 
high-energy tail. At less than TeV energies  the CR flux is larger than
$\rm 1\,m^{-2}\,s^{-1}\,sr^{-1}$ and it is possible to measure
the charge $Z$ and mass number $A$ of individual particles with, e.g., a
magnetic spectrometer in a balloon, or in orbit. Some low-energy
results for H and He are shown in fig.~\ref{AllPart}b.
They vary with solar activity.

\begin{figure}
\vspace {-.4cm}
\begin{center}
\hbox
{\epsfig{file=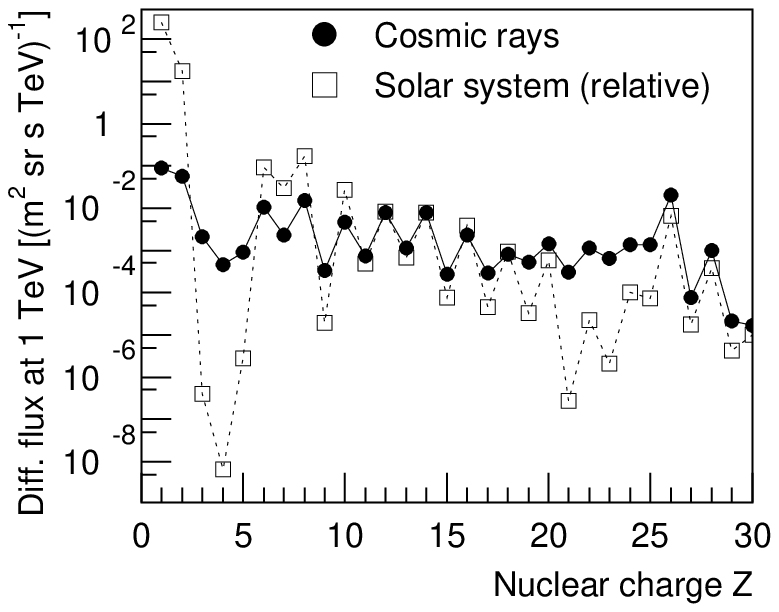,width=7.1cm,height=4.8cm}
\epsfig{file=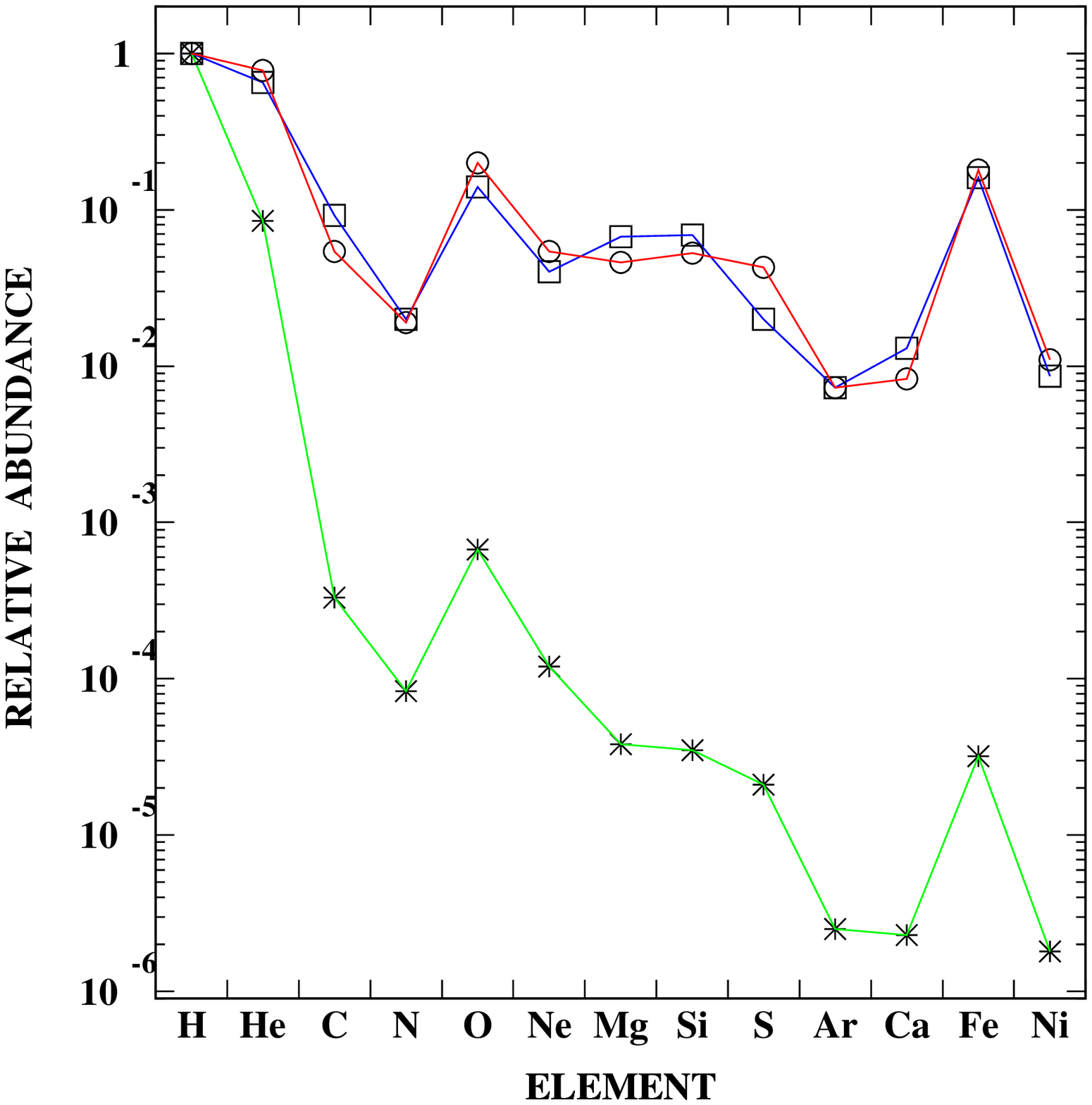, width=5.3cm,angle=0}}
\end{center}
\vspace{-1.2cm}
\caption{ The CR fluxes at 1 TeV and the relative abundances
in the ISM. {\bf Left (a)} Fluxes of elements up to Z and abundances
in the solar ISM.
{\bf Right (b)} Solar ISM abundances (stars), CR fluxes
(squares) of the primary CRs, and the corresponding CB-model predictions
(circles), all normalized to H.
}
\label{abundances}
\end{figure}

The CR fluxes of the lightest 30 elements at $E\!=\!1$ TeV
(of a nucleus, not per-nucleon) are shown in fig.~\ref{abundances}a,
and compared with the relative abundances in the interstellar medium
(ISM) of the solar neighbourhood. Elements such as Li, Be and B
are relatively enhanced in the CRs, they result from 
collisional fragmentation of heavier and abundant `primaries' such as
C, N and O. Otherwise, the solar-ISM and CR $Z$-distributions are akin,
but for H and He. In fig.~\ref{abundances}b the abundances relative
to H of the CR primary elements up to Ni ($Z\!=\!28$) are plotted as
blue squares. The stars are solar ISM abundances. 
CR positrons and antiprotons attract attention
as putative dark matter products, but it is nearly
impossible to prove that their fluxes are not entirely secondary.

The Galaxy has a complex magnetic-field structure with 
$B_{_{\rm G}}\!=\!{\cal{O}}(1\,\rm \mu G)$ 
and coherent domains ranging in size up to $\sim\!1$ kpc,
$\sim\!1/8$ of our distance to the Galactic center. In such a field, a nucleus
of $E\!\simeq\!p\!>\!Z\,(3\!\times\!10^9)$ GeV would hardly be deflected. 
For $Z\!=\!1$, this energy happens
to be the `ankle' energy, at which the flux of fig.~\ref{AllPart} bends up.
CRs originating within the Galaxy and having $E\!>\!E_{\rm ankle}$ would
escape practically unhindered. The CR flux does not bend {\it down} at that
energy, thus the generally agreed conclusion that CRs above the ankle are
mainly extragalactic. CRs of Galactic origin and $E\!<\!E_{\rm ankle}$ 
are `confined', implying that their observed and source fluxes obey:
\begin{equation}
F_{\rm o}\propto\tau_{\rm conf}\,F_{\rm s},\;\;\;
\tau_{\rm conf}\propto(Z/p)^{\beta_{\rm conf}},\;\;\;
\beta_{\rm conf}\sim\!0.6\pm\!0.1,
\label{confinement}
\end{equation}
with  $\tau_{\rm conf}$ a `confinement time', deduced from the study of
stable and unstable  CRs and their fragments.

\begin{figure}[]
\centering
\epsfig{file=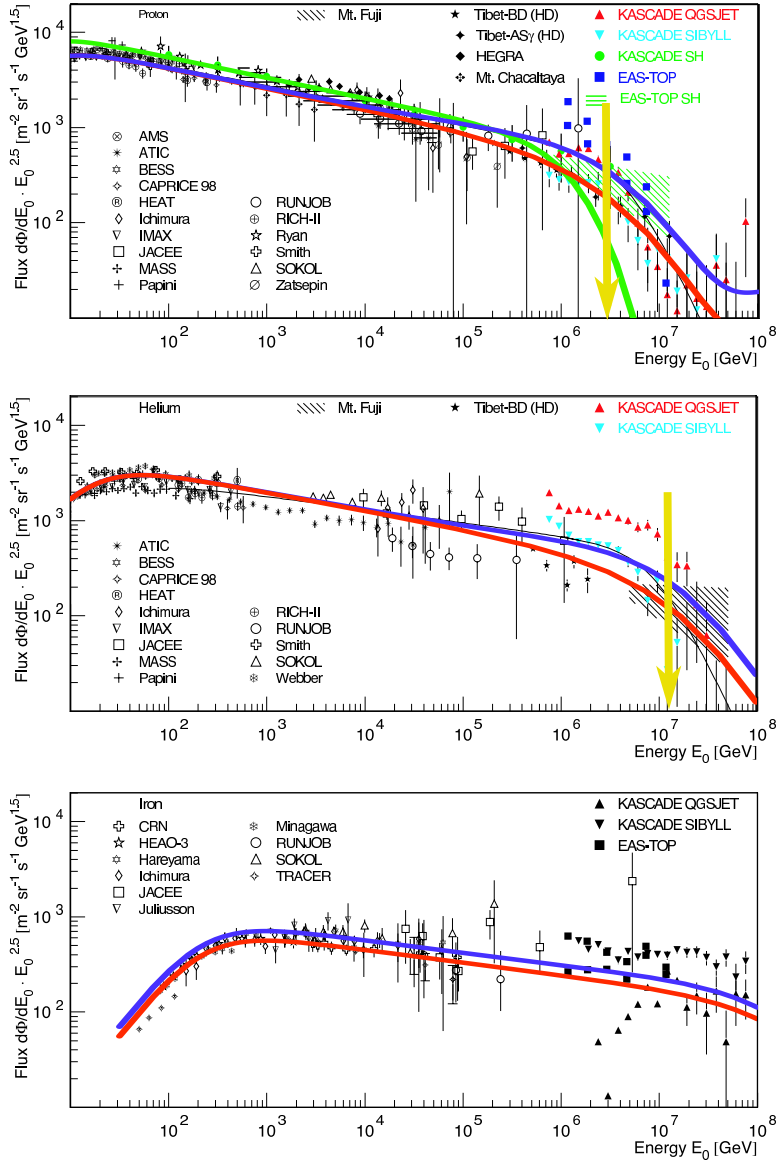,width=10.2cm}
\caption{
$E^{2.5}\,dF/dE$ CR spectra in
the `knee' region.
Top: protons; middle: $\alpha$ particles; bottom: iron nuclei.
The data ensemble was kindly provided  by K.H. Kampert. The
CB-model predictions are explained in the text. Notice the Fe
`hip', occurring at the same $\gamma$ as the H, He hips of fig.~\ref{AllPart}b.
}
\label{KASKADE}
\end{figure}

At $E\!=\!10^6$-$10^8$ GeV the all-particle spectrum of fig.~\ref{AllPart}a
bends in one or two `knees'. The knee flux  is too small to measure directly
its energy and composition, which are inferred from the properties of
the CR shower of hadrons, $\gamma$'s, $e$'s  and $\mu$'s, initiated
by the CR in the upper atmosphere. The results for H, He
and Fe are shown in fig.~\ref{KASKADE}. Note that even the same 
data leads to incompatible results, depending on the Monte-Carlo
program used to analize the showers. But 
the spectra of the various elements seem to have `knees' which
scale roughly with  $A$ or $Z$, the data not been good
enough to distinguish.

\begin{figure}
\begin{center}
\hbox
{\epsfig{file=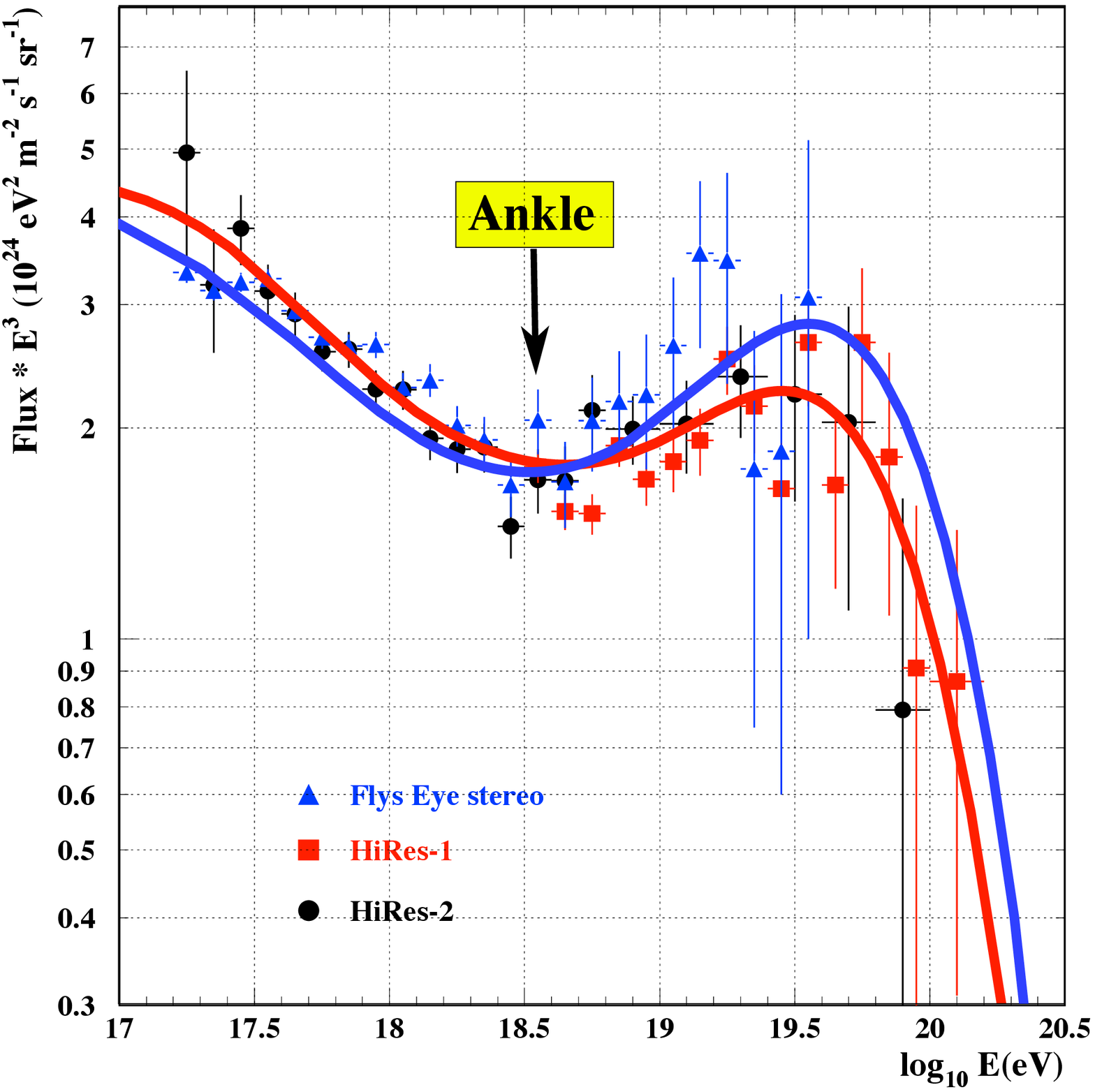,width=4.3cm,height=4.45cm}
\epsfig{file=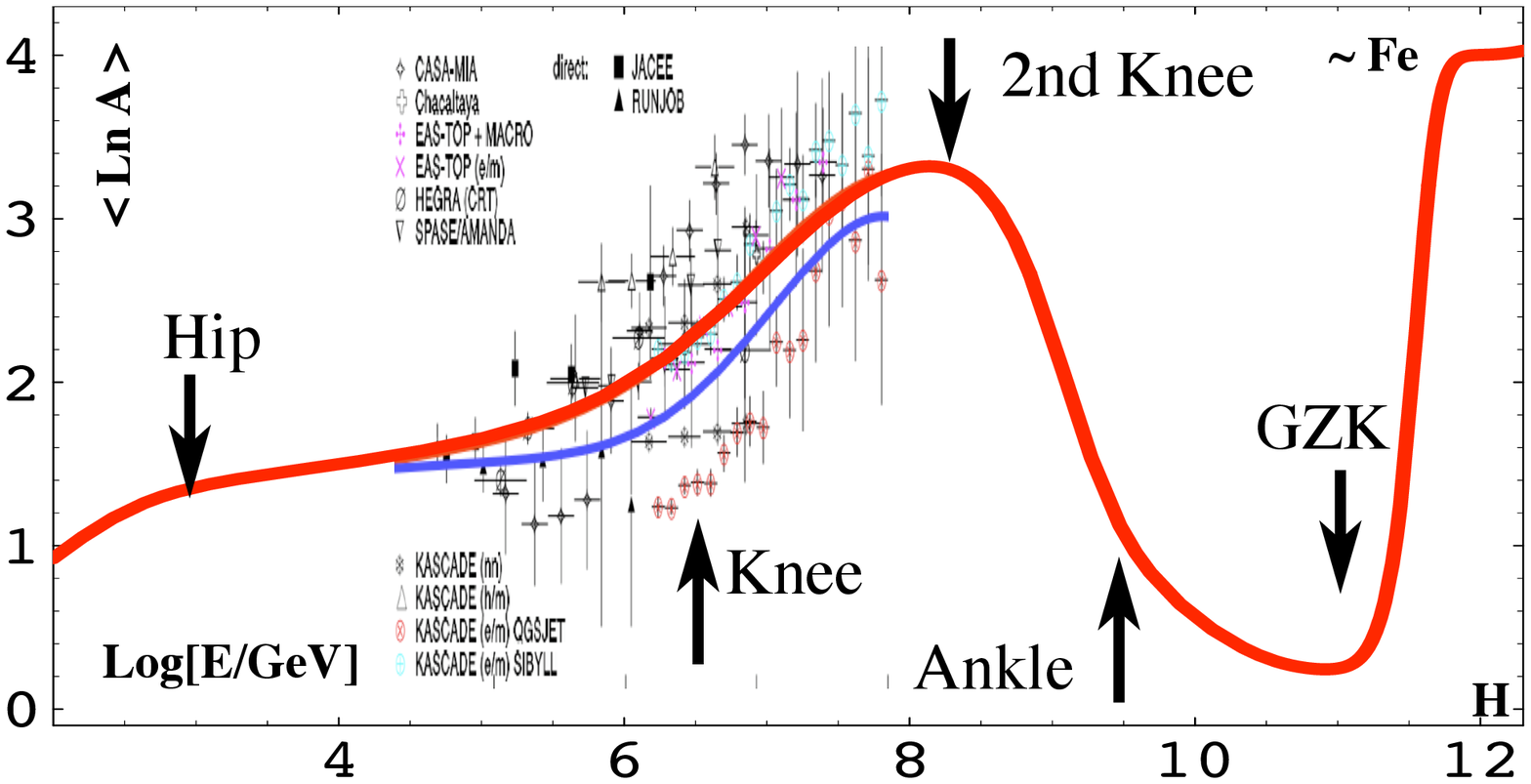, width=7.6cm,height=4.3cm}}
\end{center}
\vspace{-1.2cm}
\caption{{\bf Left (a)} The $E^3$-weighed CR flux at the highest observed energies.
The data are not the oldest, nor the most recent, but show the currently observed 
trend. {\bf Right (b)} The CR mean of the Neperian log of $A$, versus energy.
In (a) the lines are two extreme CB-model
predictions. The normalizations below and above the ankle, 
shown here to coincide with the data, are separately predicted to
within a factor of $\sim\!3$. In (b) the lines are as in (a).}
\label{highE}
\end{figure}

The high-$E$ end of the $E^3$-weighed CR spectrum is shown in fig.~\ref{highE}a.
These data  and the more recent ones of HIRES and Auger, clearly show a
cutoff, predicted by Greisen, Zatsepin and Kuzmin (GZK) as the result of the 
inevitable interactions of extragalactic CR protons with the 
microwave background radiation. The reactions
$p\!+\!\gamma\!\to\!n\!+\!\pi^+;\,p\!+\!\pi^0$ cut off the flux at 
$E\!>\!E_{_{\rm GZK}}\!\sim\!A\times 10^{11}$ GeV, from distances larger
than tens of Mpc. Similarly, extragalactic nuclei of $E\!>\!10^9$ GeV are 
efficiently photo-dissociated in the cosmic infrared radiation, the 
corresponding CR flux should not contain many.

At very high energies,
rough measures of the CR $A$-distribution are extracted
from the `depth of shower maximum', $X$, the number  of grams/cm$^2$ of
atmosphere travelled by a CR shower before its $e^\pm$/$\gamma$
constituency reaches a maximum. At a fixed energy, $X$ decreases with $A$,
since a nucleus is an easily broken bag of nucleons of energy $\sim\!E/A$.
As in fig.~\ref{highE}b, the data are often presented as $\langle ln[A](E) \rangle$,
which approximately satisfies
$X(A)\!\sim\!X(1)-x\, ln[A]$, with $x\!\sim\!37$ grams/cm$^2$ the
radiation length in air.

If CRs are chiefly Galactic in origin, their accelerators must
compensate for the escape of CRs from the Galaxy,  to sustain the
observed CR flux: it is known from meteorite records that
the flux has been steady for the past few Giga-years.
The Milky Way's luminosity in CRs must therefore satisfy:
\begin{equation}
{L_{_{\rm CR}} ={4\pi \over c} \int {1\over \tau_{\rm conf}}
\,E\,{dF_{\rm o}\over dE}\,dE\,dV}
\sim 1.5\times 10^{41}{\rm ~erg~s^{-1}}\, ,
\label{CRlum}
\end{equation}
where $V$ is a CR-confinement volume. The quoted 
standard estimate 
of $L_{_{\rm CR}}$ is very model-dependent\cite{DDLum}.

\section{More than you ever wanted to know about Gamma-Ray Bursts}

\begin{figure}
\hskip -.5truecm


\epsfig{file=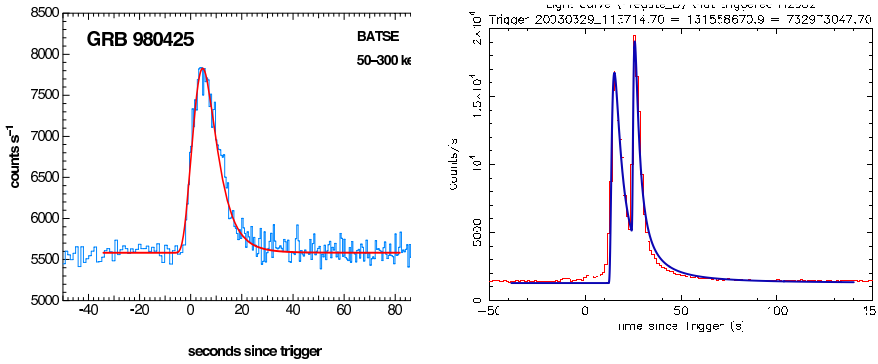,width=11.6cm}


\caption{
The $\gamma$-ray count-rate light curves of GRB 980425 (a: left) and GRB 030329
(b: right).
In the CB model, each pulse  corresponds to one cannonball. The single pulse
in (a) and the two pulses in (b) are fit with Eq.~\ref{shape}.
}
\label{try}
\end{figure}

Two $\gamma$-ray count rates of GRBs, peaking at 
$dN/dt\!=\!{\cal{O}}(10^4)\,\rm s^{-1}$, 
are shown in fig.~\ref{try}. 
The typical energy of the  $\gamma$-ray of GBBs is $\sim\!250$ keV. 
The total {\it `isotropic equivalent'} energy of a source of
 such photons at a typical redshift, $z\!=\!{\cal{O}}(1)$, 
is $E_\gamma^{\rm iso}\!\sim\!10^{53}$ erg, similar to the available energy in a core-collapse
SN explosion, i.e.~half of the binding energy of a solar-mass neutron star,
maybe a bit more for a black-hole remnant. It is hard to imagine a process
with $>\!1$\% efficiency for $\gamma$-ray production. Since GRBs are 
observed to be made by  SNe, either the parent stars are 
amazingly special, or the $\gamma$-rays are narrowly beamed.

The total-duration distribution of the $\gamma$-rays of GRBs has two peaks,
with a trough at $\sim\!2$s dividing (by definition) two distinct types. 
`Long' GRBs are more common and better measured than short ones;
one is more confident discussing mainly the former, as I shall.
The long GRB light curves of fig.~\ref{try} are not atypical. The
`pulses' of a given GRB vary in intensity,
but have similar widths, a fairly universal exponential rise, and a power decay
$\propto\!t^{-a}$, $a\!\sim\!2$.
 The number of `clear pulses' averages to
$\sim\!5$, it may reach $\sim\!12$. The 
 pulse-to-pulse delays are random, extending 
from ${\cal{O}}$(1s) to ${\cal{O}}(10^2$s).
 Put all the above in a random-generator and, concerning 
 long GRBs, `you have seen them all'.




\begin{figure}
\begin{center}



\hskip 2cm
\hbox{\epsfig{file=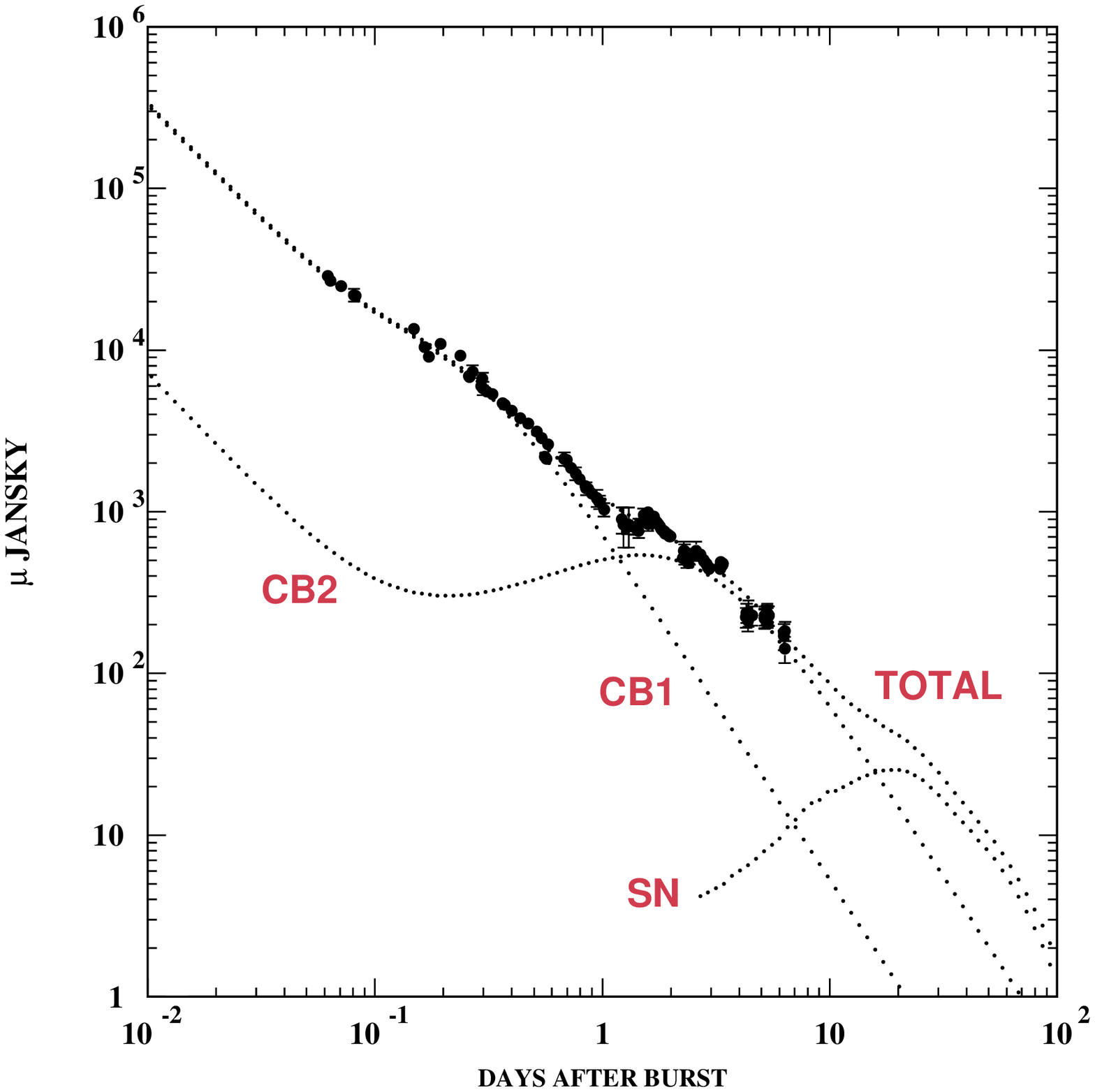,width=5.0cm}
\epsfig{file=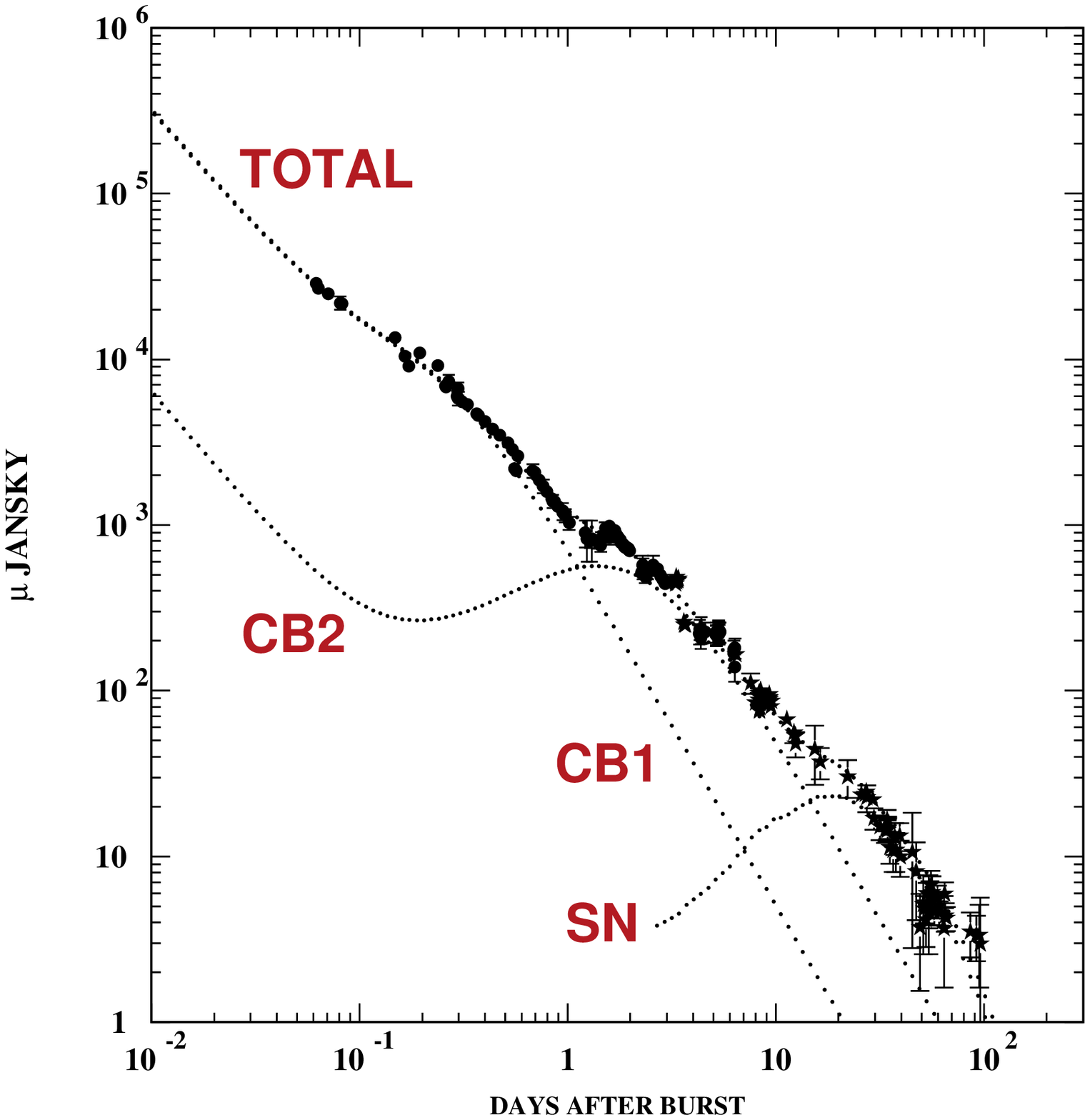,width=4.8cm}
}
\hbox{\epsfig{file=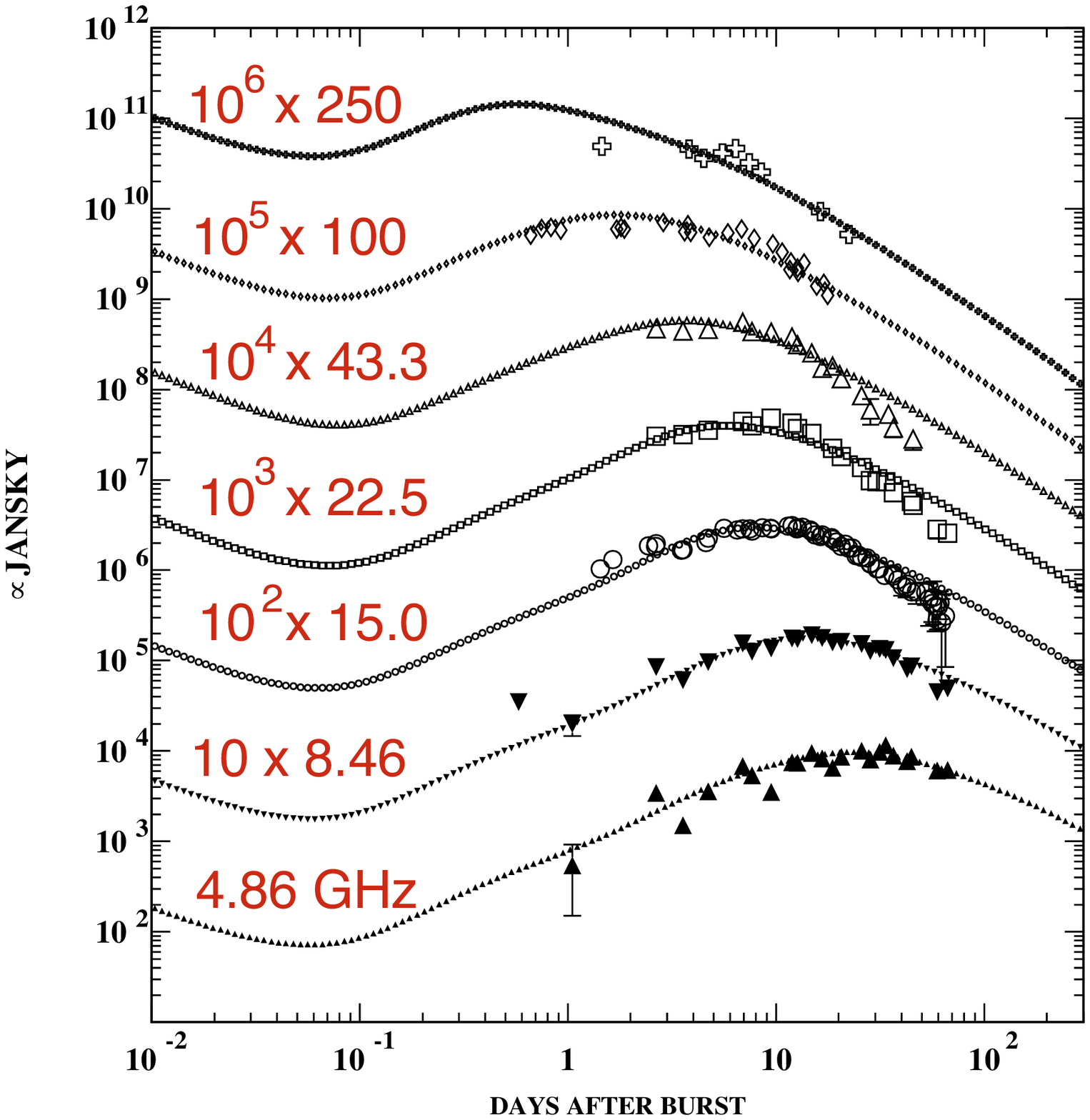,width=5.2cm}
\epsfig{file=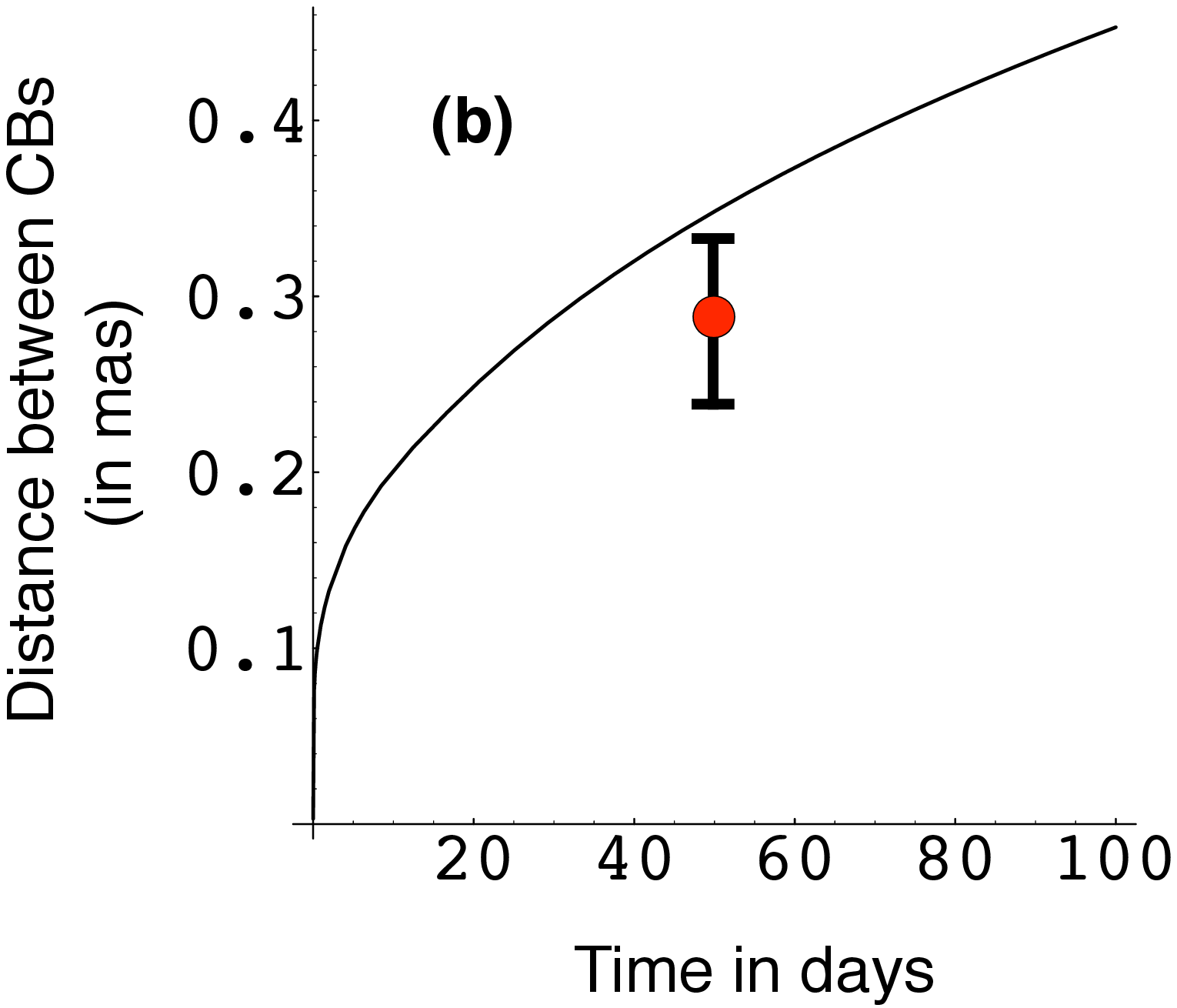,width=4.8cm}
}
\vskip .3cm 
\vbox{\epsfig{file=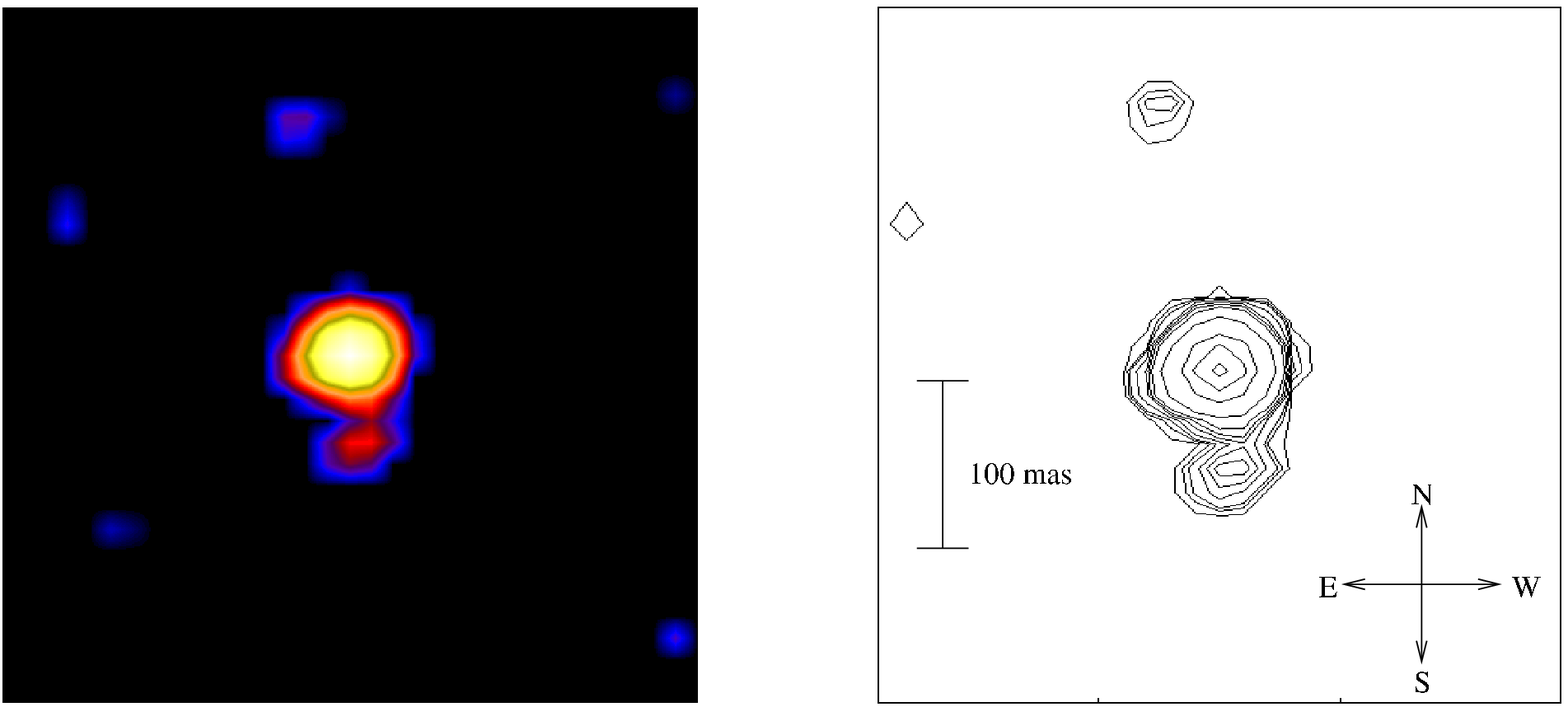,width=10cm}
}
\end{center}
\vspace{-.5cm}
\caption{Left to right and top to bottom. a) and b)
The R-band AG of GRB 030329. A micro-Jansky is $10^{-29}$ erg cm$^{-2}$ s$^{-1}$.
a) Six days of data are used to predict the next $\sim\!100$ days, and the SN
contribution. b) The SN is seen. c) The radio afterglows of GRB 030329.
 d) The predicted and observed inter-CB {\it superluminal} angle in 
this GRB. (e,f): SN1987A and its two `mystery spot' CBs\cite{Costas}.
The motion of the Northern one was superluminal.
}
\label{try2}
\end{figure}

GRBs are not often seen more than once a day,
they are baptized with their observation date. GRBs 980425
and 030329, shown in fig.~\ref{try}, 
originated at  $z\!=\!0.0085$ (the record smallest)
and $0.168$, respectively. How are the redshifts known? 
GRBs have ``afterglows" (AGs): they are observable in
 radio to X-rays for months after their $\gamma$-ray
signal peters out. The AG of GRB 030329 in the `R-band' (a red-light interval)
and radio is
shown in fig.~\ref{try2}a-c. Once the object is seen in optical or radio,
its direction can be determined with much greater precision
than via $\gamma$ rays. Very often the source is localized
within a galaxy, whose lines can be measured to determine $z$
(in some cases a lower limit on $z$ is deduced from absorption lines in intervening
material). 

GRB 980425 was {\it `associated'} with a supernova called SN1998bw: within directional 
errors and within a timing uncertainty of $\sim\!1$ day, they coincided. The 
luminosity of a 1998bw-like SN peaks at $\sim 15\,(1+z)$ days. 
The SN light competes at that time and frequency with the AG of its 
GRB, and it is not always easily detectable.  Iff one has a predictive theory
of AGs, one may test whether GRBs are associated with  `standard torch' SNe, 
akin to SN1998bw, `transported' to the GRBs' redshifts. The test was already
conclusive (to us) in 2001\cite{AGoptical}.
One could even foretell {\it the date} in which a GRB's SN 
would be discovered. For example, GRB 030329 was so
`very near' at $z\!=\!0.168$, that one could not resist posting such a  daring
prediction\cite{SN030329} during the first few days of AG observations. 
The prediction and the subsequent SN signal are shown in fig.~\ref{try2}a,b.
The spectrum of this SN was very well measured and seen to coincide snugly with
that of SN1998bw, and this is why the SN/GRB association ceased to be doubted:
{\it long GRBs are made by core-collapse SNe.}

Astrophysicists classify SNe in Types, mainly depending on the composition of their ejecta.
Within very limited statistics the SNe associated with GRBs are of Type Ib/c. These constitute
some 15\% of core-collapse SNe, the fascinating ones which beget neutrinos, neutron stars
and presumably black holes. Type Ia SNe are probably mere explosions of accreting white dwarfs,
but they are very luminous, and of cosmological standard-candle fame.

GRBs have many `typical' properties. Their spherical-equivalent number of $\gamma$-rays
is $\sim\!10^{59}$. Their spectrum at fixed $t$ is very well approximated by:
\begin{equation}
{dN\over dE}\biggm|_t
\propto
\left[{T(t)\over E}\right]^\alpha\; e^{-E/T(t)}+b\;
\left[1-e^{-E/T(t)}\right]\;
{\left[T(t)\over E\right]}^\beta
\label{totdist}
\end{equation} 
with $b\!\sim\!1$, $\alpha\!\sim\!1$, 
$\beta\!\sim\!2.1$. Early in the evolution of a pulse, the `peak energy' 
(characterizing  the photons carrying most of the GRB's energy) is
$E_p\!\sim\!T[0]\!\sim\!250$ keV, evolving later to $T(t)\!\sim\!t^{-2}$.
A pulse's shape at fixed $E$ is well fit by:
\begin{equation}
{dN\over dt}\biggm|_E\approx \Theta[t]\,e^{-[{\Delta t(E)/t}]^2} 
\left\{ 1 - e^{-[{\Delta t(E)/t}]^2}   \right\};\;{\Delta t(E_1)\over \Delta t(E_2)}
\approx \left[{E_2\over E_1} \right]^{1\over 2},
\label{shape}
\end{equation}
with $\Delta t\!\sim\!{\cal{O}}(1$s) at $E\!\sim\!E_p$. Eq.~(\ref{shape}) reflects an approximate 
spectro-temporal
correlation whereby $E\,dN/(dE\,dt)\!\approx\!F[E\,t^2]$, which we call the $E\,t^2$ `law'.

\begin{figure}[]
\vspace {-1cm}
\vbox{
\hbox{
\centering
\epsfig{file=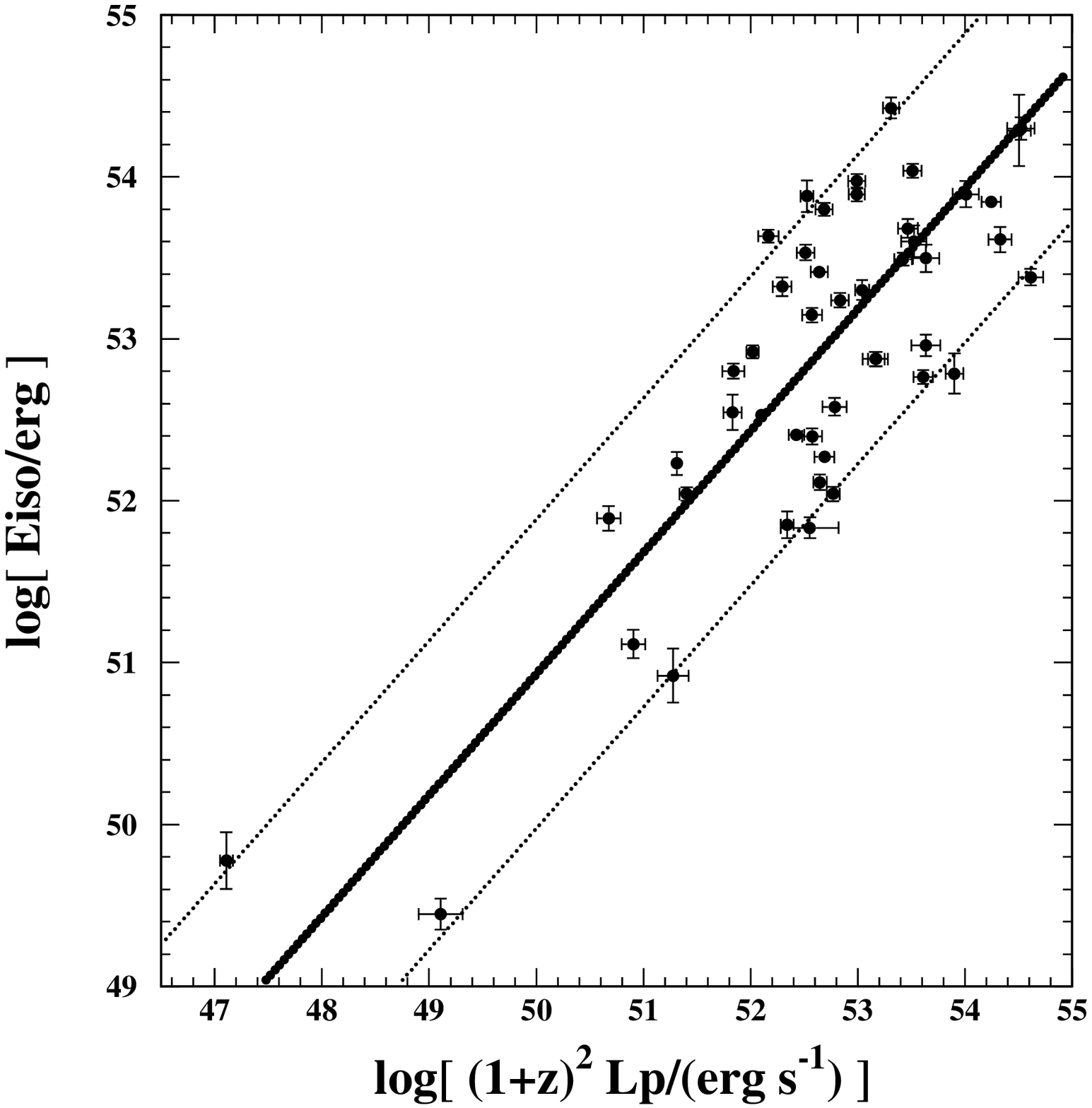,width=4.5cm}
 \epsfig{file=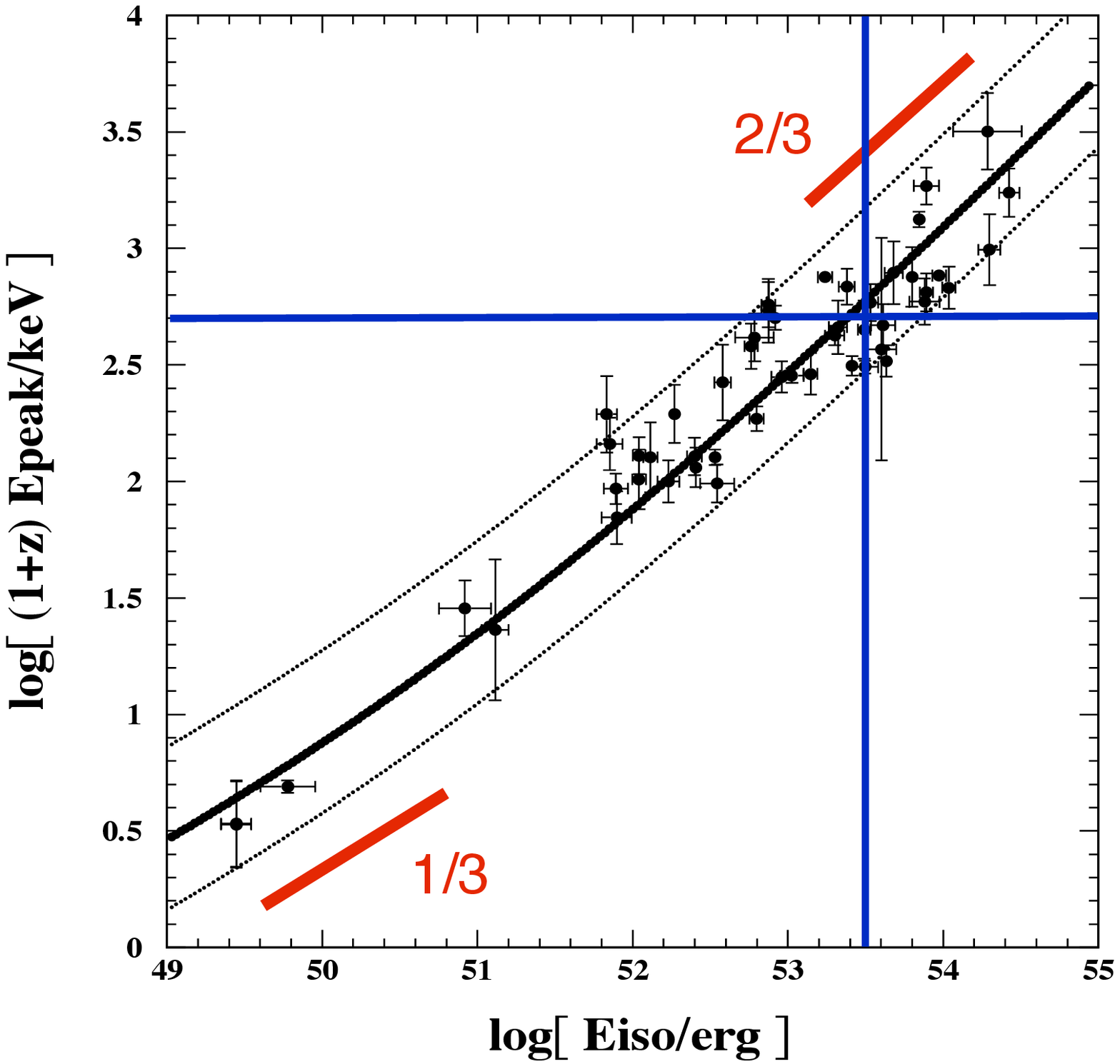,width=4.5cm,height=4.5cm}
}}
\vbox{
\hbox{
\centering
\epsfig{file=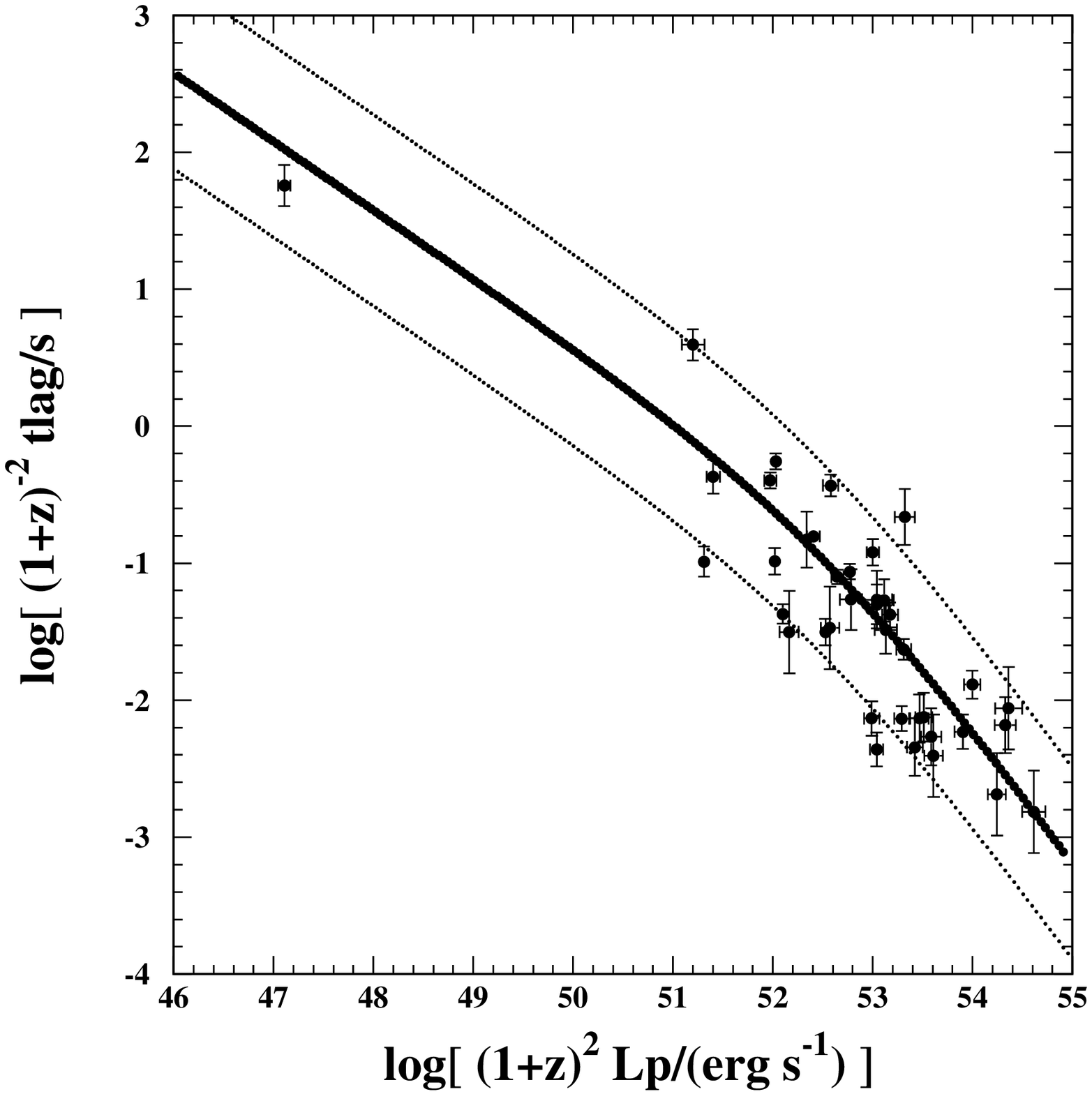,width=4.5cm}
\epsfig{file=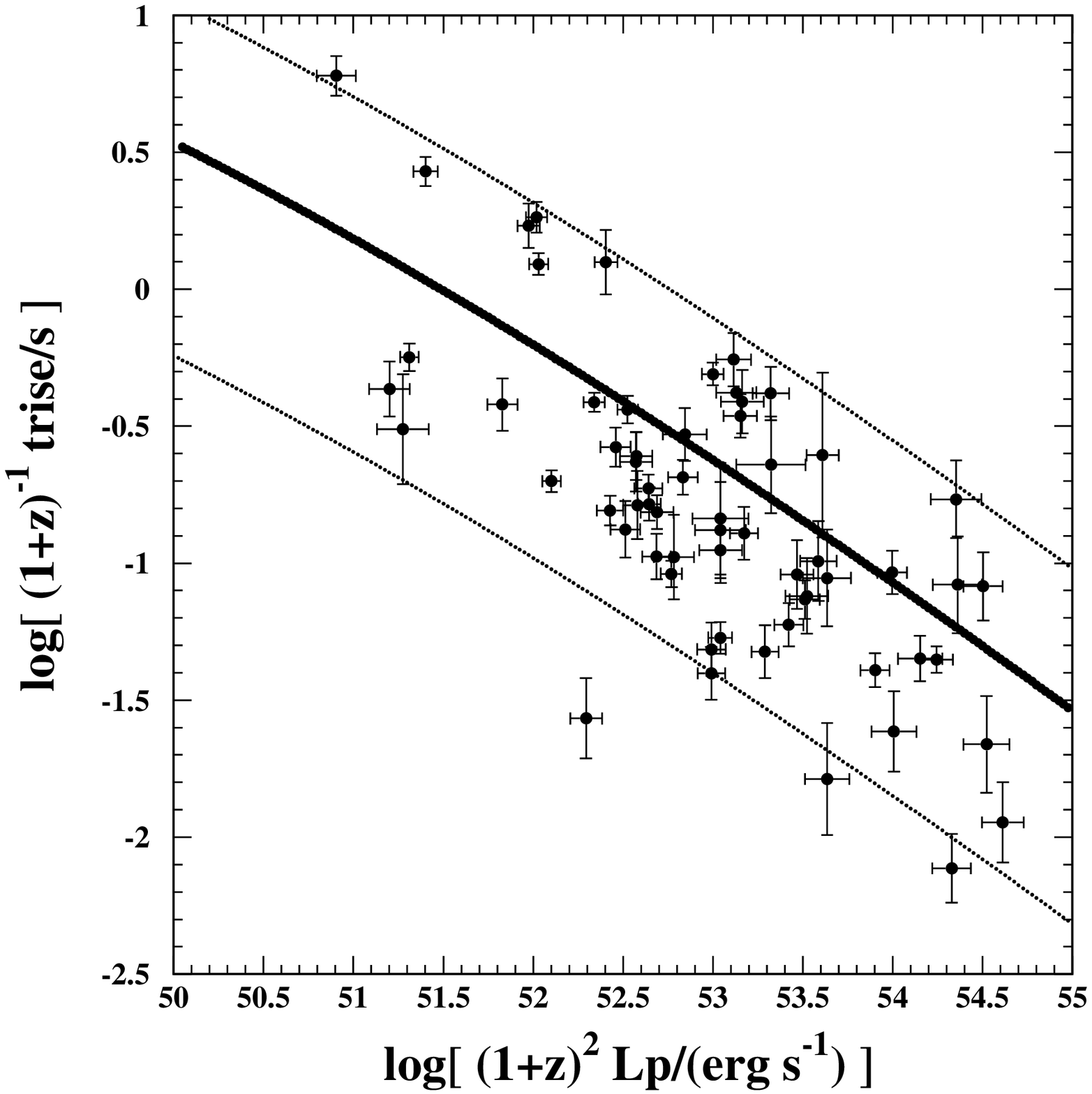,width=4.5cm}
}}
\vbox{
\hbox{
\centering
\epsfig{file=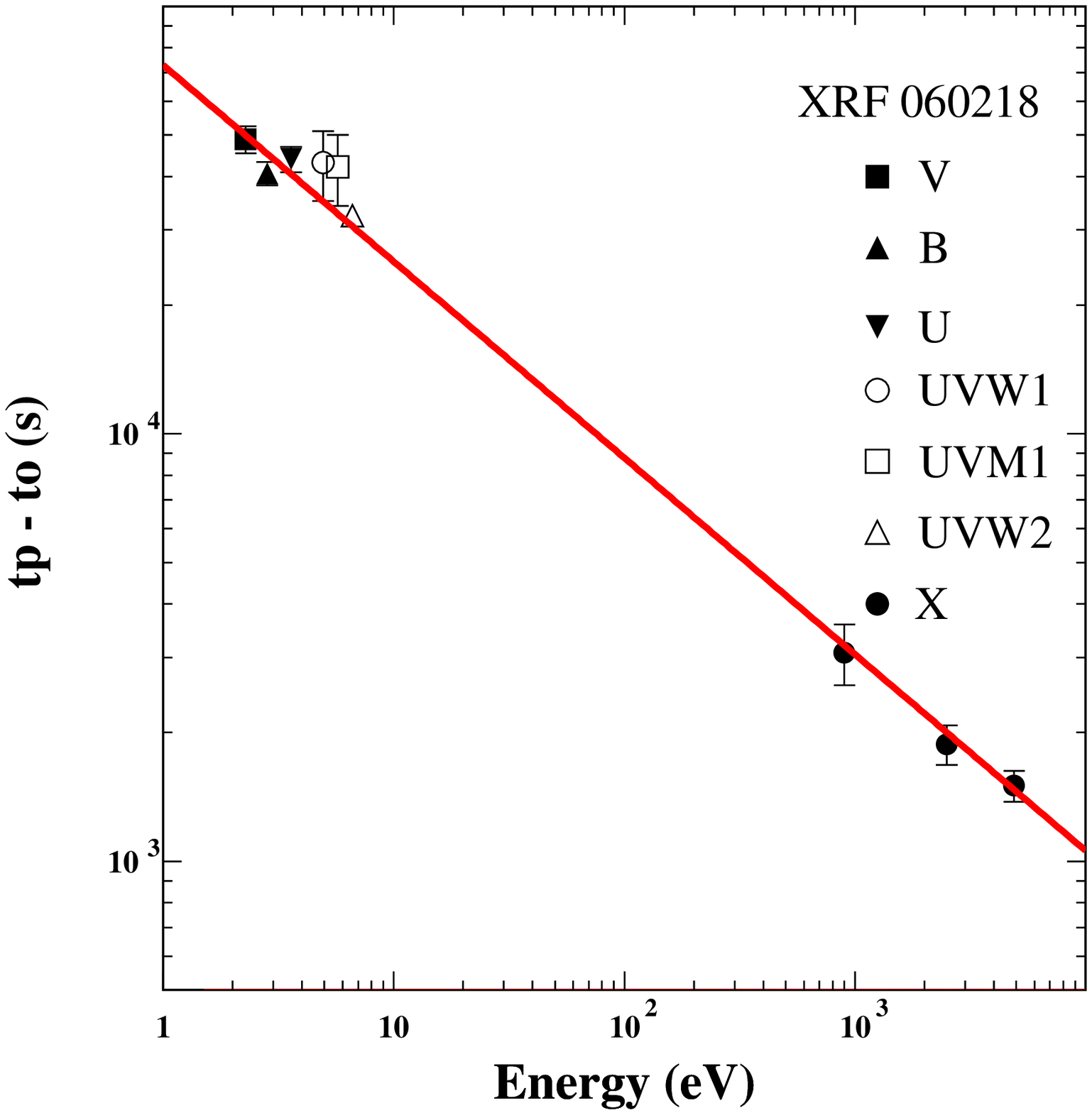,width=4.5cm}
\epsfig{file=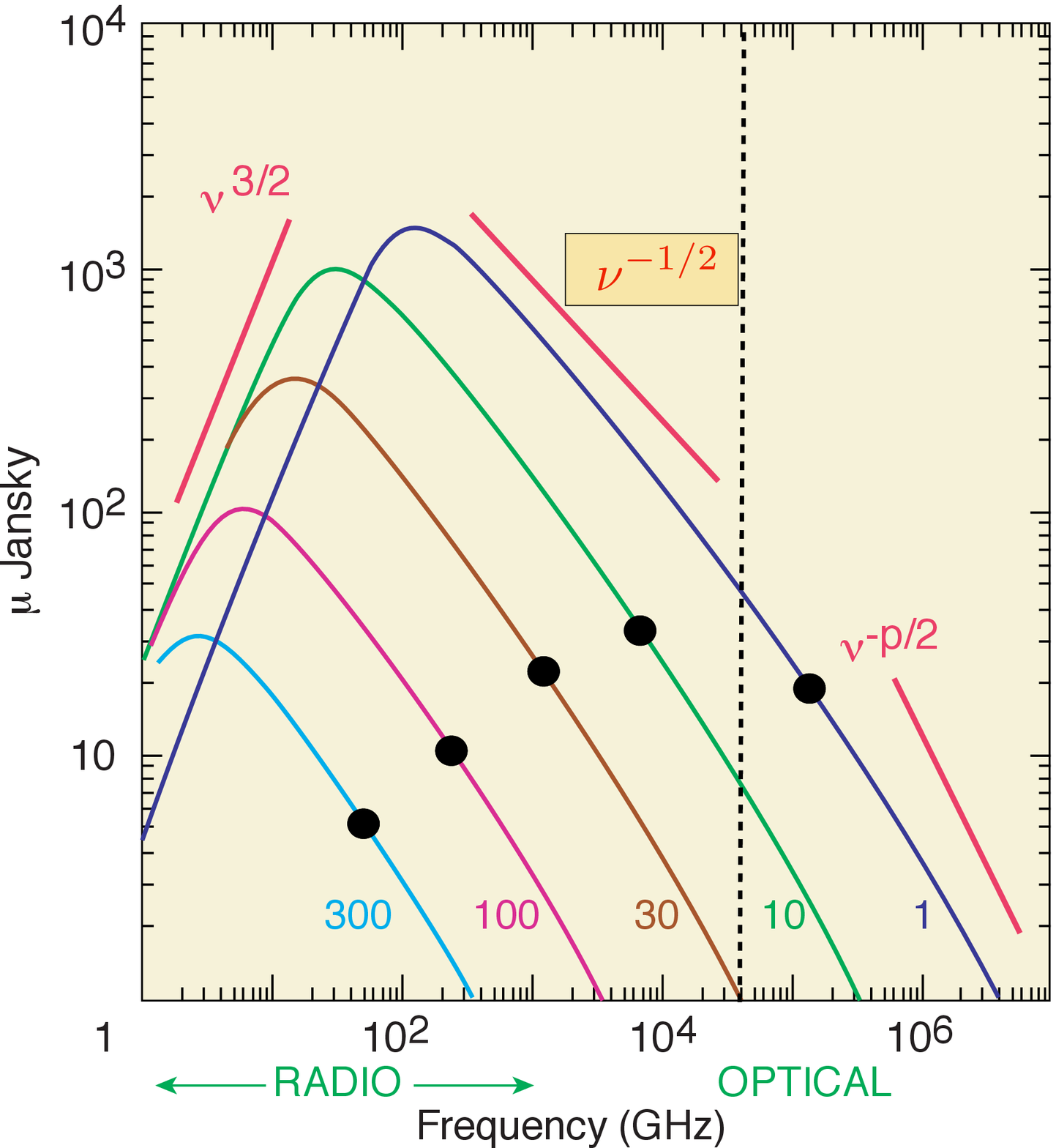,width=3.85cm}
}}
\caption{
Left to right and top to bottom:
a) The $[E_\gamma^{\rm iso},L_p^{\rm iso}]$ correlation.
b) The $[E_p,E_\gamma^{\rm iso}]$ correlation.
The limiting slopes are in red, the central predictions in blue.
c) The $[t_{\rm lag},L_p^{\rm iso}]$ correlation. 
d) The $[t_{\rm min}^{\rm rise},L_p^{\rm iso}]$ correlation.
The data are for Swift-era GRBs of known $z$.
e) The peak times of the one-pulse X-ray flash XRF 060218 at different energy intervals.
f) A typical AG spectrum at different number of days after the burst
(1, 10,... 300).  In the figure $p/2\!\sim\!1.1$.} 
 \label{f1}
\end{figure}

The values of $E_p$; of the isotropic-equivalent energy 
and luminosity, $E_\gamma^{\rm iso}$ and $L_p^{\rm iso}$; of
a pulse's rise-time $t_{\rm rise}$; or of its `lag-time'
$t_{\rm lag}$ (a measure of how a pulse peaks at a later time in a lower energy interval)
vary from GRB to GRB over orders of magnitude. But they are strongly correlated,
as shown in Figs.~(\ref{f1}a-d). It is patently obvious that such an organized set of results
is carrying a strong and simple message, which we shall decipher.

X-ray flashes (XRFs) are lower-energy kinsfolk of GRBs. They are defined
by having $E_p\!<\!50$ keV. Their pulses are wider than the ones of GRBs and
their overlap is more pronounced, since the total durations of (multi-pulse)
XRFs and GRBs are not significantly different. In fig.~\ref{f1}e I show the time at which 
the single pulse of XRF 060218 peaked (measured from the start of the count-rate rise)
as measured in different energy intervals. This is an impressive validation
of the $E\,t^2$ law (the red line), also screaming for a simple explanation.

Analytical expressions summarizing the behaviour of GRB and XRF afterglows in time 
(from seconds to months) and frequency (from radio to X-rays) do exist (DDD02/03),
but they are somewhat more complex than Eqs.~(\ref{totdist},\ref{shape}).
The typical AG behaviour is shown in fig.~\ref{f1}f, as a function of frequency, at 1, 
10...~300 days after burst (the value of $p$ is $\sim\!2.2\pm0.2$). This simple figure
reflects a rich behaviour in time and frequency.  `Chromatic bends' (called `breaks'
in the literature) are an example. At a fixed time, the spectra steepen from
$\sim\!\nu^{-0.5}$ to $\sim\!\nu^{-1.1}$ at the dots in the figure. Around a given frequency, such as 
the optical one marked by a dotted line, the optical spectrum makes this same
transition as a function of time (at $t\!\sim\!3$d, for the parameters of this example),
while the spectral shape at X-ray frequencies stays put.

\subsection{The Swift era}

Physicists, unlike ordinary year-counting mortals, live in `eras'. 
Many are waiting for the LHC era or the Plank era, GRB astronomers
are in their `Swift era'. Various satellites are currently contributing
to a wealth of new data on GRBs and XRFs. Swift is one of them.
Within 15 seconds after detection, its 15-150 keV Burst Alert Telescope
 sends to ground a 1 to 4 arcmin position 
estimate, for use by robotic optical ground telescopes. In 20 to 75 s, Swift
slews to bring the burst location into the field of view of its
0.3-10 keV X-ray Telescope and its 170-650 nm UV/Optical Telescope.
With nominal celerity, Swift has filled a gap in GRB data: the very `prompt'
X-ray and optical radiations.

\begin{figure}[]
\centering
\vbox{
\hbox{
 \psfig{file=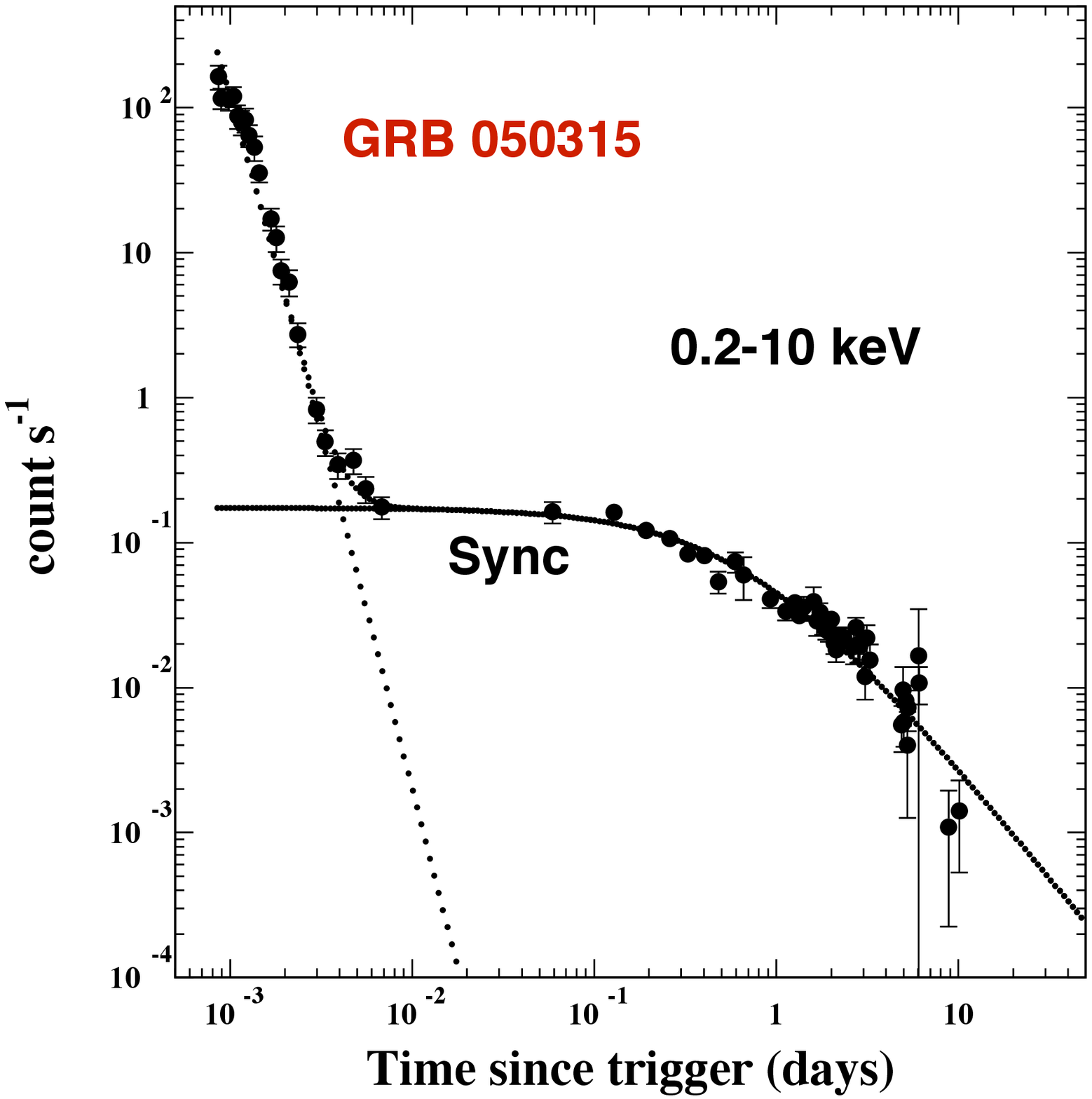,width=6cm}
 \psfig{file=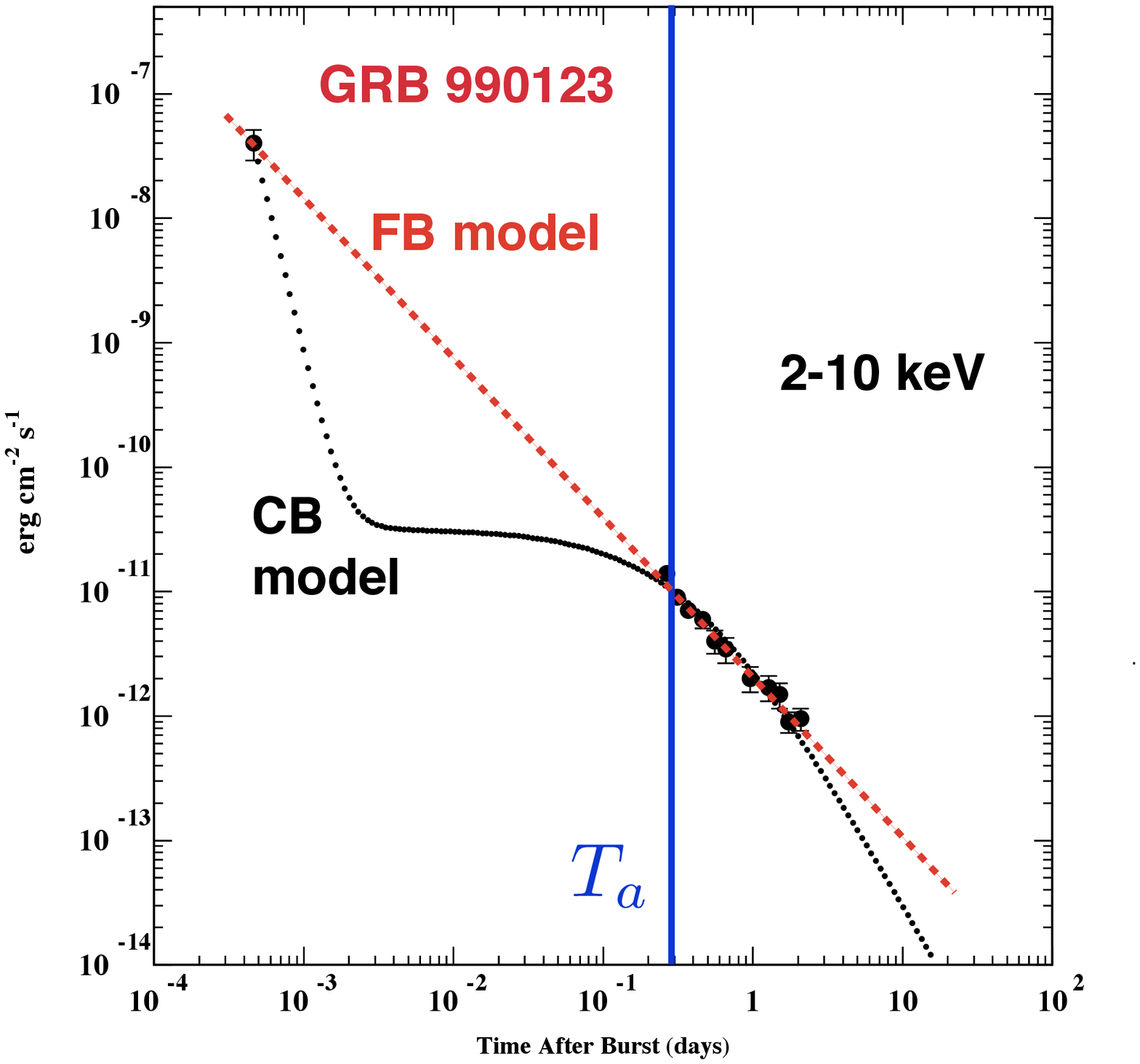,width=6cm }
}}
\centering
\vbox{
\hbox{
\psfig{file=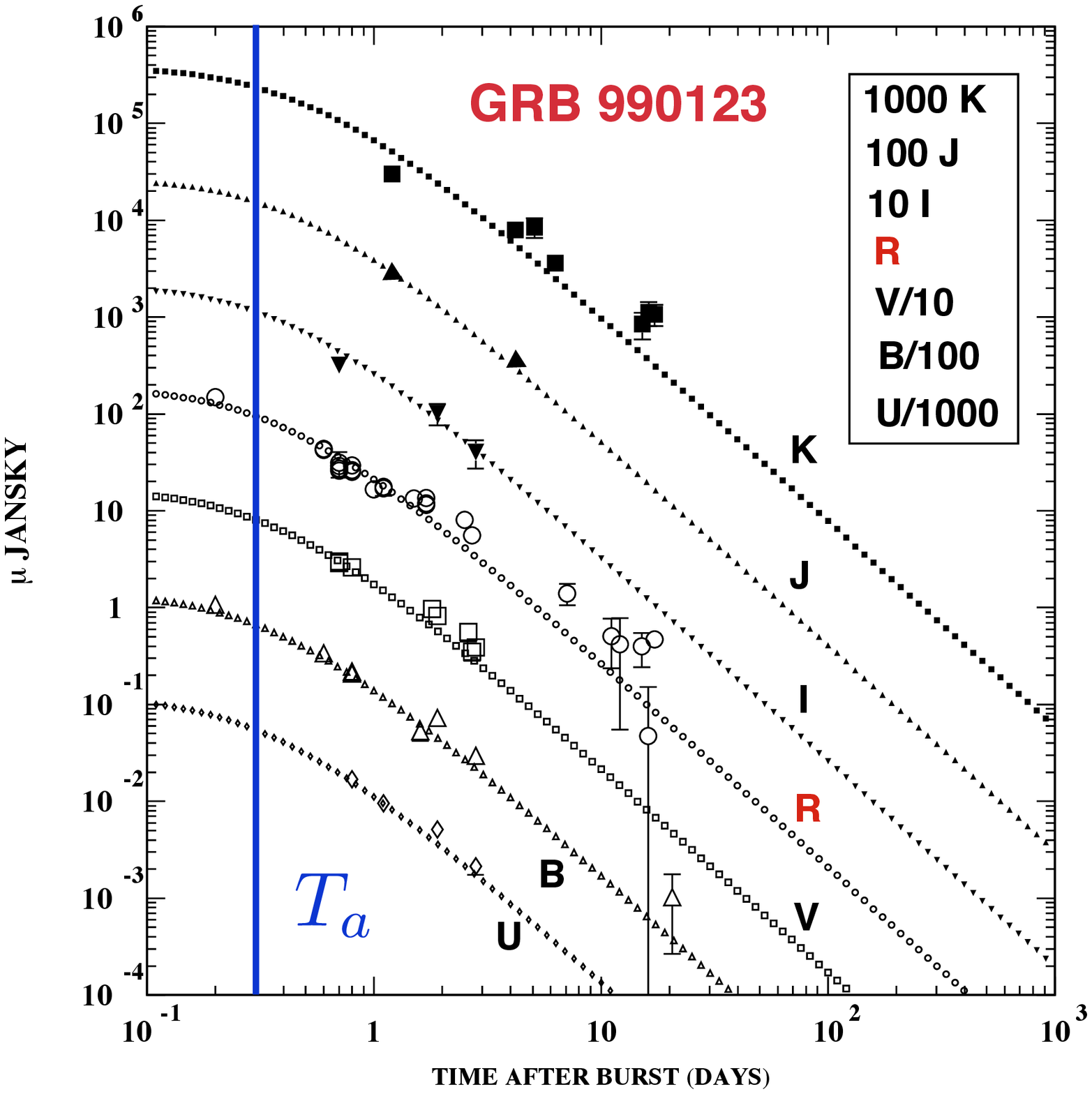,width=6cm}
\psfig{file=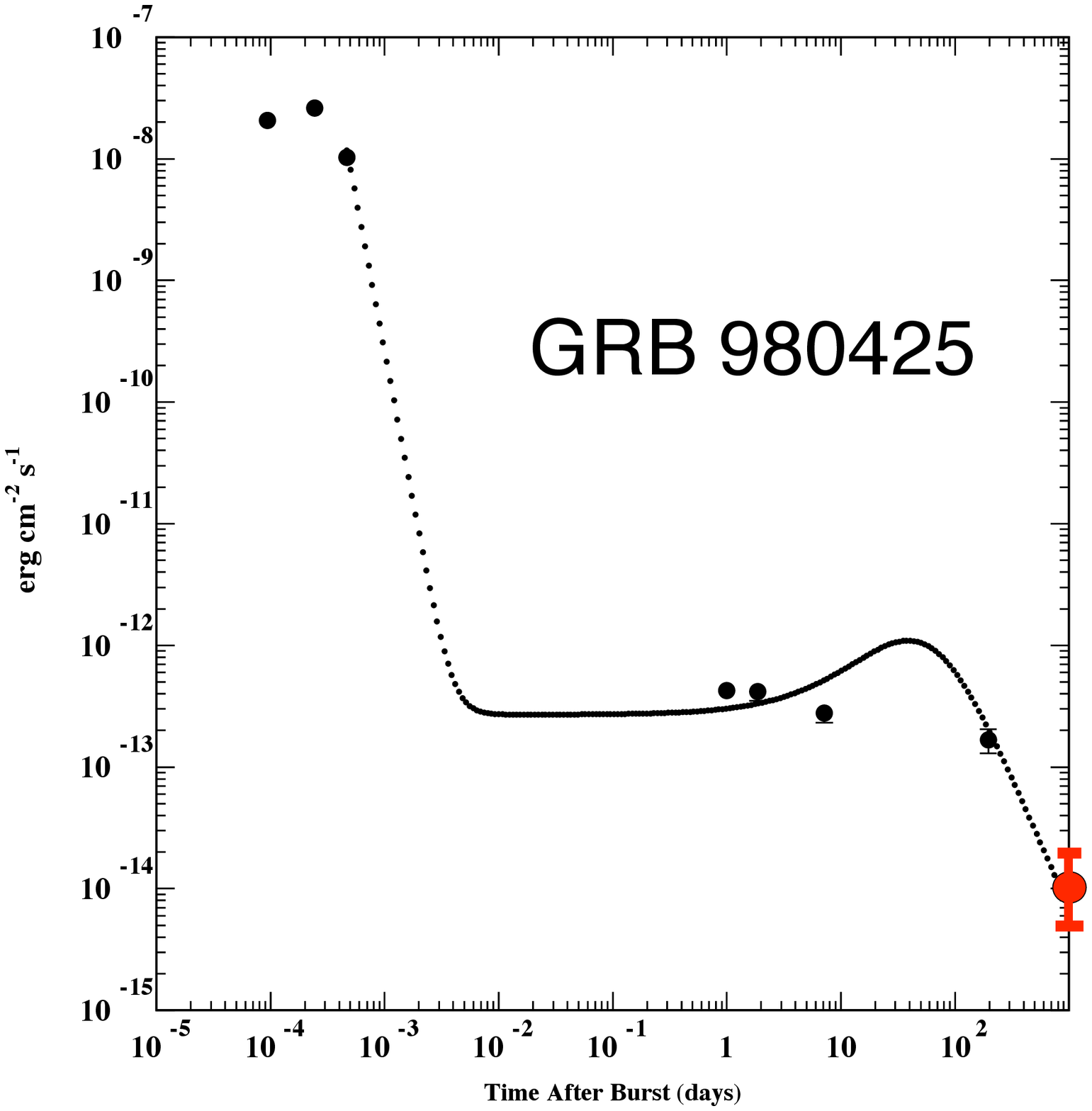,width=6cm}
}}
\vspace*{8pt}
\caption{Right to left and top to bottom.
a) The 
canonical 0.2-10 keV X-ray light curve of GRB  050315,
fit to the CB model.
b) Pre-SWIFT predictions for the 2-10 keV X-ray AG
in the CB (DDD02) and fireball\cite{Maio}
models, compared to
data for GRB 990123. 
c)
Broad band optical data on GRB 990123, fit in the CB model
(DDD03). 
The `bend' in (b) and (c) is due to the CBs' deceleration, and is achromatic.
d) The X-ray light curve of GRB 980425
 showing a very pronounced `canonical' behaviour and 
what we called (DDD02)
a long `plateau'.
The last (red) point postdates the original figure.
}
\label{fpreSwift}
\end{figure}

Swift has established a {\it canonical behaviour} of the X-ray and optical AGs of a 
large fraction of GRBs. The X-ray fluence decreases very fast from a
`prompt' maximum. It subsequently turns into a `plateau'. After a time of
${\cal{O}}(1$d), the fluence bends (has a `break', in the usual parlance)
and steepens to a power-decline. 
In fig.~\ref{fpreSwift}a, this is shown for a Swift GRB. This bend
is achromatic: the UV and optical light curves vary in proportion to it.
Although all this is considered a surprise, it is not. In
fig.~\ref{fpreSwift}b I show a pre-Swift AG and its interpretation in
two models. In fig.~\ref{fpreSwift}c one can see that the bend of this
GRB was achromatic. Even the good old GRB 980425, the first to be clearly
associated with a SN, sketched a canonical X-ray light curve, see fig.~\ref{fpreSwift}d.

\begin{figure}[]
\centering
\vbox{
\hbox{
 \psfig{file=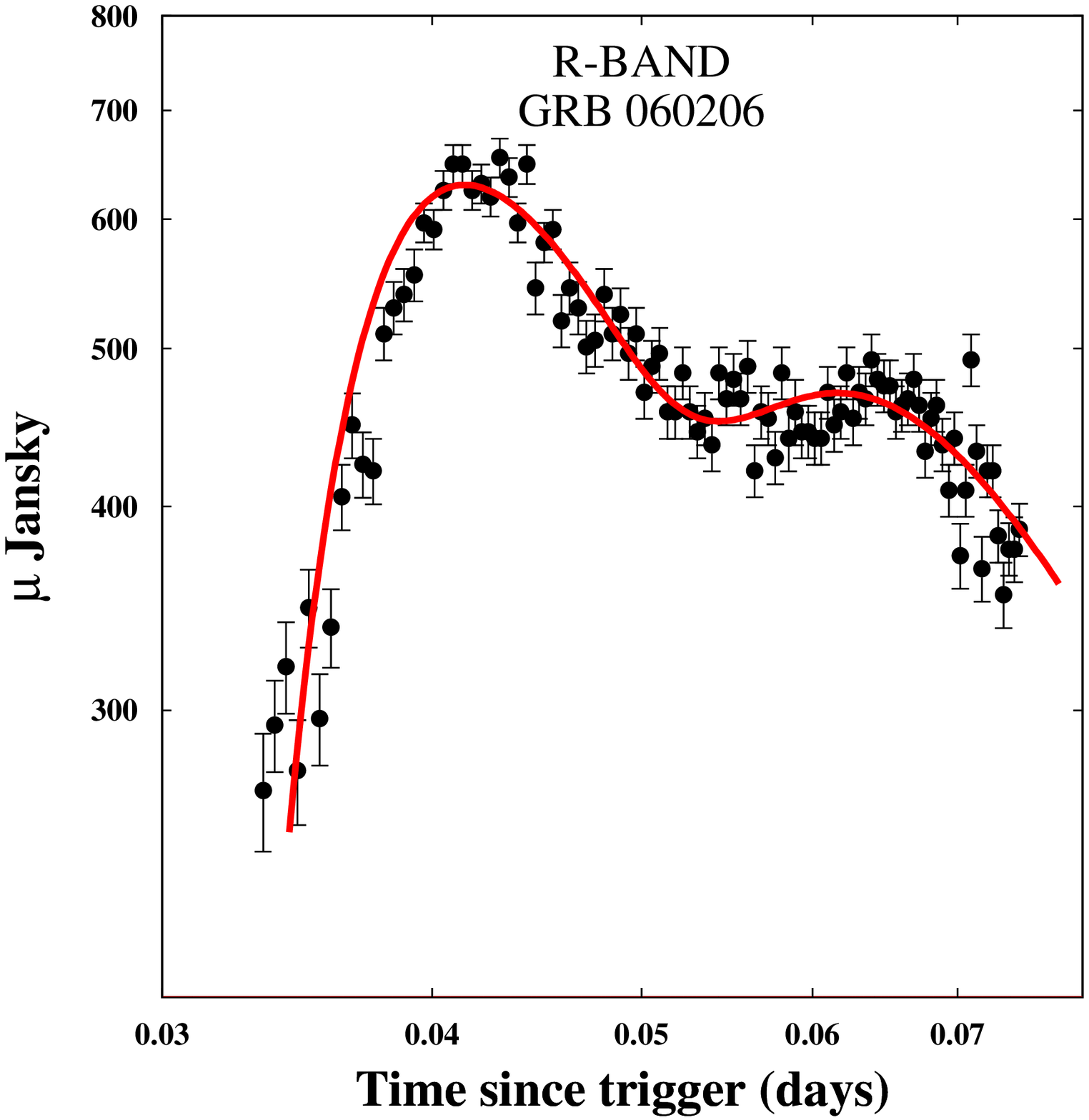,width=6cm }
 \psfig{file=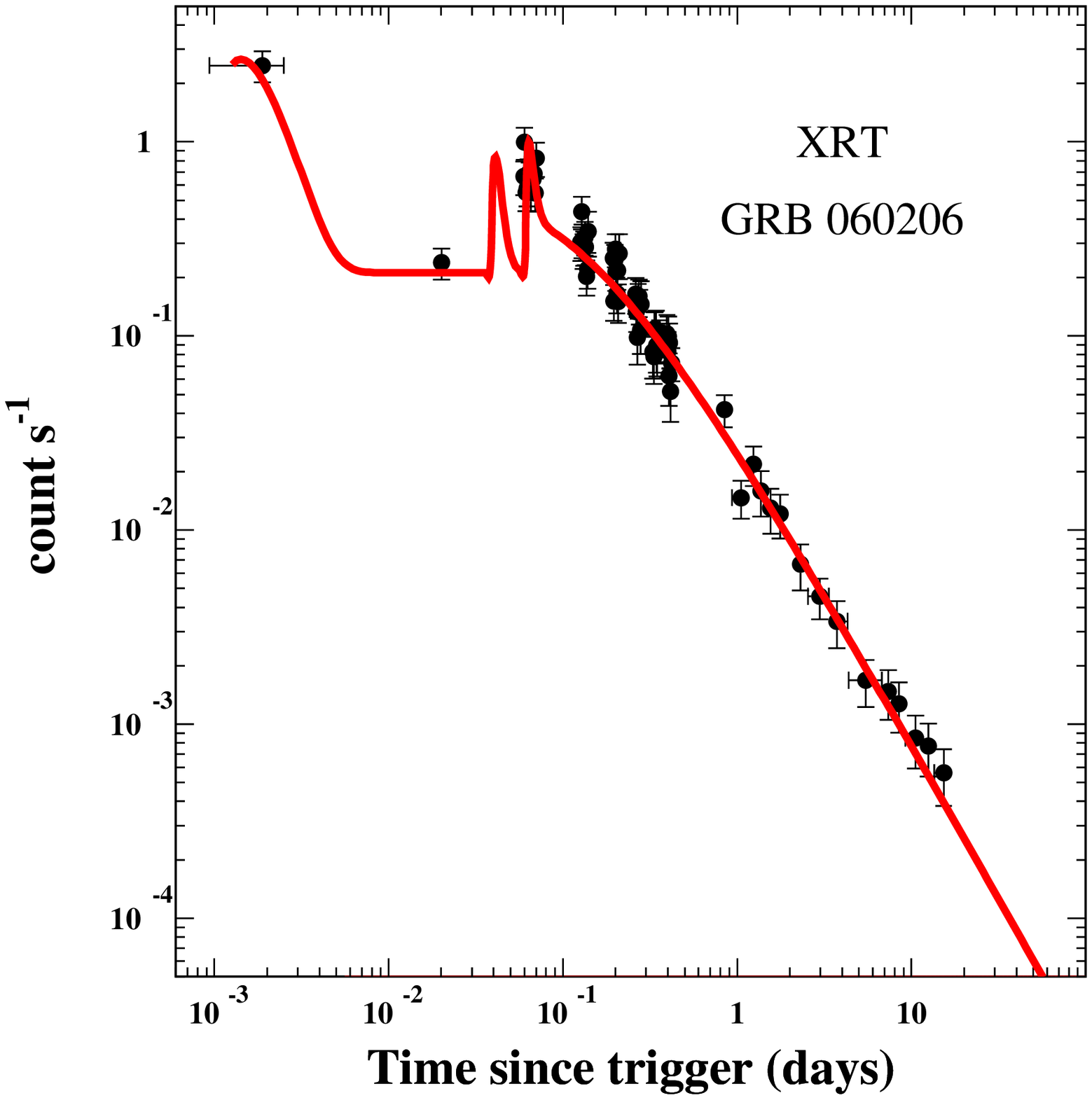,width=6cm}
}}
\centering
\vbox{
\hbox{
\psfig{file=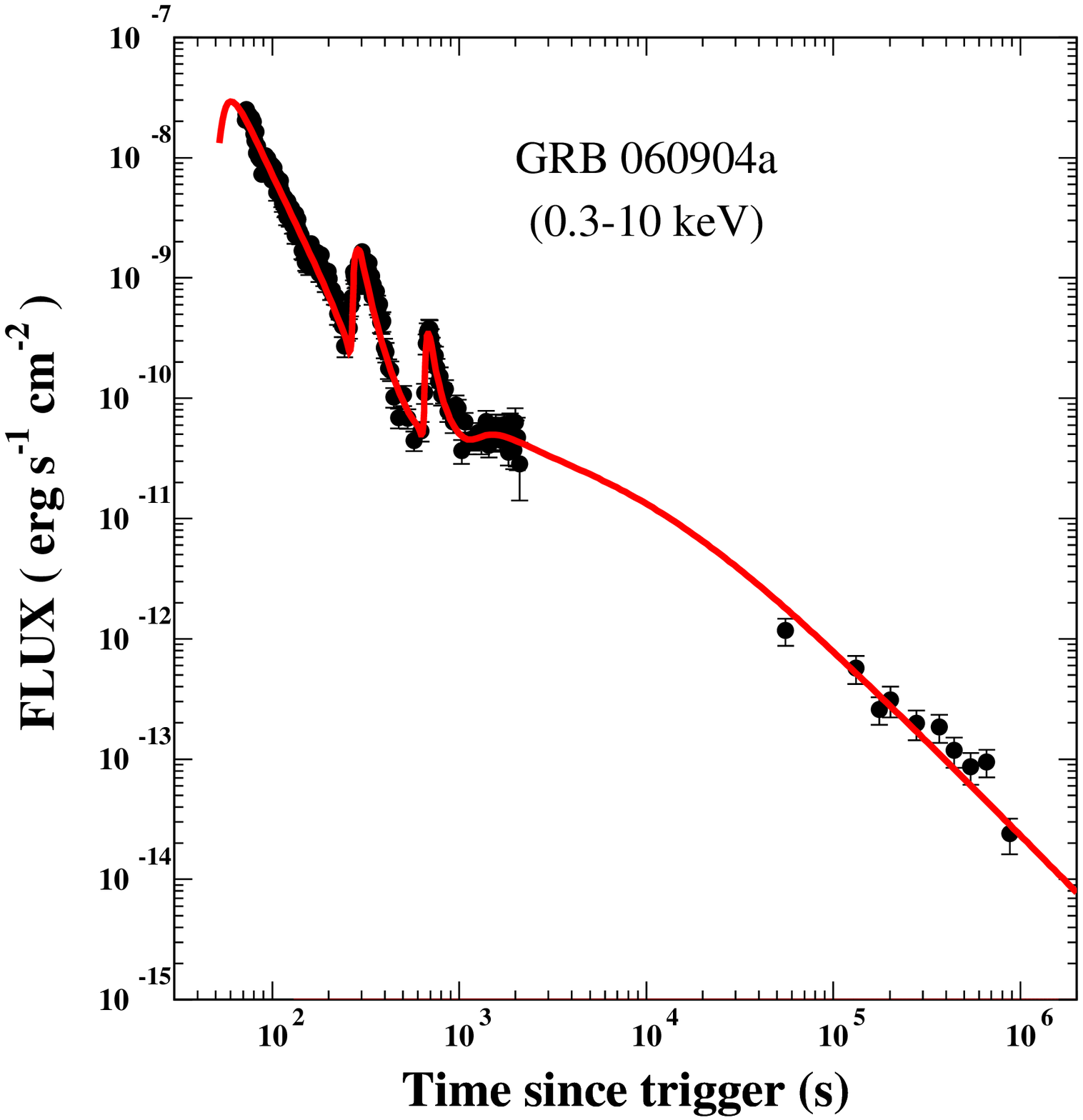,width=6cm}
\psfig{file=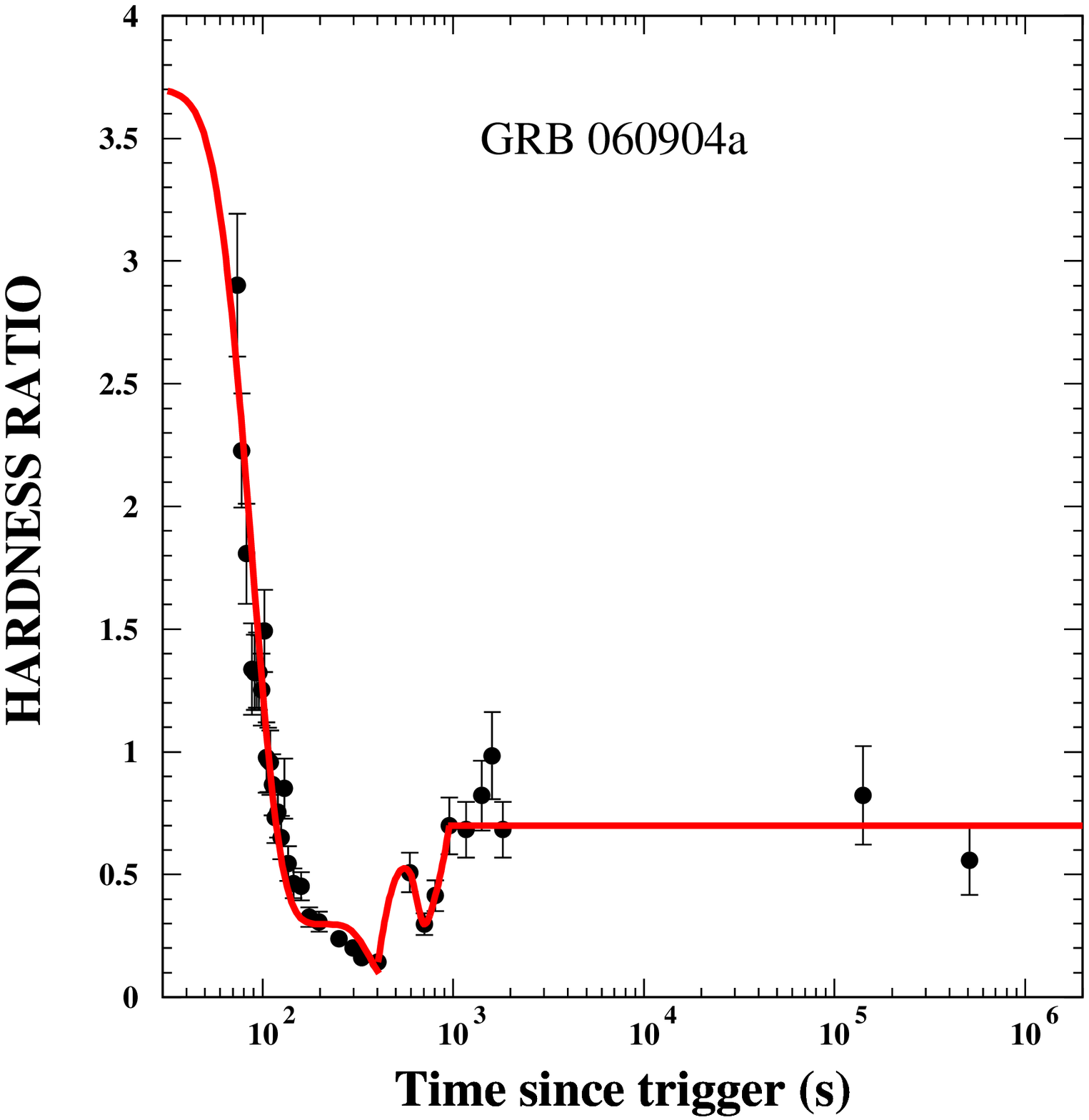,width=6cm}
\vspace{-.5cm}
}}
\vspace*{8pt}
\caption{Left to right and top to bottom:
a) Early
R-band light curves of GRB 060206.
b) Its X-ray light curve. 
c) The  X-ray light
curve of GRB 060904a with its late flares: the progressively
dying pangs of its accreting engine. 
d) The  hardness ratio of GRB 060904a traces the `prompt' ICS pulses of its light curve,
settling to a constant as SR becomes dominant in the `afterglow'. 
The above understanding of all these data is specific to the CB model.
}
\label{postSwift1}
\end{figure}

\begin{figure}[]
\centering
\centering
\vbox{
\hbox{
\psfig{file=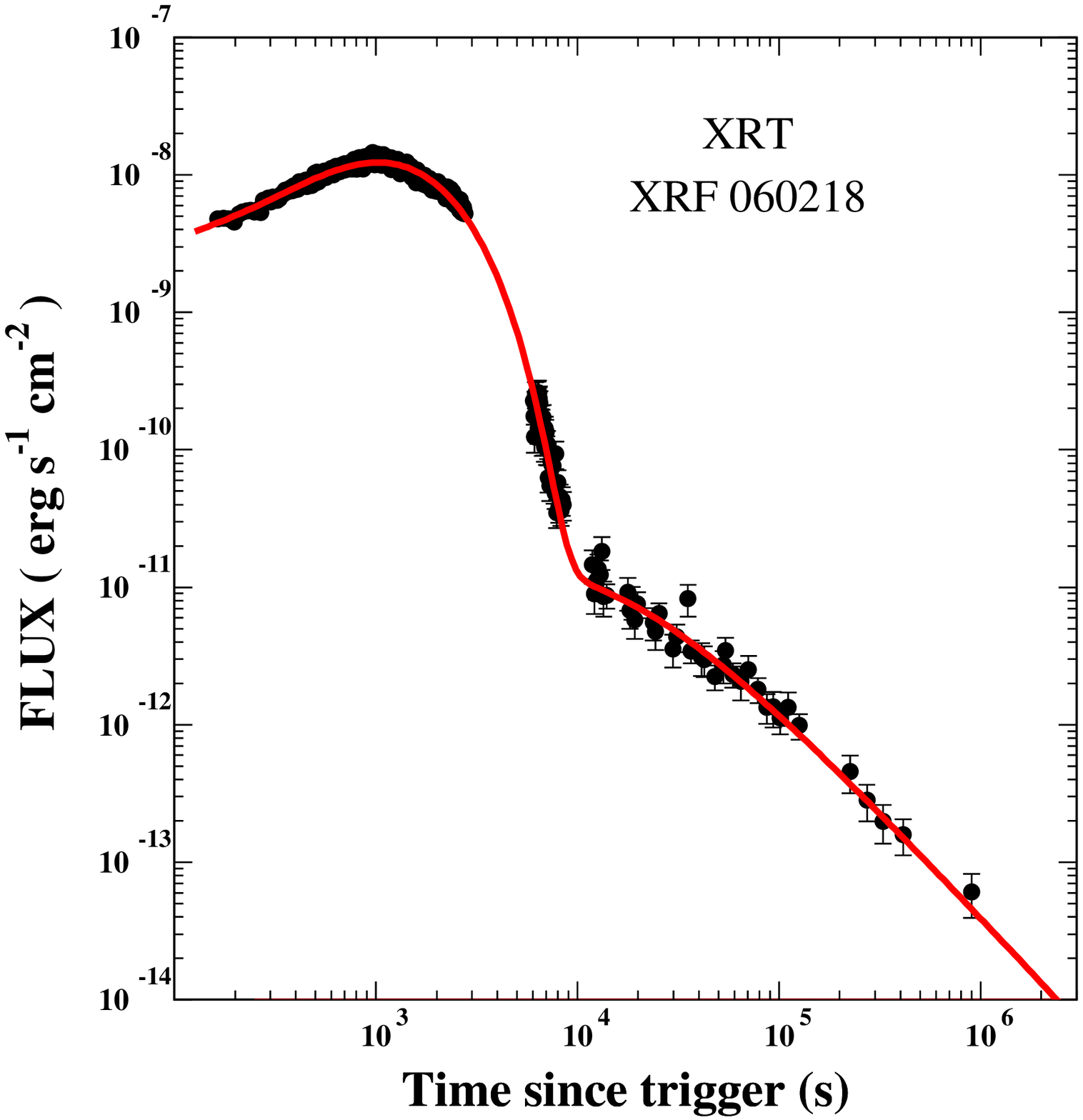,width=6cm}
\hskip -.5cm
\psfig{file=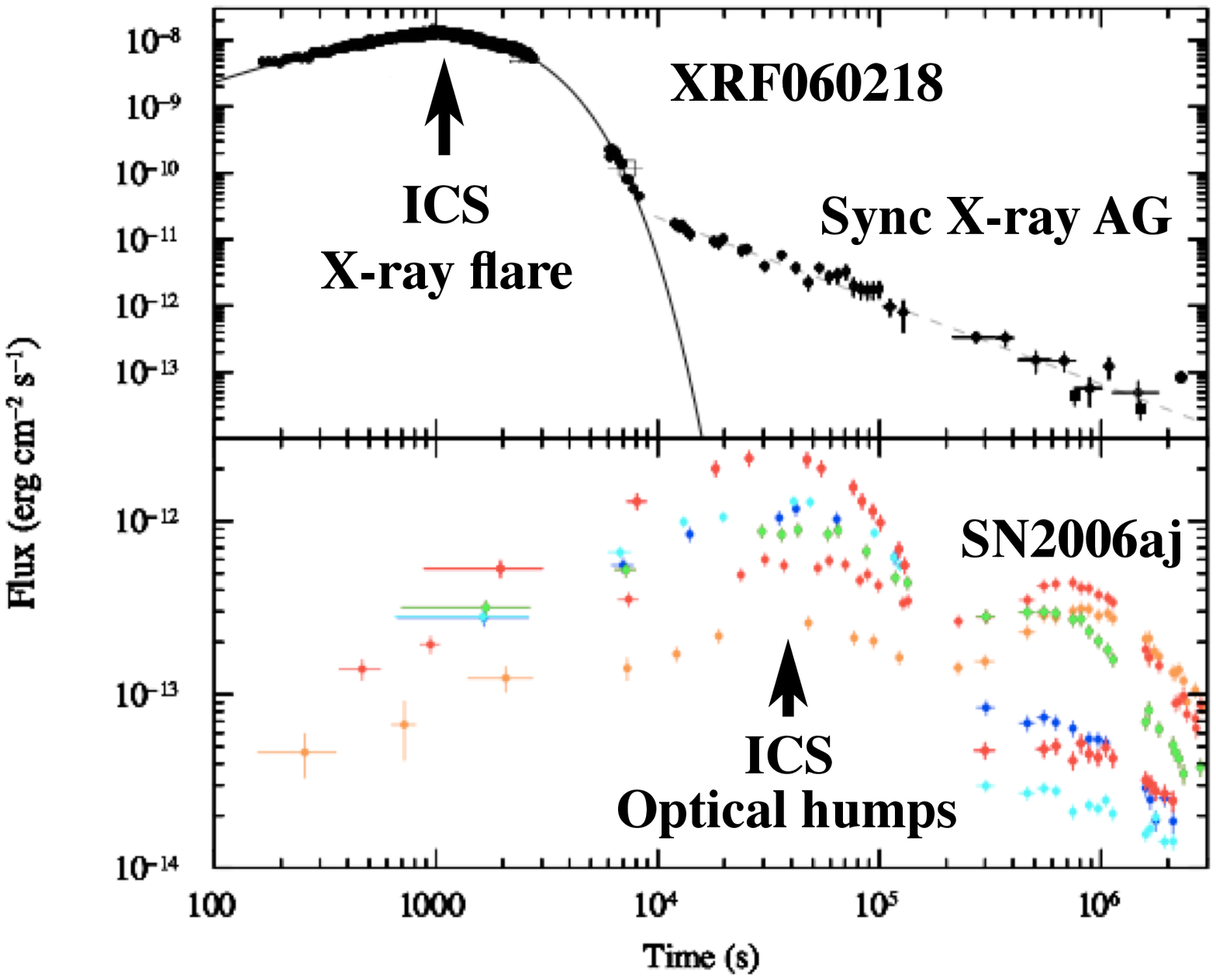,width=7.4cm}
}}
\centering
\vbox{
\hbox{
\psfig{file=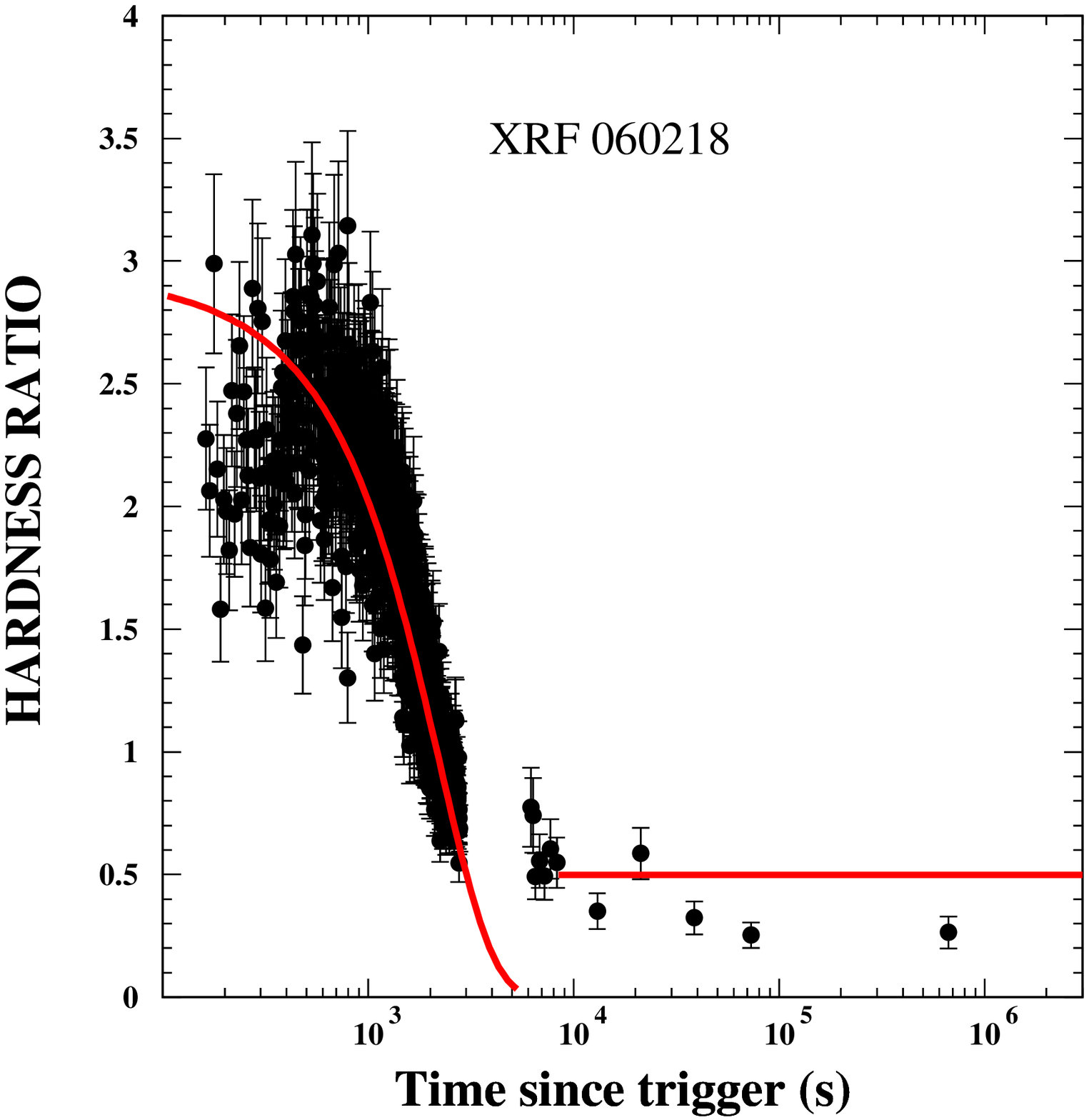,width=6cm}
\psfig{file=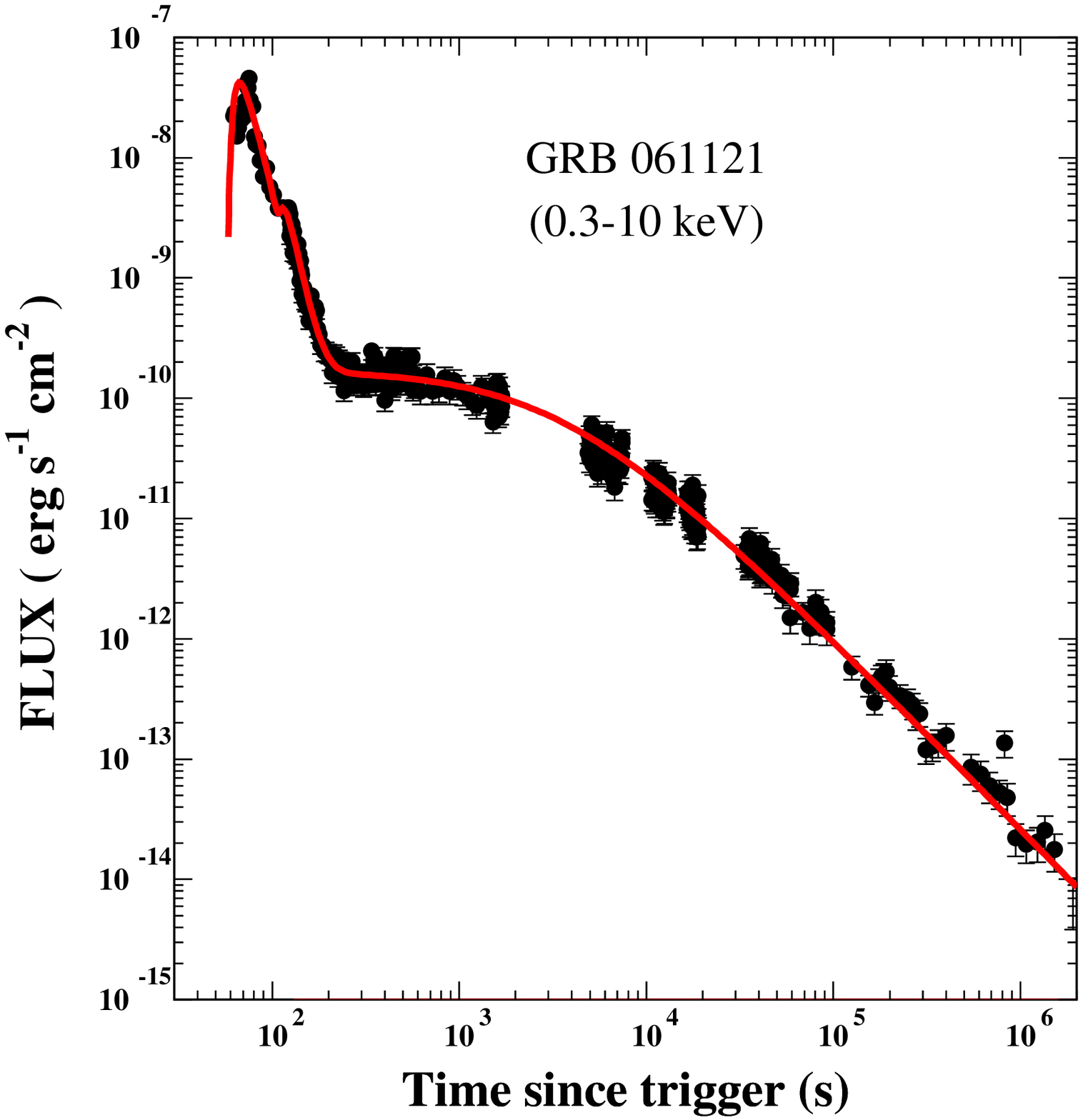,width=6cm}
\vspace{-.5cm}
}}
\vspace*{8pt}
\caption{Left to right and top to bottom:
a) The X-ray light curve of XRF 060218.
b) Data on XRF 060218/SN2006aj. Upper part:
the 0.3-10 keV SWIFT-XRT light curve, with
fits by Campana et al.\cite{Campana}. Lower part: UVO light curves.
In our model, the X-ray flare and optical humps are made by ICS by a single CB.
c) The hardness ratio of XRF 060218.
d) The extensive X-ray light
curve of GRB 061121. The lines in (a,c,d) are CB-model fits.
}
\label{postSwift2}
\end{figure}

The $\gamma$ rays of a GRB occur in a series of {\it pulses}, 1 and 2 in the
examples of fig.~\ref{try}. Swift has clearly established that somewhat
wider X-ray {\it flares} coincide with the $\gamma$ pulses, having, within errors, the
same start-up time. On occasion, even wider optical {\it humps} are seen, 
as in fig.~\ref{postSwift1}a. The X-ray counterpart of the second hump in 
this figure is
clearly seen in fig.~\ref{postSwift1}b. In an XRF the X-ray flares can be
very wide, as in the one-flare example of fig.~\ref{postSwift2}a. In such
a case, the accompanying optical `humps'  peak very late, at
$t\!=\!{\cal{O}}(1$d), as in fig.~\ref{postSwift2}b. All these interconnected 
$\gamma$-pulses, X-ray flares and optical humps are described by 
Eqs.~(\ref{totdist},\ref{shape}). They are obviously 
manifestations of a common underlying phenomenon, which we shall dig out.
Finally, Swift has discovered that not all X-ray light curves are
smooth after the onset of their fast decay, as the one in fig.~\ref{postSwift2}a
is. Well after $\gamma$ pulses are no longer seen, relatively weak
X-ray flares may still be observable, as is the case 
in figs.~\ref{postSwift1}c,\ref{postSwift2}d.

\section{Breath-taking entities: the astrophysical jets of cannonballs}

\begin{figure}[]
\centering
\vbox{\hspace {.4cm}
\hbox{
 \epsfig{file=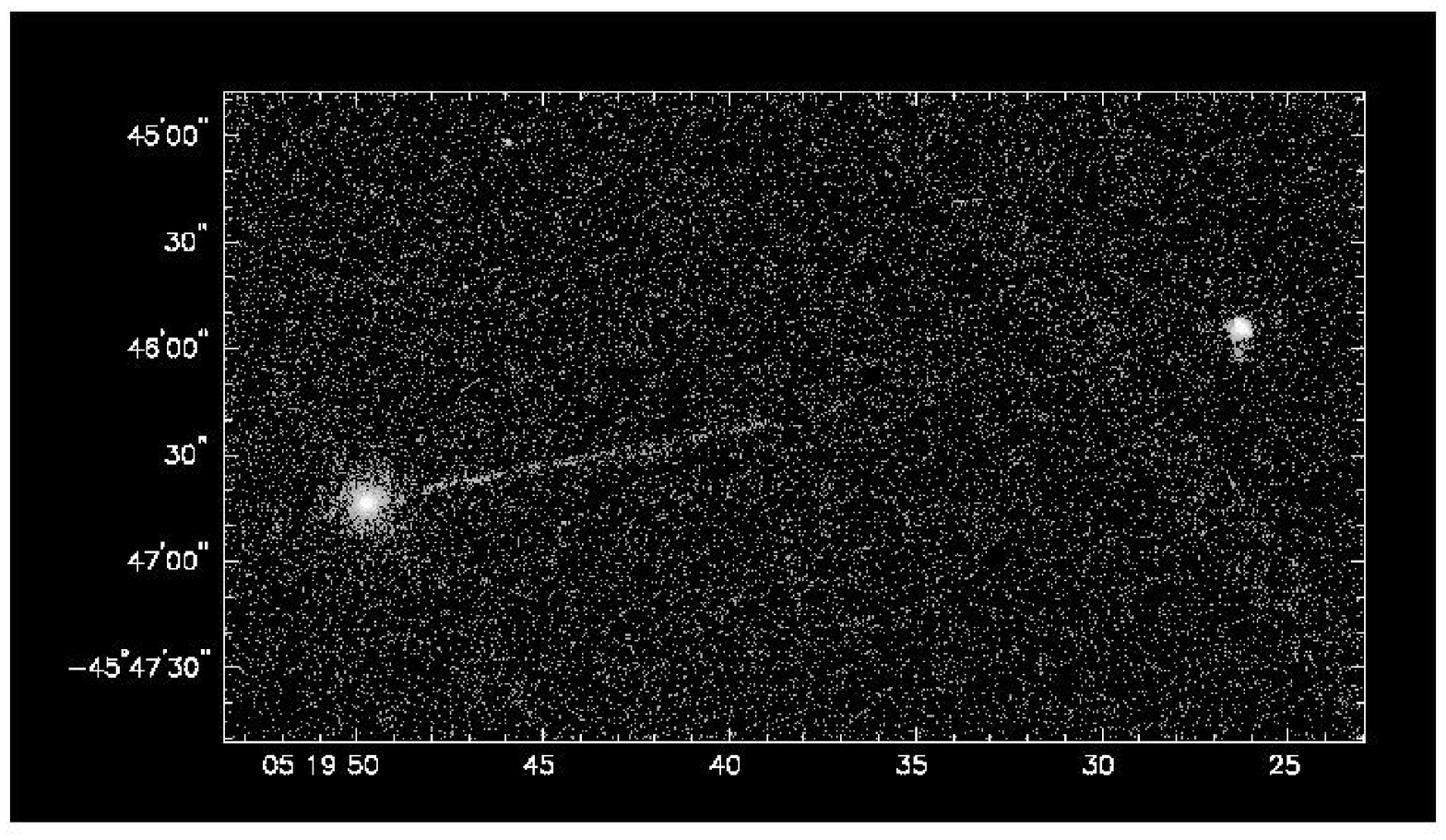,width=7.cm,angle=90}
 \epsfig{file=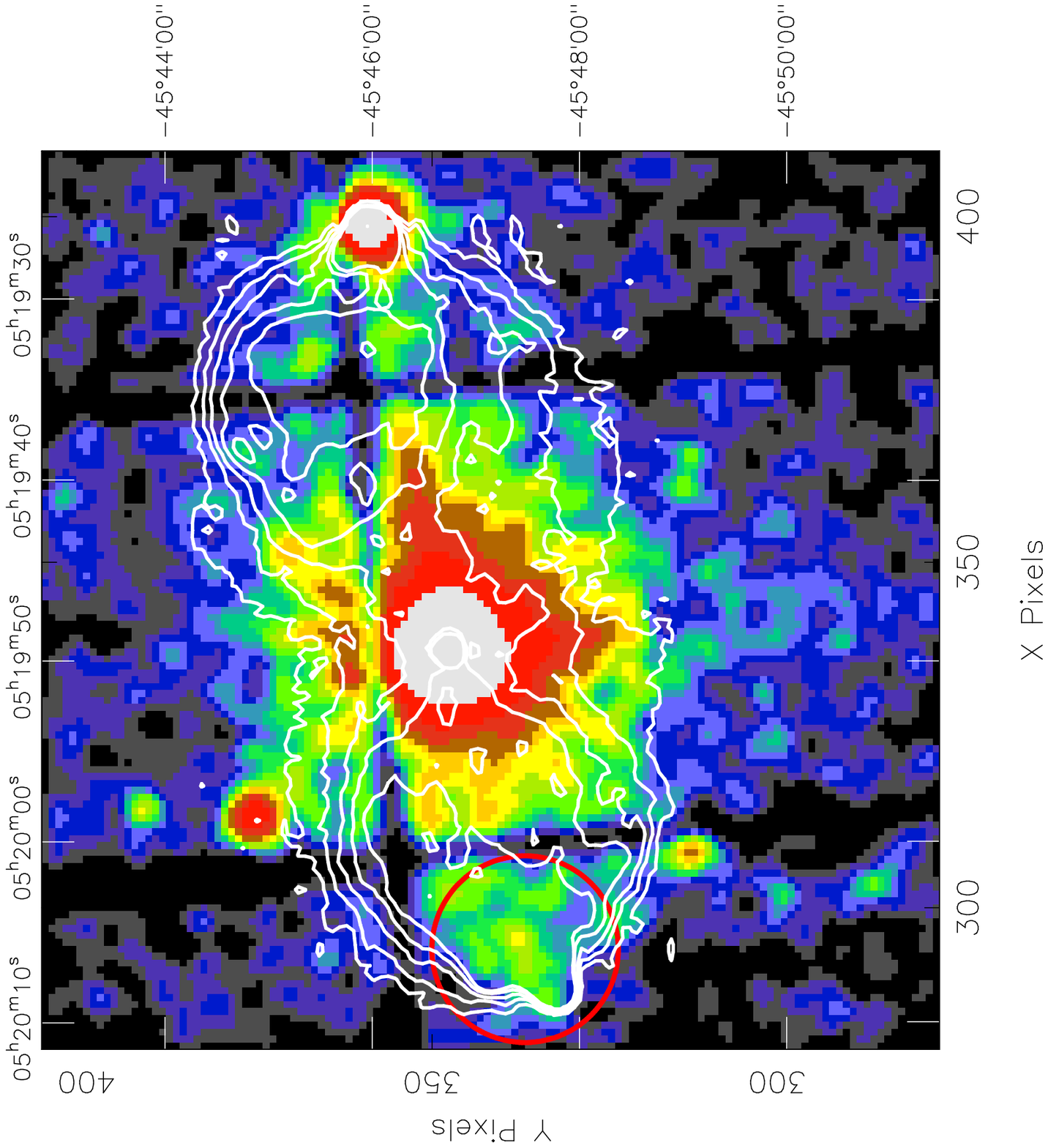,width=6.5cm}
}\hspace {.2cm}
\hbox{
\epsfig{file=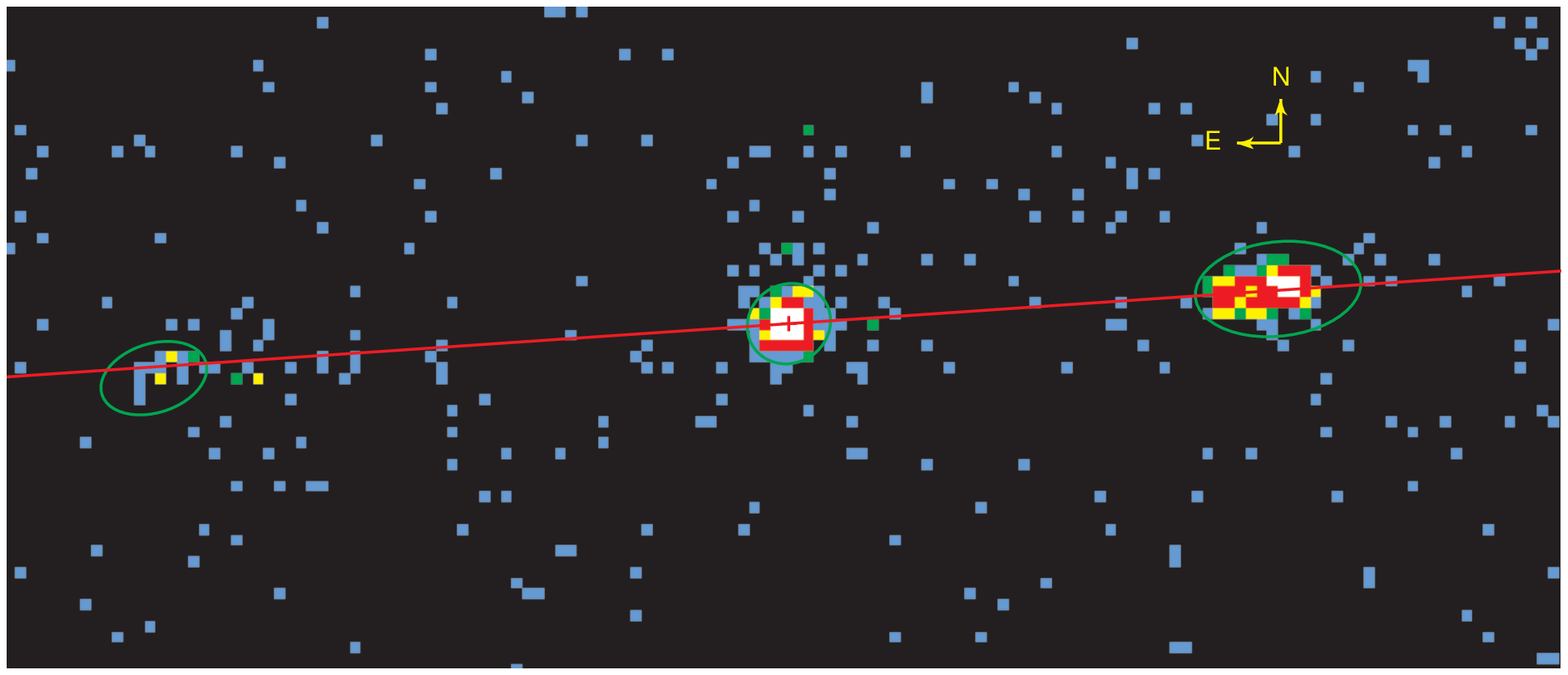,width=6.5cm,angle=90}
\epsfig{file= 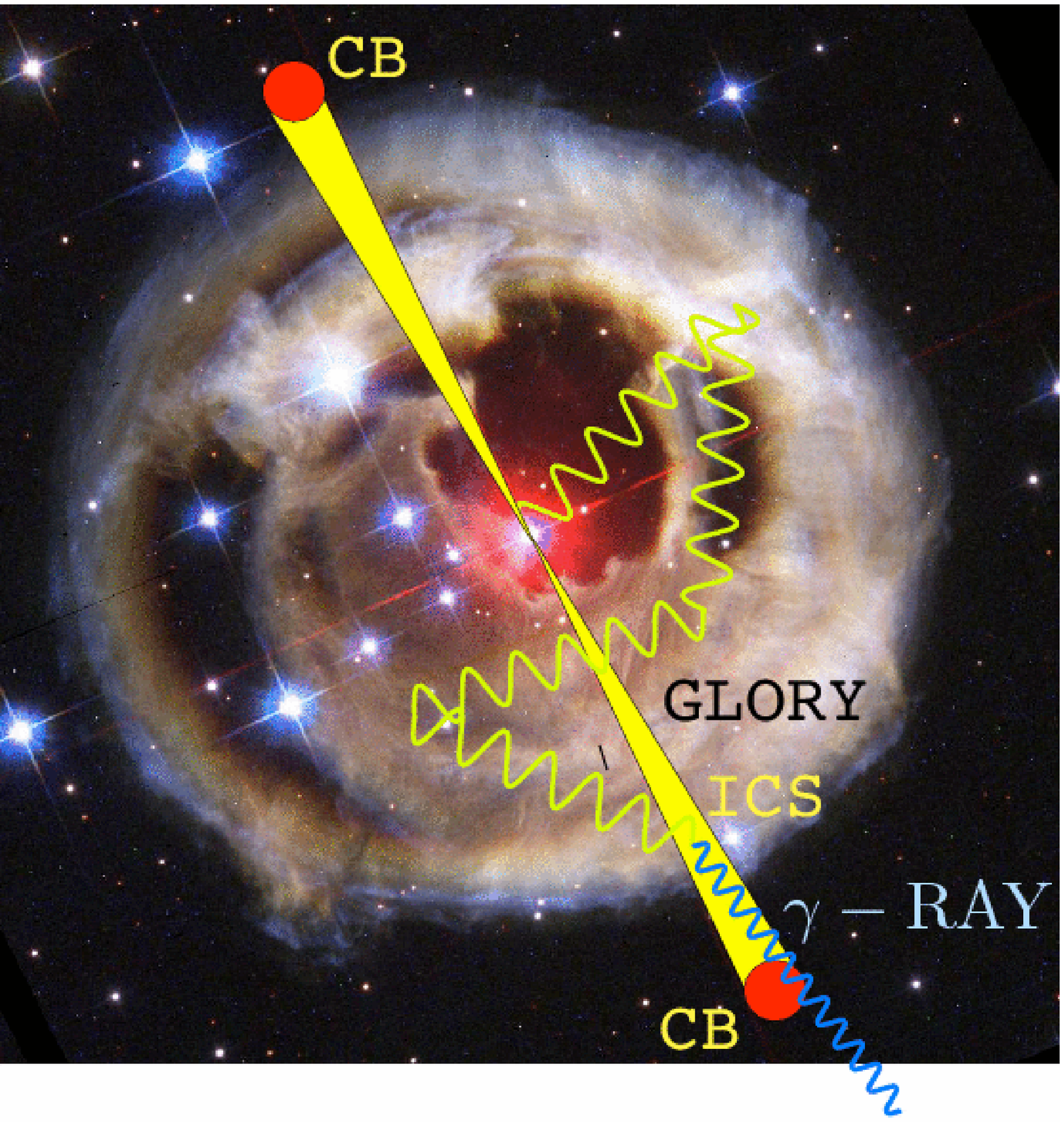, width=6.5cm}
}
}
\vspace*{-5pt}
 \caption{Left to right and top to bottom. a) and b) X-ray images of Pictor A. 
 b) Image
centred at the leftmost spot in (a) and
superimposed on VLA radio contours.
c) Two relativistic CBs emitted in opposite directions
 by the microquasar XTE J1550-564, seen in X-rays.
d)  HST picture from 28 October 2002 of the {\it glory}, or light echo,
 of a pre-supernova outburst of the red supergiant V838 Monocerotis,
doctored with some CB-model art-work. 
 }
 \label{Pictor}
\end{figure}

A look at the web --or at the sky, if you have the means-- results in the
realization that jets are emitted by many astrophysical systems.
One  impressive case
is the quasar Pictor A, shown in figs.~(\ref{Pictor}a,b). {\it Somehow}, its active
galactic nucleus is discontinuously spitting {\it something that does not
appear to expand sideways before it stops and blows up}, having by then
travelled almost $10^6$ light years. Many such 
systems have been observed. They are very relativistic: the Lorentz factors (LFs)
$\gamma\!\equiv\! E/(mc^2)$ of their ejecta are typically of ${\cal{O}}(10)$.
The mechanism responsible for these mighty ejections ---suspected
to be due to episodes of violent accretion into a very
massive black hole--- is not understood.

In our galaxy there are `micro-quasars', whose central black
hole's mass is a few $M_\odot$. The best studied
is GRS 1915+105.
In a non-periodic manner, about once a month, this object emits two
oppositely directed {\it cannonballs}, travelling at $v\sim 0.92\, c$.
When this happens, the continuous X-ray emissions
 ---attributed to an unstable accretion disk--- temporarily decrease.
Another example is the $\mu$-quasar XTE J1550-564, shown in
fig.~\ref{Pictor}c. 
The process reminds one of the blobs thrown
up as the water closes into the `hole' made by a stone
dropped onto its surface, but it is not understood; for quasars and $\mu$-quasars, 
the `cannon's' relativistic, general-relativistic, catastrophic,
magneto-hydro-dynamic details remain to be filled in!
Atomic lines of many elements have been seen
in the CBs of $\mu$-quasar SS 433. Thus, at least in this case, the
ejecta are made of ordinary matter, and not of a fancier substance,
such as $e^+e^-$ pairs.

\section{The Cannonball Model: summary}
The `cannon' of the CB model is analogous to the ones
of quasars and microquasars.
In an {\it ordinary core-collapse} SN event, due to the parent star's
rotation, an accretion disk  is produced around
the newly-born compact object, either by stellar material originally
close to the surface of the imploding core, or by more distant stellar matter
falling back after the shock's passage \cite{GRB1,ADR}. A CB 
made of {\it ordinary-matter plasma} is emitted, as
in microquasars, when part of the accretion disk
falls abruptly onto the compact object. {\it Long-duration} GRBs  
and {\it non-solar} CRs are produced by these jetted CBs. To agree with 
observations, CBs must be launched with LFs, $\gamma_0\!\sim\!10^3$,
and baryon numbers $N_{_{\rm B}}\!=\!{\cal{O}}(10^{50})$, corresponding to
$\sim\!1/2$ of the mass of Mercury, a miserable $\sim\!10^{-7}$th of a solar mass.
Two jets, each with $n_{_{\rm CB}}\!=\!\langle n_{_{\rm CB}}\rangle\!\sim\!5$ CBs, carry  
\begin{equation}
E_{\rm jets}=2\,n_{_{\rm CB}}\,\gamma_0\,N_{_{\rm B}}\,m_p\,c^2
\sim 1.5\times 10^{51}\;\rm erg,
\label{Ejets}
\end{equation}
comparable to the energy of the SN's non-relativistic shell,
that is ${\cal{O}}(1\%)$ of the explosion's energy, $\sim\!98$\% of which is carried
away by thermal neutrinos.

We have seen that long GRBs are indeed made by SNe, as advocated
in the CB model well before the pair GRB030329/SN2003dh convinced
the majority. But
do SNe emit cannonballs? Until 2003\cite{SLum030329}, there was only one
case with data good enough to tell: SN1987A, the core-collapse
SN in the LMC, whose neutrino emission was seen. Speckle interferometry
data taken 30 and 38 days after the explosion\cite{Costas}
did show two back-to-back relativistic CBs, see fig.~\ref{try2}e,f. 
The approaching one was {\it superluminal}: seemingly moving at $v\!>\!c$.

\begin{figure}
\vskip -1.cm
\hskip 2truecm
\begin{center}
{\epsfig{file=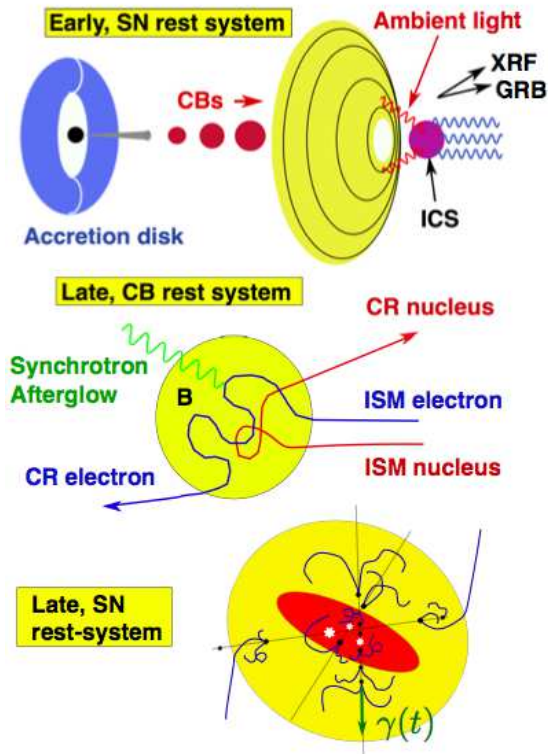, width=7.3cm}}
\end{center}
\caption{The CB model
of GRBs, XRFs and CRs. A core-collapse SN results in
a (black) compact object, a fast-rotating torus of non-ejected
material and a (yellow) shell of non-relativistic ejecta. 
Matter (not shown) episodically accreting
into the central object produces
two narrowly collimated beams of CBs;  some of
the `Northern' CBs are depicted. As the CBs move through
the {\it glory} of non-radial
 light surrounding the star, they forward Compton up-scatter
its photons to GRB or XRF energies, depending on how close
the line of sight is to the CBs' direction. 
Each CB produces a GRB `pulse'. Later, a CB gathers and
scatters ISM particles,
which are isotropized by its inner magnetic field. In the SN rest
system the particles are boosted by the CB's motion: they have become
CRs. The synchrotron radiation of the gathered electrons is the late AG
of the GRB or XRF. As the CBs'
collisions with the ISM
slow them down, the CBs generate CRs all along their 
trajectories, in the galaxy and its halo. 
CRs are also forward-produced, diffusing thereafter in the
local magnetic fields.}
\label{figCB}
\end{figure}

A summary of the CB model is given in Fig.~\ref{figCB}. 
The {\it `inverse' Compton scattering} (ICS) of light by electrons within a CB  
produces a forward beam of higher-energy photons: a pulse of a GRB or an XRF.
The target light is in a temporary reservoir: the {\it glory}, illuminated by the
early SN light and illustrated by analogy in fig.~\ref{Pictor}d.
A second mechanism, {\it synchrotron radiation} (SR), takes over later and generally
dominates the AG. The $\gamma$ rays ionize the ISM on the CB's path. The CBs
collide with the ISM electrons and nuclei, boosting them to cosmic ray status.
The ISM penetrating a CB's plasma creates turbulent magnetic fields within
it. The ISM electrons moving in this field emit the mentioned SR. 
This paradigm accounts for all properties of GRBs and CRs.

The observed
properties of a CB's radiation depend crucially on the angle $\theta$ of its
motion relative to the line of sight to the observer, via the Doppler factor
\begin{equation}
\delta  =  1/[\gamma(1  -  \beta\cos\theta)]  \approx
  2\gamma/(1  +  \theta^2\gamma^2)
\label{delta}
\end{equation}
by which a photon's energy is
boosted from the CB's rest system to that of a (cosmologically nearby) observer. For an
isotropic emission in the CB's system, the observed photon number, 
energy flux and luminosity are $\propto \delta,\,\delta^3,\,\delta^4$,
respectively, just as in a $\nu$ beam from $\pi$ decay. That makes GRBs
observable only extremely close to one of their bipolar CB axes, 
$\theta\!=\!{\cal{O}}(1/\gamma)\!\sim\!1$ mrad [typically 
$\gamma(t=0)\!=\!\gamma_0\!\sim\!\delta_0\!\sim\!10^3$;
and AGs are observed till $\gamma(t)\!\sim\!\gamma_0/2$]. 

The relation between CB
travel-time in the host galaxy, $dt_\star\!=\!dx/(\beta\,c)$,  and observer's time, $t$, is 
$dt_\star/dt\!=\!\gamma\,\delta/(1\!+\!z)$. Stop in awe at this gigantic factor:
a CB whose AG is observed for 1 day may have travelled for ${\cal{O}}(10^6)$ light days,
what a fast-motion video!
A CB with $\theta\!=\!1/\gamma\!=10^{-3}$ moves in the sky at an
apparent transverse velocity of $2000\, c$, yet another large Doppler aberration.

\section{GRB afterglows in the CB model }

Historically, two GRB phases were distinguished: a prompt one, and the 
{\it after}-glow. Swift data have filled the gap, there is no longer a very clear distinction.
Nor is there a profound difference between the CB-model's  radiation mechanisms,
since synchrotron radiation is but Compton scattering on virtual photons and,
 in a universe whose age is finite, all observed photons were virtual.

\begin{figure}[]
\centering
\vskip -1.5cm
\vbox{\hspace {.4cm}
\hbox{
 \epsfig{file=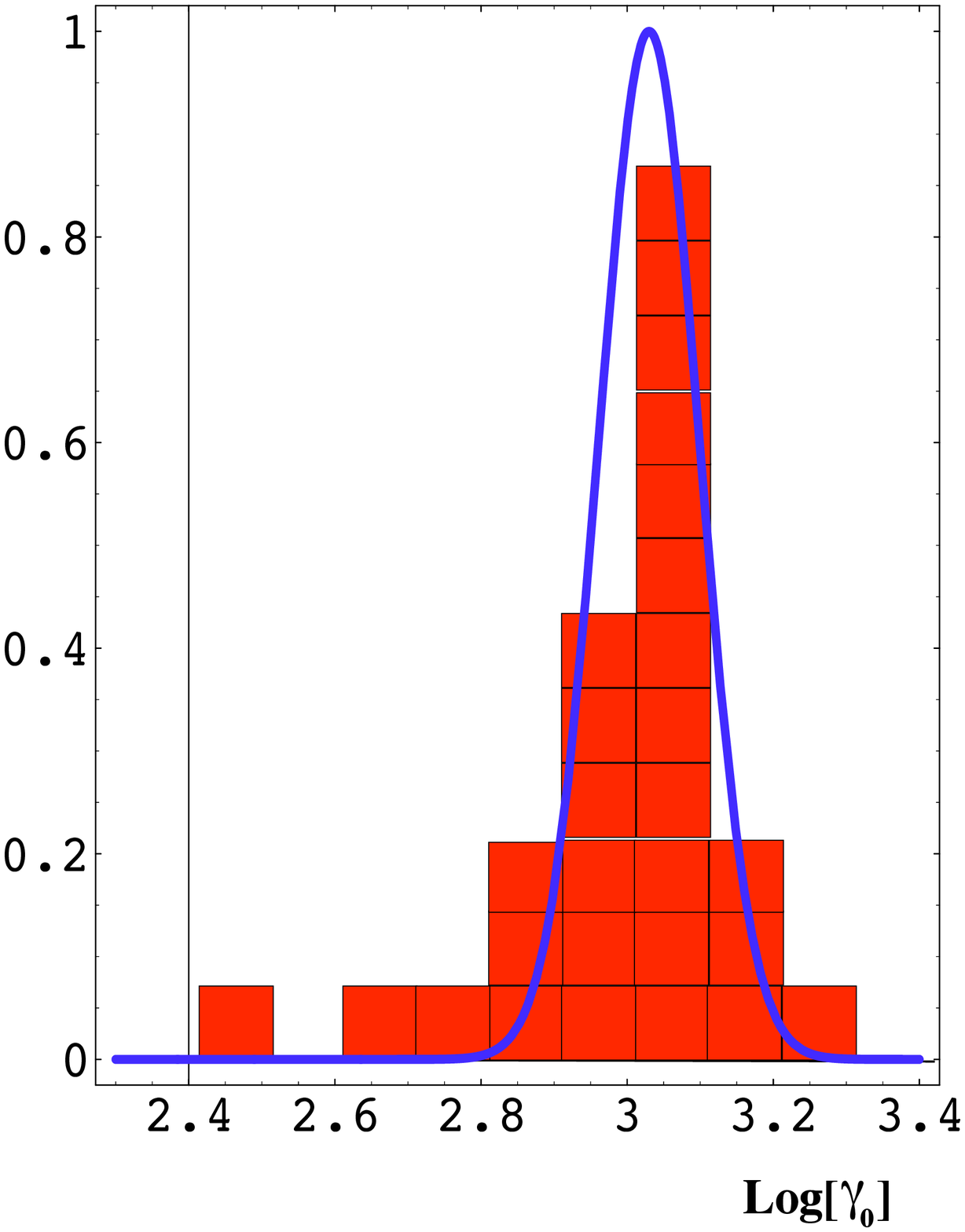,width=4.cm}
 \epsfig{file=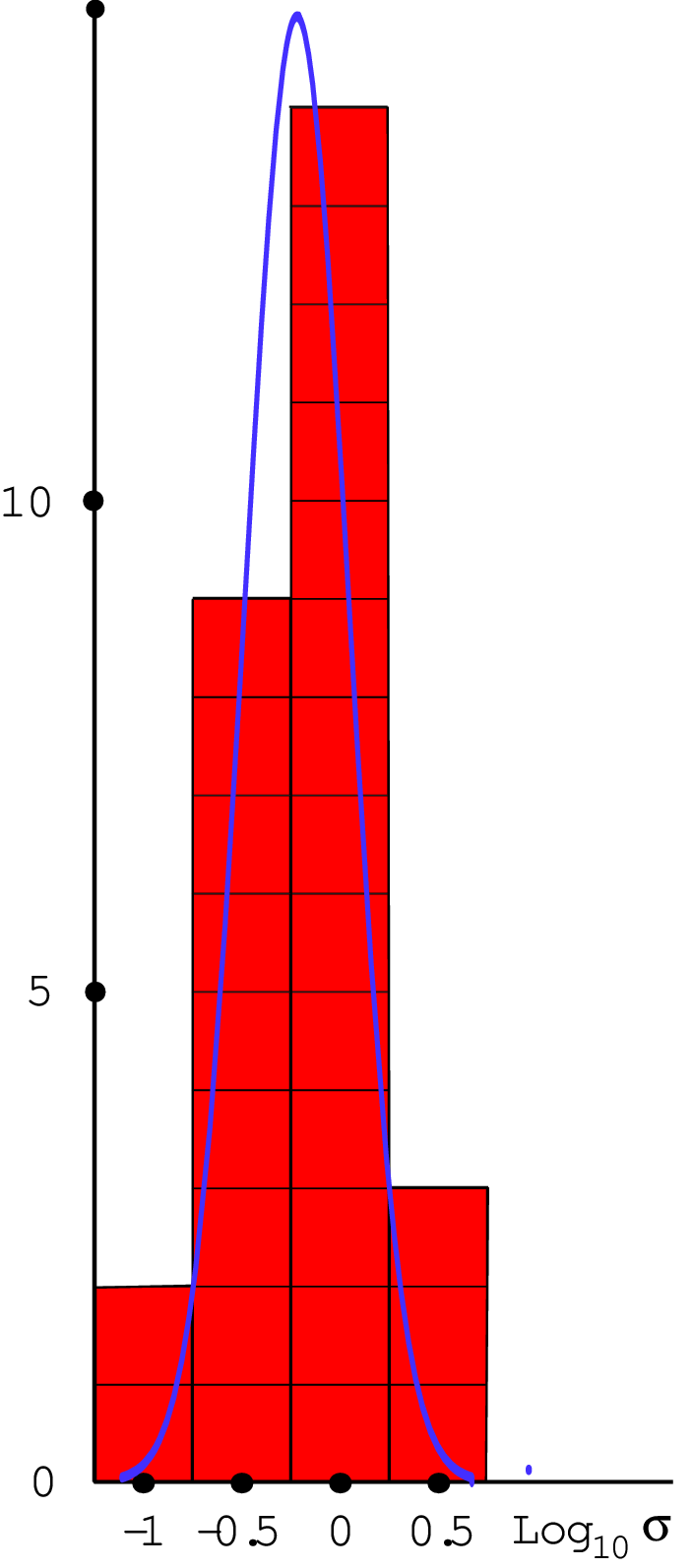,width=2.2cm}
  \epsfig{file=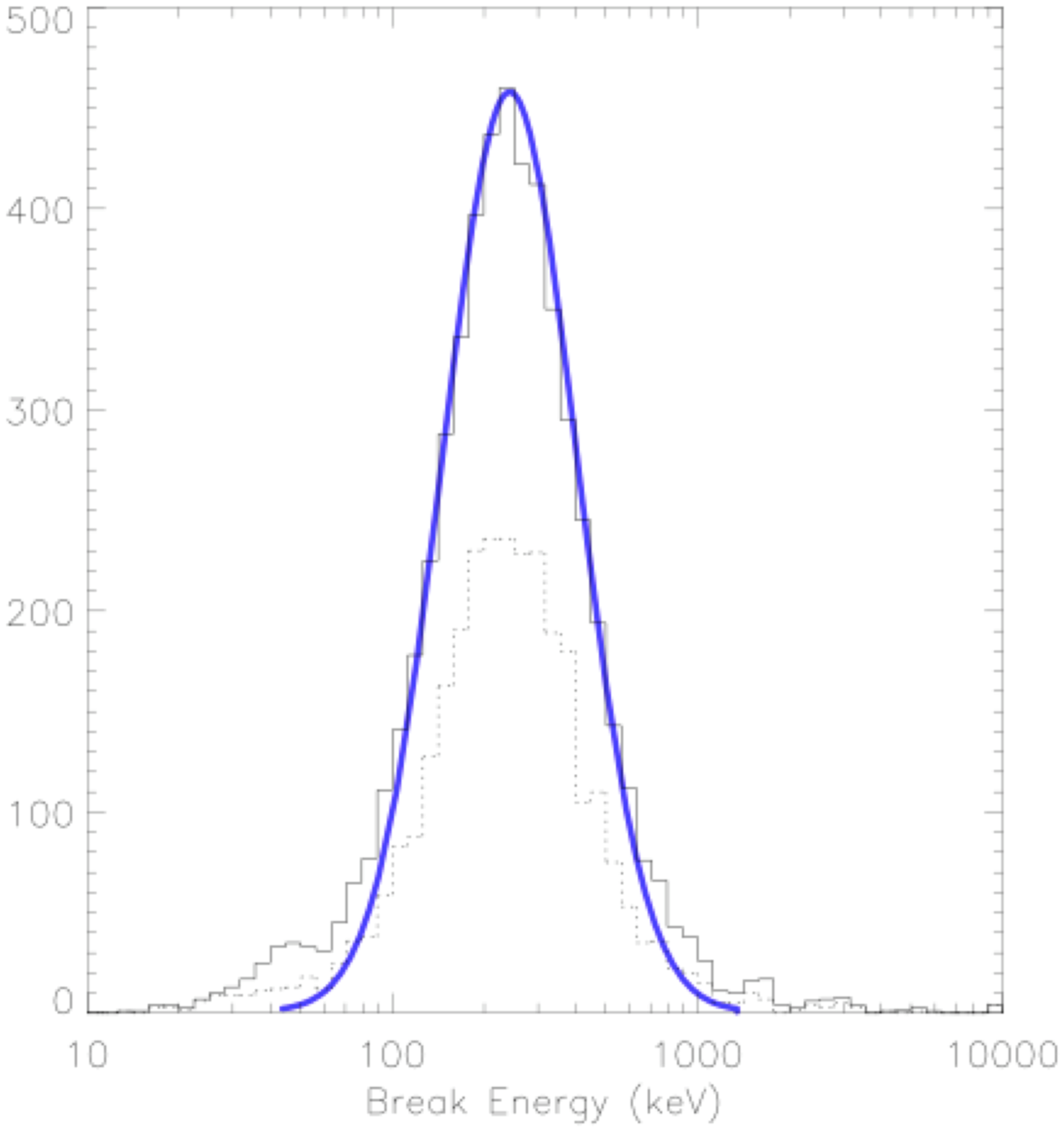,height=5.2cm,width=5.cm}
}
\hspace {.2cm}
\hbox{
\epsfig{file=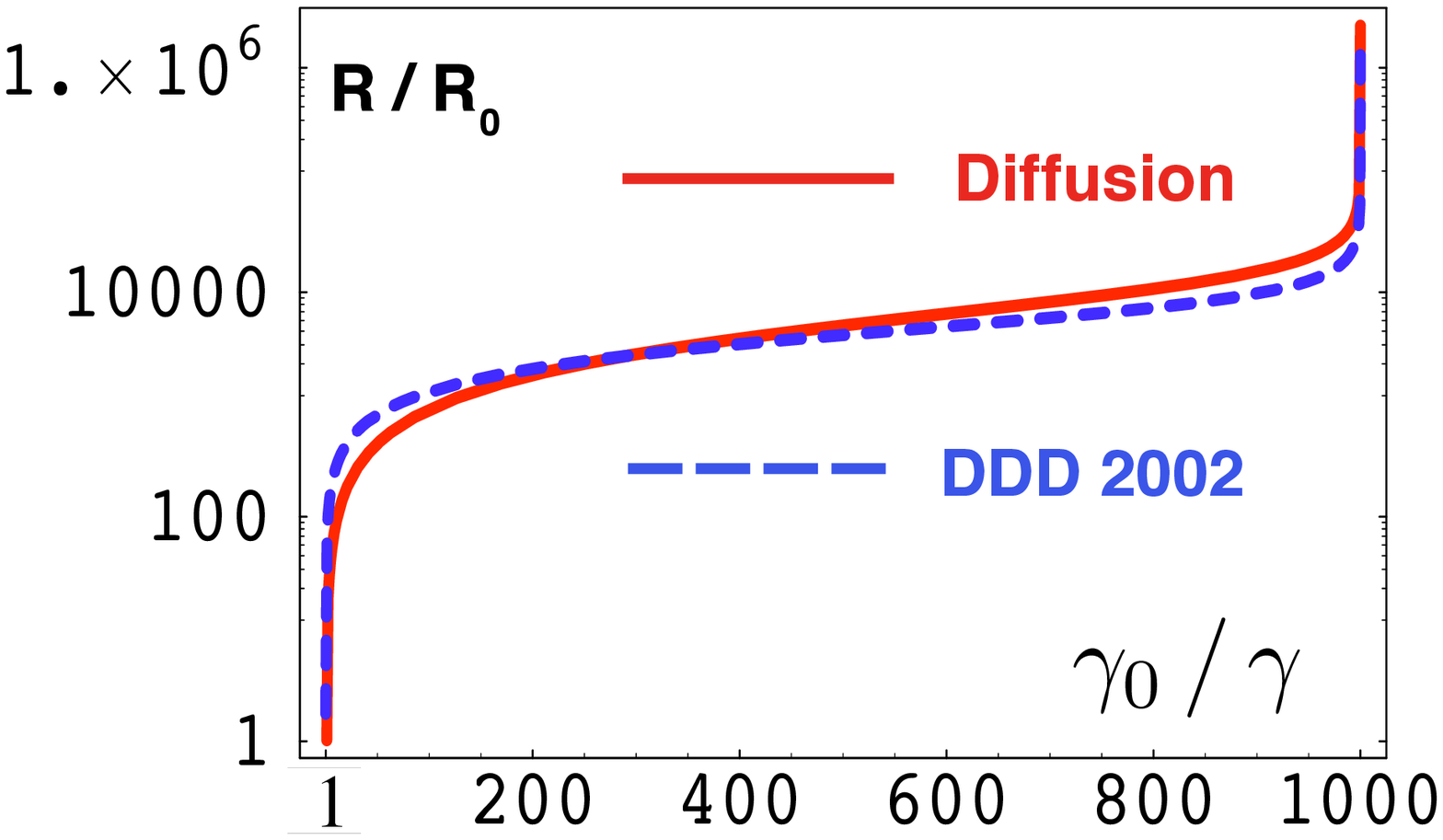,width=11cm}
}
}
\vspace*{-5pt}
 \caption{Right to left, top to bottom. a) The distribution of ${\rm Log}_{10}(\gamma)$
 for a score of  pre-Swift GRBs of known $z$. b) The distribution of 
 $\sigma\!\equiv\!{\rm Log}_{10}(\gamma\delta/10^6)$ for the same GRBs, also extracted from the
 analysis of their AGs. c) The distribution of 
peak or `break' energies in the spectrum of many  pre-Swift GRBs.
The line is the CB-model prediction, based on Eq.~(\ref{boosting}) and the observed $\sigma$ 
distribution.
d) The behaviour of $R(\gamma/\gamma_0)$ in two extremes (diffusive or instantaneous)
for the way the intercepted ISM ions reexit a CB. The coasting behaviour at $R\!\sim\!R_c$
and final blow-up at $\gamma\!\to\!1$
are of the form $R\!\approx\!R_c\,(\gamma_0/\beta\,\gamma)^{2/3}$.
 }
 \label{Distributions}
\end{figure}

In the understanding of GRBs in the CB model, SR-dominated AGs came first.
The CB-model AG analysis is strictly a `model': it contains many simplifications.
But the comparison with data determines  the  distributions
of the relevant parameters. Given these, the predictions for CRs and for
the ICS-dominated phase of GRBs (such as all properties of the $\gamma$-ray pulses)
involve only independent observations, basic physics and no `modeling'. 
For the reader who might want to move to the more decisive
sections, I anticipate the contents of this one. The distribution of
 $\gamma_0$ and $\gamma_0\,\delta_0$ values of pre-Swift GRBs 
 are shown in fig.~\ref{Distributions}a,b.
 The radius of a CB evolves as in fig.~\ref{Distributions}d. A CB does not expand 
 inertially; for most of its trajectory it has a slowly changing radius, as a common
 {\it cannonball} does. The baryon number of observed CBs is of ${\cal{O}}(10^{50})$.

To determine the fate of a CB, we make the following assumptions.
CBs initially expand at $\beta_s\,c\!=\!{\cal{O}}(c/\sqrt{3})$, the relativistic speed of sound,
swiftly becoming spherical in their rest system and losing
memory of their initial size. For the CB's baryon number returned by the
analysis, this means that CBs become `collisionless' fast: their 
nuclei and electrons do not often collide with the ISM ones 
they encounter. Hadron and Thompson cross sections being similar,
CBs also become transparent, except to long radio waves, losing
their radiation pressure.
In agreement with first-principle calculations of the relativistic
merger of two plasmas\cite{Frede}, a
chaotic magnetic field is generated within a CB by the ISM particles it sweeps in.
In accordance with observations of similar plasmas (such as the ISM itself
and the CRs it contains) the CB's magnetic field is in energy equipartition
with the impinging ISM, resulting in:
\begin{equation}
B_{_{\rm CB}}(\gamma)=3\;{\rm G}\;{(\gamma/ 10^3})\;
\left[{n/ (10^{-2}\;{\rm cm}^{-3}})\right]^{1/2}\; ,
\label{B}
\end{equation}
where $n$ is the ISM baryonic number density, normalized to a typical value in the
`superbubbles' in which most SNe and GRBs are born. 

In a CB's rest system the motion of its constituents is an inertial memory of 
the initial radial expansion, whose kinetic energy is larger than the one of
the CB's magnetic field. An ISM proton entering a CB will meander it its
magnetic field and be isotropically reemitted (in the CB's rest system). 
The rate of radial momentum loss per unit surface is a surface pressure
countering the expansion. We assume that the dominant effect of this
pressure is to counteract the expansion.
We use Newton's law to compute the ensuing radial deceleration and the CB's
radius $R(\gamma)$. The results are shown in fig.~\ref{Distributions}d.
A CB initially expands quasi-inertially. It subsequently settles into a slowly evolving radius
till it blows up as its motion becomes non-relativistic (DD06),
obeying: 
\begin{equation}
R(\gamma)\approx R_c\,(\gamma_0/\beta\gamma)^{2/3},\;\;{\rm with\,}
R_c={\cal{O}}(10^{14}\rm cm),\;for\,typical\, parameters.
\label{Radius}
\end{equation}
This is a complex problem, and ours 
 is a big simplification, once assessed by a cunning referee as
  ``almost baron Munchhausen''. Yet, the result 
describes the surprising `jet self-focusing' observed, e.g., in Pictor A, see fig.~\ref{Pictor}a.

The collisions with the ISM  continuously decelerate a
CB. For a given $R(t)$ and ISM baryon number per unit volume $n$, 
energy-momentum conservation dictates the explicit form of the
CB's diminishing Lorentz factor $\gamma(t)$. Typically $\gamma(t)$
is roughly constant for a day or so of observer's time, steepening
to $\propto\,t^{-1/4}$ thereafter. During the short $\gamma$-ray
emission time, $\gamma(t)\!\approx\!\gamma_0$.

We assume that practically all of the energy of the ISM electrons (of number density
$n_e\!\simeq\!n$) entering a CB is reemitted fast in the form of SR, so that
the corresponding observed frequency-integrated AG power per unit area is
 $dF/(dt\,d\Omega)\!=\!\pi\,R^2\,n_e\,m_e\,c^3\,\gamma^2\,\delta^4/(4\,\pi\,D_L^2)$,
 with $D_L$ the luminosity distance. The CB deceleration law, dictated
 by energy-momentum conservation, is equally simple: 
 $M_0\,\gamma_0\,d\gamma\!=\!-\pi\,R^2\,n\,m_p\,\gamma^3\,dx$, for an H-dominated ISM
 and in the extreme in which the re-emission time of protons is long on the
 scale of the CB's deceleration time. For constant 
 $n$, the distance travelled by a CB is: 
 \begin{equation}
 x(\gamma)\approx L\,\left[\left({\gamma_0/ \gamma}\right)^{2/3}-1\right],\;\;\;
 L={3\,N_{_{\rm B}}/( 2\pi\,R_c^2\,n\,\gamma_0^3)}=180\,{\rm pc,}\,
 \label{distance}
 \end{equation}
where the number is for $N_{_{\rm B}}\!=\!10^{50}$, $n\!=\!10^{-2}\,{\rm cm}^{-3}$,
$R_c\!=\!10^{14}\,{\rm cm}$, $\gamma_0\!=\!10^3$. Given all this, it appears
easy to extract from an AG's normalization and shape the values of
$\theta$, $\gamma_0$ and $N_{_{\rm B}}$, if one trusts
 the estimate of $R(\gamma)$ and uses a typical $n$. Limited observational
 information makes life a bit harder.

The spectrum of fig.~\ref{f1}f is actually the one predicted by the CB model, illustrated
for a typical choice of parameters. The chromatic
 `bends' shown as dots in this figure, for instance,
are `injection bends':  the typical SR energy, in the CB's magnetic
field, of the electrons entering it, at time $t$,
with a (relative) LF $\gamma=\gamma(t)$. The small portion of the spectrum
above the bend is emitted by a tiny fraction of electrons 
`Fermi-accelerated' in the CB's turbulent magnetic fields to a pre-synchrotron-cooling spectrum
$E_e^{-p}$, with $p\!\sim\!2.2$.
The prediction fits with no exception
the AGs of the first score of well measured GRBs (DDD02/03) of known $z$.
But only on rare occasions can one  clearly see in an AG the contributions of the various CBs
seen in the $\gamma$-ray count rate (a counterexample is the GRB of figs.~\ref{try}b, \ref{try2}a,b).
Thus, generally, the parameters extracted from AG fits refer to a dominant CB or to
an average over CBs.

After an observer's day or so, the optical and X-ray AG are typically SR-dominated, 
are above the injection bend, and are of the approximate form:
\begin{equation}
F_\nu\propto n_{_{\rm CB}}\,n^{1.05}\,R_c^2\,\gamma^{2.3}\,\delta^{4.1}\,\nu^{-1.1},
\label{SRFlux}
\end{equation}
where the unwritten proportionality factors,
such as $D_L^{-2}$, are known. From a fit to the shape of $F_\nu(t)$
one obtains $\theta\,\gamma_0$ and the combination $L$ of Eq.~(\ref{distance}). At late times
$F_\nu\!\propto\!\gamma(t)^{6.4}\!\propto\!\gamma_0^{6.4}$ with a coefficient determined
by the other fit parameters. 
The value of $\gamma_0$
is extracted from the 6.4 root of the inverse of Eq.~(\ref{SRFlux}), so that, for
a result within a factor of 2, one can tolerate large errors in the chosen $n$ or 
in the estimate of $R_c^2$. Trusting these, one can extract $N_{_{\rm B}}$
from $L$, perhaps with an uncertainty of one order of magnitude\footnote{This is what we
did in DDD02/03 but not quite what we wrote. 
I am indebted to J.~Steinberger for noticing this error.}. 
Eq.~(\ref{SRFlux}) has been used to fit, after the early fast fall-off, the X-ray and
optical data of figs.~\ref{fpreSwift},\ref{postSwift1},\ref{postSwift2}. The required form of 
$\gamma(t)$ is Eq.~(\ref{fpreSwift}), supplemented by 
the relation between CB's mileage and observer's time, see the end of Sec. 5.

\section{A GRB's $\gamma$ rays in the CB model (DD06)}

A pulse of a GRB is made by a CB crossing the parent star's {\it glory}.
The glory is a reservoir of non-radially directed light, fed by the parent
star's luminosity, as in the artist's view of fig.~\ref{Pictor}d. 
For the best studied GRB-associated
SN, 1998bw, and for ${\cal{O}}(1\rm d)$ after the explosion,
the luminosity was $L_{_{\rm SN}}\sim\!5\!\times\!10^{52}$ erg/s, in photons
of typical energy $E_i\!\sim\! 1$ eV. We adopt these values as `priors'
(parameters to be used in calculations, but independent of the CB model).
Massive stars destined to `go supernova' eject solar-mass amounts of matter 
in successive explosions during their last few thousand pre-SN years.
At the `close' distances of ${\cal{O}}(10^{16}$ cm) relevant here, these
stellar coughs generate a thick layer of `wind-fed' material with an 
approximate density profile $\rho\!\propto\!r^{-2}$ and normalization
$\rho\,r^2\!\sim \! 10^{16}$ g cm$^{-1}$, the last prior we need.
The very early UV flash of the SN suffices to ionize the wind-fed 
matter. The Thompson cross section $\sigma_{_{\rm T}}$ is such that
this matter is semitransparent: 
$\sigma_{_{\rm T}}\,\rho\,r^2/m_p\!=\!4\!\times\!10^{15}$ cm. This means
that the number of times a SN photon reinteracts on its way out 
--becoming `non-radial'-- is of ${\cal{O}}(1)$, and that the number density
of such photons is $n_\gamma(r)\!\sim\!L_{_{\rm SN}}/(4\pi\,r^2\,c\,E_i)$.
From emission-time to the time it is still one $\gamma$-ray interaction
length inside the `wind', a CB has travelled for
\begin{equation}
 t_{\rm tr}^{\rm w}=(0.3 \,{\rm s})\,{\rho\,r^2\over 10^{16}\,{\rm g\,cm}^{-1}}
 \,{1+z\over 2}\,{10^6\over \gamma_0\,\delta_0}
 \label{ttrans}
 \end{equation}
of observer's time. That is a typical $\gamma$-ray pulse rise-time 
in a GRB, and the reason why,
closing the loop, distances of ${\cal{O}}(10^{16}$ cm) were relevant.

In the collapse of a rotating star, material from `polar' directions should
fall more efficiently than from equatorial directions. The CBs would then 
be emitted into relatively empty space. We assume that the wind-material
is also under-dense in the polar directions. This is not the case for the
glory's photons, which have been scattered by the wind's matter, and 
partially isotropised.  
During the production of $\gamma$-rays by ICS, $\gamma\!\simeq\!\gamma_0$.

Consider an electron, comoving with a CB at $\gamma\!=\!\gamma_0$, and
a photon of energy $E_i$ moving at an angle $\theta_i$ relative to $\vec r$.
They Compton-scatter. The outgoing photon is viewed at an angle
$\theta$. Its energy is totally determined:
\begin{equation} 
E_p \!=\!  {\gamma\,\delta\over 1\!+\!z}\,
(1\!+\!\cos\theta_i)\, E_i \!=\!  (250\;{\rm keV})\;
\sigma\;{1\!+\!\cos\theta_i\over 1/2}\,
{E_i\over 1\;\rm eV},\;
\sigma\! \equiv\! {\gamma\,\delta\over 10^6}\,{2\over 1\!+\!z}\; , 
\label{boosting} 
\end{equation}
where I set $\beta\!\approx\! 1$ and, for a {\it semi}~transparent wind, 
$\langle\cos\theta_i\rangle\!\sim\!-1/2$. For pre-Swift GRBs $\langle z \rangle\!\approx\! 1$
and, for the typical $\gamma$ and $\delta$,
$E\!=\!250$ keV, the average peak or `break' energy 
in Eq.~(\ref{totdist}). From the fits to the
AGs of the subset of known $z$, we could determine the distribution
of $\sigma$ values, see fig.~\ref{Distributions}b. Its fitted result is
used in Eq.~(\ref{boosting}) to predict the overall $E_p$ distribution,
see fig.~\ref{Distributions}c.

The rest of the properties of a GRB's pulse can be derived on similarly
trivial grounds and with hardly more toil. During the GRB phase a CB
is still expanding inertially at a speed $\beta_s\,c$. It becomes transparent
when its radius is $R_{\rm tr}\!\sim\![3\,\sigma_{_{\rm T}}\,N_{_{\rm B}}/(4\pi)]^{1/2}$,
at an observer's time very close to that of Eq.~(\ref{ttrans}), for typical parameters.
One can simply count the number of ICS interactions of a CB's  electrons
with the glory, multiply by their energy, Eq.~(\ref{boosting}), and figure out
the isotropic-equivalent energy deduced by an observer at an angle $\theta$:
\begin{equation} 
E_\gamma^{\rm iso} \simeq 
{\delta^3\, L_{_{\rm SN}}\,N_{_{\rm CB}}\,\beta_s\over 6\, c}\,
                      \sqrt{\sigma_{_{\rm T}}\, N_{_{\rm B}}\over 4\, \pi}\sim
                      3.2\! \times\! 10^{53}\,{\rm erg},
\label{eiso} 
\end{equation} 
where the number is for our typical parameters, and agrees with observation.

\subsection{?` Is it Inverse Compton Scattering ...}

The $\gamma$ and $\delta$ dependance in Eqs.~(\ref{boosting},\ref{eiso})
is purely `kinematical', but
specific to ICS: it would be different for self-Compton or synchrotron radiation.
To verify that the $\gamma$ rays of a 
GRB are made by ICS, as proposed\cite{SD} by Shaviv and Dar,
we may look at the correlations\cite{corr1,corr2} between GRB observables. 

In the CB model, the $(\gamma,\delta,z)$ dependences of  
 the peak isotropic luminosity of a GRB, $L_p^{\rm iso}$;
its pulse rise-time, $t_{\rm rise}$; and
the lag-time between the peaks of a pulse at different energies, $t_{\rm lag}$; 
are also simply derived\cite{corr2} to be:
\begin{equation}
(1+z)^2\,L_p^{\rm iso}\propto \delta^4,\;\;\;
t_{\rm rise}\propto (1+z)/(\gamma\,\delta),\;\;\;
t_{\rm lag}\propto (1+z)^2/(\delta^2\,\gamma^2).
\label{brief}
\end{equation}
I have not specified the numerical coefficients in Eqs.~(\ref{brief}), which 
are explicit, as in  Eqs.~(\ref{boosting},\ref{eiso}). Of all the parameters and priors
in these expressions, the one explicitly varying by orders of magnitude by simply
changing the observer's angle is $\delta(\gamma,\theta)$, making it the prime putative
cause of case-by-case variability. For such a cause, Eqs.~(\ref{eiso}) and the first
of (\ref{brief}) imply that
$E_\gamma^{\rm iso}\!\propto\! [(1+z)^2\,L_p^{\rm iso}]^{3/4}$. This is tested in
fig.~\ref{f1}a. A most celebrated correlation is the $[E_p\,,E_\gamma^{\rm iso}]$ 
one, see Fig.~\ref{f1}b. It evolves from
$E_p\!\propto\![E_\gamma^{\rm iso}]^{1/3}$ for small $E_p$, to  
$E_p\!\propto\![E_\gamma^{\rm iso}]^{2/3}$ for large $E_p$.
This is because the angle subtended by a moving CB from its place of origin is
$\beta_s/\gamma$, comparable to the beaming aperture, $1/\gamma$, of the radiation
from a point on its surface. Integration over this surface implies that, for 
$\theta\!\ll\! 1/\gamma $, $\delta\!\propto\!\gamma$,
while in the opposite case $\delta$ varies independently.
The straight lines in fig.~\ref{f1}b are the central expectations of 
Eqs.~(\ref{boosting},\ref{eiso}), the data are fit to the predicted evolving power law.
The  predicted $[t_{\rm lag},L_p^{\rm iso}]$ and $[t_{\rm rise},L_p^{\rm iso}]$ correlations 
are tested in figs.~\ref{f1}c,d. The seal of authenticity of inverse Compton 
scattering ---by a quasi-point-like
electron beam--- is unmistakable in all of this, QED.

\subsection{... on a Glory's light ?}

The `target' photons subject to ICS by the CB's electrons have very specific
properties. Their number-density, 
$n_\gamma(r)\!\propto\!L_{_{\rm SN}}/r^2$, translates into the $\sim\!t^{-2}$
late-time dependence of the number of photons in a pulse since, once a CB
is transparent to radiation, ICS by its electrons simply `reads' the target-photon 
distribution. As a CB exits the wind-fed domain, the photons it
scatters are becoming more radial, so that $1\!+\!\cos\theta_i\!\to\!r^{-2}\!\propto\! t^{-2}$
in Eq.~(\ref{boosting}). For a semi-transparent wind material, which we have studied
in analytical approximations and via Montecarlo, this asymptotic behaviour is
reached fast and is approximately correct at all $t$. This means that the energies
of the scattered photons evolve with observer's time as $t^{-2}$: the
`$E\,t^2$ law' of Eq.~(\ref{shape}) and fig.~\ref{f1}e.

\subsection{The pulse shape and the spectrum}

The spectrum of a GRB, Eq.~(\ref{totdist}),  and the time-dependence of 
its pulses, Eq.~(\ref{shape}), describe the data well, and
are actually analytical approximations to the
results of ICS of an average CB on a typical glory. The spectrum of a
semitransparent glory has a `thermal bremsstrahlung' shape,
$dn_\gamma/dE_i\!\propto\!(T_i/E_i)^\alpha\,{\rm Exp}[-E_i/T_i]$, with $\alpha\!\sim\!1$
and $T_i\!\sim\! 1$ eV.
The first term in Eq.~(\ref{totdist}) is this same spectrum, boosted by ICS
as in Eq.~(\ref{boosting}), by electrons comoving with the CB, $E_e\!=\!\gamma\,m_e\,c^2$.
The second term is due to ICS by `knock-on' electrons (generated while the
CB is not yet collisionless) and electrons `Fermi-accelerated' by the CB's 
turbulent magnetic fields. They both have a spectrum $dn_e/dE_e\!\propto\!E_e^{-\beta}$, 
with $\beta\!\sim\!2$ to 2.2. They are a small fraction of the CB's electrons,
reflected in the parameter $b$, which we cannot predict. The temporal shape of a
pulse has an exponential rise due to the CB and the windy material becoming transparent
at a time  $\sim\!t_{\rm tr}^{\rm w}$, see Eq.~(\ref{ttrans}), the width of pulse
(in $\gamma$ rays) is a few $t_{\rm tr}^{\rm w}$, the subsequent decay is 
$\propto\!t^{-2}$. The time-energy correlations obey the `$E\,t^2$ law'.
All as observed.

\subsection{Polarization}

A tell-tale signature of ICS is the high degree of polarization.
For a pointlike CB the prediction\cite{SD} is 
$\Pi\!\approx\!2\,\theta^2\,\gamma^2/(1+\theta^4\,\gamma^4)$,
 peaking at 100\% at $\theta=1/\gamma$, 
the most probable $\theta$, corresponding to $90^{\rm o}$ in the CB's system. 
For an expanding CB, $\Pi$
is a little smaller. For SR, which dominates
the AGs at sufficiently late times, the expectation is $\Pi\!\approx\!0$.
The $\gamma$-ray polarization has been measured, with considerable toil, 
in 4 GRBs. It is always compatible, within very large errors, with 100\%.
The situation is unresolved\cite{Pol}. I shall not discuss it.

\section{Detailed Swift light curves and hardness ratios}

Swift has abundantly filled its goal to provide X-ray, UV and optical data
starting briefly after the detection of a GRB:
compare the Swift result of fig.~8a to the pre-Swift data in fig.~8b.
In the CB-model description of the data in figs.~8,9,10,
the abruptly falling signal is the tail of one or several
$\gamma$-ray pulses or X-ray flares, produced by ICS and jointly described 
by Eqs.~(\ref{totdist},\ref{shape}). The following `afterglow', its less pronounced
decay and subsequent achromatic `bend' are due to the CBs' synchrotron
radiation, described by Eq.~(\ref{SRFlux}). Thanks to the quality of SWIFT data one 
can proceed to test these CB-model predictions in detail.

The two prompt optical `humps' of GRB 060206 in fig.~\ref{postSwift1}a are the ICS
low-energy counterparts of its two late X-ray flares of fig.~\ref{postSwift1}b,
simultaneoulsly fit by Eqs.~(\ref{totdist},\ref{shape}).
Swift provides a rough  measure of a GRB's spectrum: the {\it hardness
ratio} of count rates in the [1.5-10] keV and [0.3-1.5] keV intervals. Given the
case-by-case parameters of a CB-model fit to the [0.3-10] keV
light curve, one can estimate the corresponding hardness ratio\cite{DDDDecline}.
This is done in figs.~\ref{postSwift1}c,d and \ref{postSwift2}a,c for GRB 060904
and XRF 060218, respectively. This last XFR is observed at a `large' angle,
$\theta\!\sim\!5$ mrad and a correspondingly small $\delta_0$,
its single X-ray pulse is, in accordance with Eq.~(\ref{ttrans}), relatively wide.
The optical and UV counterparts of the X-ray pulse are clearly
visible as the `humps' in the optical data of fig.~\ref{postSwift2}b. 
Given the `$E\,t^2$ law' of Section 7.3, the pulse peak times at different frequencies
are simply related: $t_{\rm peak}\!\propto\!E^{-1/2}$. The prediction, an example
of the ubiquitous $1/r^2$ law of 3-D physics, is tested
in fig.~\ref{f1}e. The peak fluxes at all frequencies are also related as dictated\cite{XSwift}
by Eq.~(\ref{totdist}). The adequacy of the CB model over many decades in flux and time is 
exemplified by the X-ray light curve of GRB 061121 in fig.~\ref{postSwift2}d.

The predictions for the peak $\gamma$-ray energy of Eq.~(\ref{boosting}),
its distribution as in fig.~\ref{Distributions}c,
the GRB spectrum of Eq.~(\ref{totdist}),
and the correlations of figs.~\ref{f1}a-d are clear proof that ICS is the prompt
GRB mechanism. The test of the $E\,t^2$ law in fig.~\ref{f1}e corroborates that
the `target light' becomes increasingly radially directed with distance:
{\it Inverse Compton  Scattering on a `glory's light' by the electrons in CBs is responsible for the
$\gamma$-ray pulses of a GRB and their sister X-ray flares and optical humps.}
The properties of the subsequent {\it synchrotron-dominated afterglows} are also
in accordance with the CB model.

\section{The GRB/SN association in the CB model}

We have gathered very considerable evidence that the LFs and viewing angles
of {\it observed} GRBs are $\gamma_0\!=\!{\cal{O}}(10^3)$ and $\theta\!=\!{\cal{O}}(1)$
mrad. The fraction of GRBs beamed towards us is $\sim\!\theta^2\!=\!{\cal{O}}(10^{-6})$.
The number of such observed GRBs (with a hypothetical $4\,\pi$ coverage) is a few a day.
The same coverage would result in the observation of a few million core-collapse
SN per day, in the visible Universe. These numbers are compatible with the extreme
conclusion that $all$ these SNe emit GRBs, but the estimates and errors are sufficient
to accommodate a one order of magnitude smaller fraction, which would be compatible
with most Type Ib/c emitting (long) GRBs. 

\section{Short Hard $\gamma$-ray Bursts (SHBs)}

SHBs share with (long) GRBs the properties not reflected in their name.
A good fraction of SHBs have `canonical' X-ray light curves. 
The origin of SHBs is not well established, 
in contrast to that of 
GRBs and XRFs. Clues to the origin and 
production mechanism of SHBs are provided by their similarity to long GRBs. 
The X-ray light curves of some well-sampled SHBs are `canonical'.
The similarities suggest common mechanisms generating the GRB and SHB 
radiations. This is expected in the CB model, wherein both burst types are 
produced by highly relativistic, narrowly collimated, bipolar jets of CBs,
ejected in stellar processes\cite{SD}. The
mechanisms for their prompt and AG emissions (ICS and synchrotron)
coincide with the ones of GRBs. 
The `engine' is different; it is a core-collapse supernova for GRBs and XRFs, in SHBs it may be
a merger (of two neutron stars or a neutron star and a black hole),
the result of mass accretion episodes on compact objects in 
close binaries (e.g.~microquasars), or phase transitions in increasingly 
compactified stars (neutron stars, hyper-stars or quark stars), induced by accretion,
cooling, or angular-momentum loss.

In the CB model, the `master formulae'
describing  prompt and afterglow 
emissions in long GRBs are directly applicable to SHBs, provided the 
parameters of the CBs, of the glory, and of the circumburst environment, are 
replaced by those adequate for SHBs. This results in a good description of the data\cite{DDDSHB}.

\section{Cosmic Rays in the CB model}

In the CB model, CRs are as simple to understand as GRBs. 
If relativistic CBs are indeed ejected by a good fraction of core-collapse SNe, it
is inevitable to ask what they do as they travel in the ISM. The answer 
is that they make CRs with the observed properties, simply by interacting
with the constituents of the ISM, previously ionized by the $\gamma$ rays of 
the accompanying GRB. Early in their voyage, 
CBs act as {\it Compton relativistic rackets}, in boosting
a glory's photon to $\gamma$-ray status. Analogously, all along their
trajectories, CBs act as {\it Lorentz
relativistic rackets,} in boosting an ISM nucleus or electron to CR status.
Once again, the necessary input is two-fold. On the one hand, there are the properties 
of CBs: the average number of significant GRB pulses (or CBs)
per jet (5),  the $\gamma_0$ distribution
of fig.~\ref{Distributions}a, and the $N_{_{\rm B}}\!\sim\! 10^{50}$ estimate.  On the
other hand, there are a few `priors', items of information independent of the
CR properties: the rate of
core-collapse SNe, the relative abundances, $n_{_{A}}$ (of the elements of atomic
number $A$) in the ISM, and the properties of Galactic magnetic fields.

We shall see that the CB-model predictions for the normalization of CR spectra
are correct to within a factor of ${\cal{O}}(3)$, while the ratios between
elements are correct within errors. In figs.~1, 3 and 4a, the predictions 
have been made to adjust the data, not reflecting the
common overall normalization uncertainty. 

\subsection{Relativistic rackets: The knees}

Our  simplest result concerns the `knees' of the all-particle spectrum in fig.~\ref{AllPart}a
and of the main individual elements in fig.~\ref{KASKADE}.  
The essence of  their understanding is kinematical and trivial.
In an {\it elastic} interaction of a CB at rest 
with ISM electrons or ions of LF $\gamma$,  
the light recoiling particles (of mass $m\!\approx\!A\,m_p$) retain their incoming energy.
Viewed in the ISM rest system, they 
have, for large $\gamma$, a flat spectrum extending 
up to $E\simeq 2\,\gamma^2\,m\,c^2$ [this is recognizable as the forward, 
massive-particle, $z\!=\!0$,
 analog of Eq.~(\ref{boosting})]. Thus, a moving CB
is a gorgeous {\it Lorentz-boost accelerator:}
 the particles it elastically scatters
reach up to, 
for $\gamma=\gamma_0\!=\!(1\,{\rm to}\,1.5)\!\times\!10^3$,
an $A$-dependent {\bf knee}  energy
$E_{{\rm knee}}(A)\!\sim\! (2\,{\rm to}\,4)\!\times\! 10^{15}\,A\,{\rm eV.}$ 
If this trivial process is the main accelerator of CRs, there must be
a feature in the CR spectra: endpoints at $E_{{\rm knee}}(A)$. 
The arrows in fig.~\ref{KASKADE}
show that the H and He data are compatible with this prediction. So does
the second knee of fig.~\ref{AllPart}a, the predicted Fe knee. 
The CR flux above the H knee, to which we shall return, is $\sim\!10^{-15}$ of the
total.

\subsection{The spectra below the knee}

The `elastic' scattering we just described is  dominant below
the knees. To compute the resulting spectrum, we assume
that the ISM particles a CB intercepts, trapped in its magnetic mesh,  reexit
it by diffusion, isotropically in the CB's system, and with the same `confinement' law, 
Eq.~(\ref{confinement}), as in the Galaxy
(the opposite assumption, that they are immediately elastically scattered,
yields a slightly different spectral index). The CB deceleration law is 
Eq.~(\ref{distance}), its radius evolves as in Eq.~(\ref{Radius}). 
A modest amount of algebra gives a simple result (DD06), which, for
$\gamma\!>\! 2$ and to a good approximation, reads\footnote{I am keeping
factors of $A/Z$ for kicks. Numerically, they are irrelevant: the theory
and data are not so precise, and $(A/Z)^{0.6}$ is 1 for H, 
1.6 for Fe.}:
\begin{equation}
{dF_{\rm elast}\over d\gamma_{_{\rm CR}}} \propto  n_{_A}
\left({A\over Z}\right)^{\beta_{\rm conf}}
\int_1^{\gamma_0}{d\gamma\over \gamma^{7/3}}
\,
\int_{\rm max[\gamma,\gamma_{_{\rm CR}}/(2\,\gamma)]}
^{\rm min[\gamma_0,2\,\gamma\,\gamma_{_{\rm CR}}]}
{d\gamma_{\rm co}\over \gamma_{\rm co}^{4}}\; ,
\label{NRFlux}
\end{equation}
where $\gamma_{_{\rm CR}}$ is the CR's
LF, and $\beta_{\rm conf}$ is the same confinement index as in
Eq.~(\ref{confinement}). The flux
$dF_{\rm elast}/d\gamma_{_{\rm CR}}$
depends on the priors $n_{_A}$, ${\beta_{\rm conf}}$, 
and $\gamma_0$, but not on any
parameter specific to the mechanism of CR acceleration. 
But for the normalization, this flux is $A$-independent. In the large
range  in which it is roughly a power
law, $dF_{\rm elast}/d\gamma_{_{\rm CR}}\!\propto\![\gamma_{_{\rm CR}}]^{-\beta_{_{\rm CR}}}$,
with $\beta_{_{\rm CR}}\!=\!13/6\!\approx\!2.17$.

The H, He and Fe fluxes of fig.~\ref{KASKADE} are given by Eq.~(\ref{NRFlux}),
modified by the Galactic confinement $\tau$-dependence of Eq.~(\ref{confinement}), with
$\beta_{\rm conf}\!=\!0.6$. The fastest-dropping curve in fig.~\ref{KASKADE}a
corresponds to a fixed $\gamma_0$. The other two curves are for the
$\gamma_0$ distribution of fig.~\ref{Distributions}a, and one twice as wide.
The low-energy data of fig.~\ref{AllPart}b are also described by Eq.~(\ref{NRFlux}), 
whose shape in this region (the `hip', also visible in fig.~\ref{KASKADE}c for
Fe) is insensitive to $\gamma_0$ and, thus, parameter-independent. 

\subsection{The relative abundances}

It is customary to discuss the composition of CRs at a fixed
energy $E_{_A}=1$ TeV.
This energy is relativistic, below the 
corresponding knees for all $A$, and in the domain wherein the 
fluxes are dominantly elastic and well approximated by a power law 
of index $\beta_{\rm th}\!=\!\beta_{\rm elast}\!+\!\beta_{\rm conf}\!\simeq\!2.77$.
Expressed in terms of energy ($E_{_A}\!\propto\! A\,\gamma$), and modified 
by confinement as in Eq.~(\ref{confinement}), Eq.~(\ref{NRFlux}) becomes:
\begin{equation}
{dF_{\rm obs}/ dE_{_A}}\propto \bar{n}_{_A}\,A^{\beta_{\rm th}-1}
\,E_{_A}^{-\beta_{\rm th}},\;\;\;
X_{_{\rm CR}}(A)=(\bar{n}_{_A}/ \bar{n}_p)\,A^{1.77},
\label{compo}
\end{equation}
with $\bar{n}_{_A}$ an average ISM abundance and $X_{_{\rm CR}}(A)$
the CR abundances relative to H, at fixed $E$.
The results, for input $\bar{n}_{_A}$'s in the `superbubbles' wherein most
SNe occur, are shown in Fig.~\ref{abundances}b. In these regions, the
abundances are a factor
$\sim\!3$ more `metallic' than solar (a `metal' is anything with $Z\!>\!2$).
Eq.~(\ref{compo})
snugly reproduces the large enhancements of the heavy-CR relative abundances,
in comparison with solar or superbubble abundances
(e.g.~$A^{1.77}\!=\! 1242$ for Fe). The essence of this result is deceptively simple:
in the kinematics of the collision of a heavy object (a CB) and a light one (the ISM nucleus),
their mass ratio ($N_{_{\rm B}}/A\sim\!\infty$!) is irrelevant.

\subsection{Above the knees}

We discussed around Eq.~(\ref{B}) the generation of
turbulently-moving magnetic fields (MFs) in the merger of two plasmas.
Charged particles interacting with these fields tend
to gain energy: a relativistic-injection, `Fermi' acceleration process, for which
numerical analyses\cite{Frede} result in a spectrum $dN/dE\!\propto\!E^{-2.2}$,
$p\!\sim\!2.2$. For the ISM/CB merger, we (DD04) approximate the spectrum of 
particles accelerated within a CB, in its rest system, as:
 \begin{equation}
{dN/d\gamma_{_A}}\propto\gamma_{_A}^{-2.2}\,
\Theta(\gamma_{_A}\!-\!\gamma)\,\Theta[\gamma_{\rm max}(\gamma)\!-\!\gamma_{_A}],
\,
\gamma_{\rm max}
 \simeq  10^5\;\gamma_0^{2/3}\;(Z/A)\; \gamma^{1/3},
\label{gammaA}
\end{equation}
The first  $\Theta$ function reflects the
fact that it is much more likely for the light particles to gain than to lose
energy in their elastic collisions with the heavy `particles' (the CB's
turbulent MF domains). The second $\Theta$  is 
the Larmor cutoff implied by the finite radius and
MF of a CB, with a numerical value given for the typical
adopted parameters. But for the small dependence of $\gamma_{\rm max}$
on the nuclear identity (the factor $Z/A$), the spectrum of Eq.~(\ref{gammaA}) is
universal. 
Boosted by the CB's motion, an accelerated and re-emitted particle may
reach a Larmor-limited
$\gamma_{_{\rm CR}}{[\rm max]}\!=\!2\,\gamma\,\gamma_{\rm max}$,
a bit larger, for $\gamma\!=\!\gamma_0\!\sim\!1.5\!\times\!10^3$, than the 
corresponding GZK cutoffs. Our model has a {\it single source}, CBs,
for the acceleration of CRs from the lowest to the largest observed energies.

\begin{figure}
\begin{center}
\epsfig{file=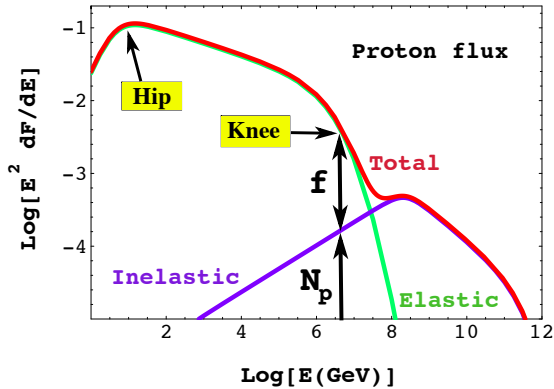, width=7.5cm,angle=-0}
\end{center}
\vspace{-.4cm}
\caption{Contributions to the $E^2$-weighed proton source spectrum.}
\vspace {-.4cm}
\label{DDelinel}
\end{figure}

The calculation of the `elastic' spectrum of Eq.~(\ref{NRFlux}) was done for
the bulk of the ISM particles entering a CB, assuming that they
were not significantly Fermi-accelerated, but kept their
incoming energy, i.e.~$dN/d\gamma_{_A}\!\propto\!\delta(\gamma_{_A}-\gamma)$.
The `inelastic' spectrum, with $dN/d\gamma_{_A}$ as in Eq.~(\ref{NRFlux}),
yields an equally simple result. The two $E^2$-weighed
spectra are shown (for H) in fig.~\ref{DDelinel}.
The inelastic contribution is a tiny fraction of the total, and
is negligible below the knee, a point at which
we may compare the ratio of fluxes, $f$, the only parameter freely fit to
the CR data. The boost of ISM particles by a CB and their acceleration within it
are mass-independent, so that the ratio $f$ is universal.

\begin{figure}
\vspace {-1.8cm}
\begin{center}
\epsfig{file=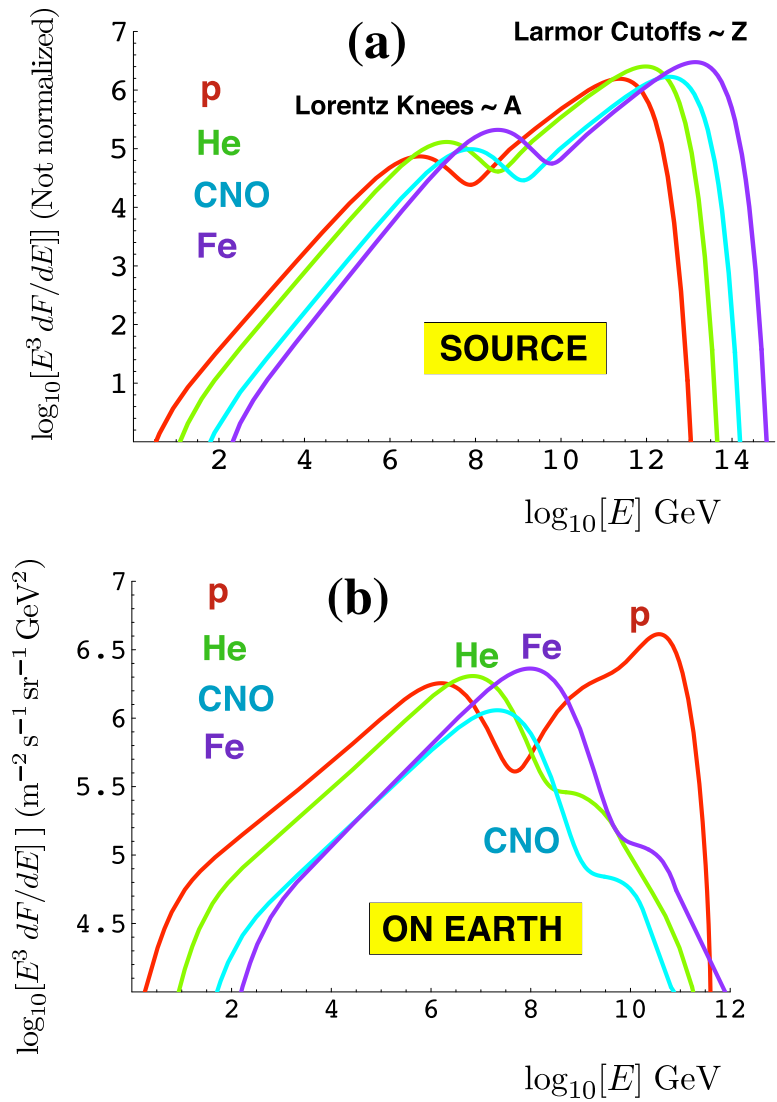, width=10.cm,angle=-0}
\end{center}
\vspace{-.5cm}
\caption{Predicted spectra for the abundant elements
and groups. The vertical scales are $E^3\,dF/dE$. (a): The source
spectra, with a common arbitrary normalization. (b): The normalized CR spectra 
at the location of the Earth. Notice that the horizontal and
vertical scales are different in (a) and (b).}
\vspace{-.8cm}
\label{Groups}
\end{figure}

The $E^3$-weighed {\it source} spectra for the main elements 
are shown in fig.~\ref{Groups}a. They are very different from the {\it observed}
spectra of fig.~\ref{Groups}b, for many reasons. Below the ankle(s) the slopes
differ due to Galactic confinement, see Eq.~(\ref{confinement}).
Above the ankles the flux from Galactic sources is strongly suppressed: we 
would see their straight-moving CRs only for CB jets pointing to us.
The CRs above the ankle are mainly extragalactic in origin, and they also cross 
the Galaxy just once. Extragalactic CRs of $A\!>\!1$ are efficiently photo-dissociated
by the cosmic infrared light.
Extragalactic CRs are GZK-cutoff. All this can be modeled
with patience and fair confidence. Below the ankle
extragalactic CRs may have to fight the CR `wind' of our Galaxy,
analogous to that of the Sun. We have covered our lack of information on
this subject by choosing two extreme possibilities (DD06), resulting in the two
curves of figs.~\ref{AllPart}a and \ref{highE}a,b.

In fig.~\ref{highE}b I have converted the results of fig.~\ref{Groups}b
into a prediction for $\langle{\rm Ln}\,A(E)\rangle$. The flux at the second
knee is dominated by Galactic Fe at its knee. Thereafter this flux
decreases abruptly to let extragalactic H dominate all the way from
the ankle to the nominal position of the proton's GZK cutoff. Above
that point the high-energy tail of Galactic Fe may dominate again.

\subsection{The CR luminosity, and the overall normalization of the CR flux}

The rate of core-collapse SNe in our Galaxy is $R_{_{\rm SN}}\!\sim\!2$ per
century. In the CB model, we contend that $\sim\!50$\% of the energy of CRs
is transfered to the magnetic fields {\it they} generate\cite{DDMag}. If all
core-collapse SNe emit CBs, the Galactic CR luminosity should be
$L_{_{\rm CR}}\!\sim\!R_{_{\rm SN}}\,E_{\rm jets}/2\!\sim\!4.7\!\times\!10^{41}$ 
erg s$^{-1}$, with $E_{\rm jets}$ as in Eq.~(\ref{Ejets}). This  is
3 times larger than the rhs of Eq.~(\ref{CRlum}). 
The `discrepancy' is not worrisome. A smaller fraction of SNe may generate
high-$\gamma_0$ CBs.  The rhs of
Eq.~(\ref{CRlum}) is for `standard' CRs, but the
confinement volume
and time of the CB model are non-standard by factors of $\sim\!10$.
All inputs are fairly uncertain.

The calculation of the
flux above the ankle is lengthy but  straightforward. But for the GZK effect, its
shape is that of the source H flux, since protons at that energy
should escape other galaxies directly, and
enter ours unhindered.
Its normalization per SN is fixed.
The SN rate per unit volume is measured in the local Universe.
The overall flux is the result of the integration over redshift of the flux 
from past SNe. The integrand must be  properly red-shifted and
weighed with the star-formation rate as a function of $z$ (SN progenitors have 
short lives on Hubble-time scales).
The integration in $z$ is an integration over look-back time,
as opposed to distance, since CRs do not travel straight. The error in the result
is hard to estimate,  its central value is within a factor of 2 of the 
observations (DD06). This explains the coincidence that the ankle is 
the escape energy of protons from the Galaxy {\it and} the place where
the extragalactic flux --not enhanced by confinement and thus less steep--
begins to dominate.

\subsection{CR diffusion, CR electrons, and the $\gamma$ background radiation}

In the standard paradigm CRs are accelerated by
the nonrelativistic ejecta of SNe. SNe occur mainly in the central realms
of the Galaxy, so that CRs must diffuse to arrive to our location. A directional
asymmetry  is predicted, and not observed\cite{Plaga}.
For CR electrons the problem is even more severe: their cooling time in the
Galaxy's light, and magnetic fields, is so short that they should have lost
all their energy on their way here.

Our source distribution is totally non-standard, CBs generate CRs all
along their many-kpc-long trajectories, see Eq.~(\ref{distance}) and fig.~\ref{figCB}c.
Depending on the ISM density profile they encounter, CBs may travel for
up to tens of kpc before they become nonrelativistic. It takes some 
$6\!\times\! 10^4$ years
to travel 20 kpc at $v\!\sim\!c$. If a Galactic SN occurs every 50 years, and
emits an average of 10 CBs, there are currently several thousand CBs 
in  the Galaxy and its halo. This is a very diffuse CR source, satisfactory
in view of the previous paragraph. We have not yet studied 
the CR source-distribution and diffusion in detail.

Below their knee at $2\,\gamma_0^2\,m_e\,c^2\!\sim\!2.3$ TeV, the source
spectrum of CB-accelerated electrons has the same index as that
of nuclei: $dN/dE_e\!\propto\!E^{-\beta_{_{\rm CR}}}$,  $\beta_{_{\rm CR}}\!\approx\!2.17$.
The predicted spectrum\cite{GBR}, steepened by radiative energy loses, has an index
$\beta_e\!=\!\beta_{_{\rm CR}}+1\!\approx\!3.17$. Its observed slope\cite{Slope} is 
$3.2 \pm 0.1$ above $E_e\!\sim\!10$ GeV, an energy below which other losses should
dominate (DD06).

The Gamma Background Radiation (GBR), measured by EGRET from a few MeV
to $\sim\!10^5$ MeV, was argued to be dominantly of cosmological origin, in directions
above the disk of the Galaxy and away from its bulge\cite{EGRET}. A more careful analysis
reveals a significant correlation of its intensity with our position relative
to the Galactic centre\cite{GBR}. The CB-model reproduces this correlation, provided
a good fraction of the GBR is generated by CR electrons at high galactic 
latitudes, as they cool radiatively by the very same process that steepens their
spectrum. The predicted index of the radiated GBR photons is 
$\beta_\gamma\!=\!(\beta_e\!-\!1)/2\!=\!2.08$. The observed one\cite{EGRET}
 is $2.10\!\pm\!0.03$.

\section{If CBs are so pervasive, why are they not readily observed?}

The answer is simple. The CBs of SNe
are tiny astrophysical objects: their typical mass is 
half of the mass of Mercury.
Their energy flux at all frequencies is $\propto\!\delta^3$,
large only when their
Lorentz factors are large. But then, the radiation is also
extraordinarily collimated, it can only be seen nearly on-axis.
Typically, observed SNe are too far to {\it photograph} their CBs
with sufficient resolution.

Only in two SN explosions that took place close enough, the
CBs were in practice observable.  One case
was SN1987A, located in the LMC,
whose approaching and receding CBs were
photographed, see fig.~\ref{try2}e,f. 
The other case was SN2003dh, associated with GRB030329,
at $z=0.1685$. In the CB model interpretation,
its two approaching CBs were first `seen', and fit,
as the two-peak $\gamma$-ray light curve of fig.~\ref{try}b and the two-shoulder
AG of fig.~\ref{try2}a,b. This allowed us 
to estimate the time-varying angle of their superluminal
motion in the sky\cite{SLum030329}. Two sources or `components'
were indeed clearly seen in radio observations
at a certain date, coincident
with an optical AG rebrightening. We claim
that the data agree with our expectations\footnote{The
size of a CB is small enough to expect its radio image to
scintillate, arguably more than observed\cite{Taylor}.
Admittedly, we only realized a posteriori that the ISM electrons a CB
scatters, synchrotron-radiating in the ambient magnetic field, would
significantly contribute at radio frequencies, somewhat blurring the 
CBs' radio image\cite{SLum030329}.
},  including 
the predicted inter-CB separation\cite{SLum030329} of fig.~\ref{try2}d.
The observers claimed the contrary, though the 
evidence for the weaker `second component' is $>20\sigma$.
They report\cite{Taylor} that this component is 
`not expected in the standard model'.
The unpublished and no-doubt spectacular-discovery 
picture of the two superluminally moving sources would have
been worth a thousand words... in support of the CB model.

\section{Other CR sources}

We have defended the simplistic view that CBs from SNe is all one needs to 
generate CRs at all energies. The recent data of Auger\cite{Auger}
show a significant correlation between the arrival directions of ultra-high energy 
CRs (UHECRs) and Active Galactic Nuclei (AGNs) located within a distance of 
75 Mpc ($z\!\sim\!0.02$), 
comparable to the GZK `horizon' at the observed energies, $E\!>\! 56$ EeV.
As the authors discuss, this
does not mean that AGNs are the actual sources\cite{Rus}, for AGNs are 
themselves correlated
with  matter  and with active regions of enhanced stellar birth and death.
More work on correlations is no doubt in progress. A search for correlations
with GRBs is less hopeful, for the fraction of them observed from within
$z\!\sim\!0.02$ is negligible, and the observed `correlated' CRs may
have been bent by magnetic fields up to a few degrees, implying a 
possible delay of millenia between the arrival times of $\gamma$'s and CRs.

Observations of AGNs are an ingredient of the `inspiration' of the
CB model, as we have discussed in connection with Pictor A, see fig.~\ref{Pictor}a,b.
Naturally, we have estimated their contribution to the UHECR proton flux,
concluding that they may constitute at most 1 to 10\% of the flux generated by
extragalactic SNe (DD04). The estimate is for the energy-integrated
flux; in applying it to UHECRs we assumed the same energy dependence
for the flux generated --by the same mechanisms-- by the CBs of SNe
and AGNs (a small difference of spectral index implies an enormous
uncertainty). In a subsequent study\cite{DDGBR2} of the CR electron flux, 
assumed to be in a fixed proportion to the proton flux, we used more recent
inputs, and simplified and modified our upper limit to $40$\%. 
But we forgot\footnote{In these
days of large experimental collaborations percolated by theorists,
rumours herald publications. It might have been possible to turn
this comment into a renewed\cite{Rus}
timely `prediction', prior to the Auger announcement.} to extract the
putative consequences for the AGN contribution to UHECR protons!

\section{Discussion and conclusion}

We do not have a solid understanding of accretion onto black holes or neutron
stars. But such processes are observed to result in the ejection of relativistic
and highly collimated jets. We assumed that a similar process takes place
as a stellar core collapses, leading to a supernova event. We posited that the SN's
relativistic ejecta --two jets of cannonballs-- are the sources of GRBs and CRs. The 
association between SNe and (long) GRBs is now established. We argued 
that the electrons in a CB, by inverse Compton scattering on the illuminated
surroundings of the exploding star, generate the  $\gamma$ rays, X-rays, 
UV and optical light of the `prompt' phase of a GRB. The ensuing results
are the simplest and most predictive, they are a firm `theory'. 
In this paper I have, however, followed the historical development, in which the CB
parameters were first extracted from the observations of the afterglow of GRBs. 
This involves a `model', a set of arguable but simple hypothesis leading to 
the prediction of the properties of the AG --dominated by synchrotron
radiation by the ISM electrons that a CB intercepts-- as a function of frequency 
and time. In the historical order 
the `prompt' results for GRBs are predictions of the theory. Some results for
Cosmic Rays --the ISM particles that CBs scatter in their journey-- are also 
 `theory', others can be viewed as further tests of the `model'.

The results for GRB afterglows may be based on a simplified model, but they work with 
no exception all the way from radio to X-rays (DDD02,03). In particular, they describe 
correctly GRB 980425, located at a redshift two orders of magnitude closer than average.
Its associated SN is the one we `transported' to conclude  --thanks to the  
reliability of our AG model-- that core-collapse SNe generate long GRBs (DDD02). 
The X-ray light curve of GRB 980425 and a few others, with extremely scarce data,
was described with the `canonical' properties later observed in detail in many
SWIFT-era GRBs. It is not recognized that the two CBs of GRB 030329 were seen,
or that their separation in the sky was the predicted `hyper-luminal' one. In view 
of the overall success of the
CB model, this is a durable hurdle: GRBs so close and luminous are very rare.

The accuracy of the predictions for the prompt phase of GRBs amazes even the 
CB-model's proponents. The typical values and the correlations between
the $\gamma$-ray prompt observables leave little doubt that the production
mechanism is inverse Compton scattering on `ambient' light of $\sim\!1$ eV
energy. The approximate scaling law $E\,dN_\gamma/dE dt\!\propto\!F(E\,t^2)$
--spectacularly  confirmed in the case of XRF 060218--
demonstrates that the light is that of a `glory': the early SN light scattered by the
`windy' pre-SN ejecta. A GRB spectrum that works even better\cite{HessLady}
than the phenomenological `Band' expression is also predicted.
The flux and its spectral evolution during the prompt and rapid-decline phases
are the expected ones, as we tested in minute detail with SWIFT data.

In the internal-external fireball model of GRBs, highly relativistic thin
conical shells of $e^+e^-$ pairs, sprinkled with a finely tuned baryon `load',
collide with each other generating a shock that accelerates their constituents
and creates magnetic fields. Each collision of two shells 
produces a GRB pulse by synchrotron radiation. The ensemble
of shells collides with the ISM to produce the AG by the same
mechanism. The energy available to produce the GRB pulse --as 
two shells moving in the {\it same} direction collide-- is more than one order
of magnitude smaller than that of the merged shells as they collide with 
the ISM at rest. The ratio of observed GRB and AG energies is more
than one order of magnitude, but in the opposite direction. 
This `energy crisis' in the comparison 
of bolometric prompt and AG fluences\cite{Pir} is not resolved.
Moreover, the GRB spectrum cannot be accommodated on grounds of synchrotron 
radiation\cite{GCL}, the `standard' prompt mechanism. 
The SWIFT-era observations also
pose decisive problems to the standard model, 
whose microphysics\cite{Petal},
reliance on shocks\cite{Ketal} 
and correlations based on the jet-opening angle\cite{Setal}
have to be abandoned, according to the cited authors.

In spite of the above, the defenders of the fireball model are not
discouraged. Their attitude
towards the CB model, whose observational support is so remarkable,
 is not equally supportive. This may be due to cultural differences.
Particle physicists believe that complex phenomena may have
particularly simple explanations. They thrive on challenging their standard views. 
Doubting or abandoning a previous consensus in astrophysics is less easy.

The CB-model description of Cosmic Rays is also simplistic: there is only one
source of (non-solar) CRs at all energies, and only one parameter to be fit.
 The model has a certain inevitability: if CBs with the properties deduced
from GRB physics are a reality, what do they do as they scatter the 
particles of the interstellar medium? We have argued that they
transmogrify them into CRs with all of their observed properties. The mechanism
is entirely 
analogous to the ICS responsible for a GRB's prompt radiation. Suffice it to substitute
the CB's electron, plus the ambient photon, by a moving-CB's inner magnetic field, 
plus an ambient nucleus or electron. 

After a century of CR measurements, the CB-model
results lack the glamour of predictions. Yet,
the expectations for the knee energies, and for the relative abundances of CRs, are 
`kinematical', simple, and  verified. They constitute evidence,
in my opinion, that the underlying model is basically correct.
The prediction for the shape of the spectra: the low energy hips, the large energy stretch
very well described by a power-law of (source) index $\beta_s\!=\!13/6$, and the
steepening at the knees, are also verified. The index $\beta_s$ is measured well enough
for the adequacy of the prediction  to be sensitive to the details of the
underlying model, such as the form of the function $R(\gamma)$ in Eq.~(\ref{Radius}).
I cannot claim that the fact that the prediction is right on the mark is much more
than a consistency test, for the physics underlying this aspect of the problem may be
terrifyingly complex. The same CR source --cannonballs from supernovae, this time
extragalactic-- satisfactorily describes the CR data above the ankle. Finally, the
properties of CR electrons, and of the high-latitude Gamma `Background' Radiation,
are also correctly reproduced.

Most CR scholars agree with the `standard' paradigm
that the flux well below the knee is produced by the
acceleration of the ISM in the frontal shocks of the nonrelativistic ejecta of SNe.
In spite of recent observations of  large magnetic fields\cite{Yasu}
in collisions of SN shells and molecular clouds,
nobody has been able to argue convincingly that this process can accelerate
particles up to the (modest) energy of the knee, {\it and} to show
that the number and efficiency of the putative sources suffices to generate the
observed CR luminosity (to my satisfaction, I  add, to make these statements
indisputable). 
From this point on, there is no `standard' consensus on the origin of CRs, e.g.,~of the
highest-energy ones. In this sense, the CB model is regarded as
yet another model, which it is. After all, we are only saying that CRs are accelerated
by the jetted relativistic ejecta of SNe, as opposed to the quasi-spherical, non-relativistic
ones. Yet, the CB model is also rejected by the CR experts, sometimes even in 
print\cite{Hillas}, though it survives the critique\cite{HResponse}. But, concerning
CRs, the model does not trigger the same
indignant wrath as in the GRB realm. 

I have shown that the problem of GRBs is convincingly --i.e.~predictively--
solved and that, on the same simple basis, all properties of CRs can be easily derived.
Only an overwhelmed minority recognizes these facts, in
contradiction with Popper's and Ockham's teachings. I would conclude with
a dictum attributed to Lev Landau: 
{\it `In astrophysics, theories never die, only people do.'}

\vskip .4cm
\noindent
{\bf Acknowledgement}

\noindent
I thank Shlomo Dado, Arnon Dar and Rainer Plaga for our
collaboration.

\end{document}